\documentclass[10pt,aps,showpacs,nofootinbib,prd,aps,epsf,floats,
               amsmath,amssymb,amsfonts,axodraw,twocolumn]{revtex4}
\usepackage{amsmath, amssymb}
\newcommand{\mathsym}[1]{{}}

\usepackage{graphicx}
\usepackage{amsmath}
\usepackage{amssymb}
\usepackage{bm}
\newcommand{\rem}[1]{}
\newsavebox{\PSLASH}
 \sbox{\PSLASH}{$p$\hspace{-1.8mm}/}
 
\renewcommand{\theequation}{\thesection.\arabic{equation}}
\newcounter{saveeqn}
\newcommand{\add}{\addtocounter{equation}{1}}
\newcommand{\alpheqn}{\setcounter{saveeqn}{\value{equation}}%
\setcounter{equation}{0}%
\renewcommand{\theequation}{\mbox{\thesection.\arabic{saveeqn}{\alph{equation}}}}}
\newcommand{\reseteqn}{\setcounter{equation}{\value{saveeqn}}%
\renewcommand{\theequation}{\thesection.\arabic{equation}}}

 \newsavebox{\notrightarrow}
 \sbox{\notrightarrow}{$\to$\hspace{-4mm}/}
 
 \newsavebox{\PARTIALSLASH}
 \sbox{\PARTIALSLASH}{$\partial$\hspace{-1.6mm}/}
 
 \newsavebox{\ASLASH}
 \sbox{\ASLASH}{$A$\hspace{-2.1mm}/}
 
 \newsavebox{\KSLASH}
 \sbox{\KSLASH}{$k$\hspace{-1.8mm}/}
 
 \newsavebox{\LSLASH}
 \sbox{\LSLASH}{$\ell$\hspace{-1.8mm}/}
 
 \newsavebox{\QSLASH}
 \sbox{\QSLASH}{$q$\hspace{-1.8mm}/}
 
 \newsavebox{\DSLASH}
 \sbox{\DSLASH}{$D$\hspace{-2.2mm}/}
 
 \newsavebox{\DbfSLASH}
 \sbox{\DbfSLASH}{${\mathbf D}$\hspace{-2.8mm}/}
 
 \newsavebox{\DELVECRIGHT}
 \sbox{\DELVECRIGHT}{$\stackrel{\rightarrow}{\partial}$}
 
 \newcommand{\blue}{\IfColor{\textCadetBlue}{}}
\newcommand{\black}{\IfColor{\textBlack}{}}
\newcommand{\red}{\IfColor{\textRed}{}}
\newcommand{\green}{\IfColor{\textOliveGreen}{}}
\newcommand{\lila}{\IfColor{\textRedViolet}{}}

\begin{document}
\title{Weak decay constant of neutral pions in a hot and magnetized quark matter}
\author{Sh. Fayazbakhsh$^{a}$}\email{shfayazbakhsh@ipm.ir}
\author{N. Sadooghi$^{b}$}\email{sadooghi@physics.sharif.ir}
\affiliation{ $^{a}$Institute for Research in Fundamental Sciences
(IPM), School of Particles and Accelerators, P.O. Box 19395-5531,
Tehran-Iran\\
$^{b}$Department of Physics, Sharif University of Technology, P.O.
Box 11155-9161, Tehran-Iran}
\begin{abstract}
\noindent The directional weak decay constants of neutral pions are
determined at finite temperature $T$, chemical potential $\mu$ and
in the presence of a constant magnetic field $B$. To do this, we
first derive the energy dispersion relation of neutral pions from
the corresponding effective action of a two-flavor, hot and
magnetized Nambu--Jona-Lasinio model. Using this dispersion
relation, including nontrivial directional refraction indices, we
then generalize the PCAC relation of neutral pions and derive the
Goldberger-Treiman (GT) as well as the Gell--Mann-Oakes-Renner (GOR)
relations consisting of directional quark-pion coupling constant
$g_{qq\pi^{0}}^{(\mu)}$ and the weak decay constant
$f_{\pi^{0}}^{(\mu)}$ of neutral pions. The temperature dependence
of $g_{qq\pi^{0}}^{(\mu)}$ and $f_{\pi^{0}}^{(\mu)}$ are then
determined for fixed chemical potential and various constant
background magnetic fields. The GT and GOR relations are also
verified at finite $T,\mu$ and $eB$. It is shown that, because of
the explicit breaking of the Lorentz invariance by the magnetic
field, the directional quark-pion coupling and decay constants of
neutral pions in the longitudinal and transverse directions with
respect to the direction of the external magnetic field are
different, i.e., $g_{qq\pi^{0}}^{\|}\neq g_{qq\pi^{0}}^{\perp}$ and
$f_{\pi^{0}}^{\|}\neq f_{\pi^{0}}^{\perp}$. As it turns out, for
fixed $T,\mu$ and $B$, $g_{qq\pi^{0}}^{\|}> g_{qq\pi^{0}}^{\perp}$
and $f_{\pi^{0}}^{\|}< f_{\pi^{0}}^{\perp}$.
\end{abstract}
 \maketitle
\section{Introduction}\label{sec1}
\par\noindent
Today's theory of strong interactions is quantum chromodynamics
(QCD). At momentum scales of several GeV, QCD is a theory of weakly
interacting quarks and gluons. At low momentum scales, however, QCD
is governed by color confinement, which is accompanied by
spontaneous breaking of chiral symmetry. According to the Goldstone
theorem, spontaneous chiral symmetry breaking (S$\chi$SB) implies
the existence of Nambu-Goldstone (NG) bosons. For two quark flavors,
up and down, the isospin triplet of neutral and charged pions is
identified as the NG bosons of S$\chi$SB. Thus, at low momentum
scales, the weakly interacting pions build up the main degrees of
freedom of QCD. There is strong theoretical and experimental
evidence that chiral symmetry is restored at temperatures higher
than $200$ MeV. Crucial information on the restoration of chiral
symmetry is provided by the change in the pion properties in the
vicinity of the chiral transition point. Among these properties, the
pions' mass, $m_{\pi}$, and weak decay constant, $f_{\pi}$, are the
most important quantities, since they are deeply related to
S$\chi$SB: $m_{\pi}$ and $f_{\pi}$ are related to the quark bare
mass and the chiral condensate via the Gell--Mann-Oakes-Renner (GOR)
relation \cite{weise2012, nam2008}. To study the properties of
pions, models of Nambu--Jona-Lasinio (NJL) type have been quite
useful \cite{buballa2004}. One of the basic properties of these
models is that they include a gap equation that connects the chiral
condensate and the dynamical quark mass. This provides a mechanism
for dynamical chiral symmetry breaking and the generation of quark
quasiparticle masses \cite{weise2009}. It is known that strong
magnetic fields enhance this dynamical mass generation via the
phenomenon of magnetic catalysis \cite{klimenko1992, miransky1995}.
Different aspects of the effect of strong magnetic fields on the
properties of quark matter were recently summarized in
\cite{kharzeev2012} (see, in particular, \cite{catalysis}).
\par
In our previous paper \cite{fayazbakhsh2012}, we have investigated
the effect of external magnetic fields on the properties of neutral
pions in a hot quark matter. In particular, the neutral pion mass
and its directional refraction indices have been computed by making
use of the effective action of a bosonized two-flavor NJL model in
the presence of a constant magnetic field. We have shown that, since
the presence of constant background magnetic fields breaks the
Lorentz invariance, a certain anisotropy emerges in the refraction
indices of neutral pions in the transverse and longitudinal
directions with respect to the direction of the external magnetic
field. In the present paper, we will mainly focus on the effect of
external magnetic fields on the weak decay constant of neutral pions
at finite temperature. We will show that this quantity exhibits the
same anisotropy in the presence of constant background magnetic
fields. Modifying the PCAC relation,\footnote{The partially
conserved axial vector current (PCAC) relation combined with the
Feynman integral corresponding to the one-pion-to-vacuum matrix
element is usually used to determine the weak decay constant of
pions.} we will also prove the Goldberger-Treiman (GT) and GOR
relations in the presence of external magnetic fields. The effect of
a constant background magnetic field on the low energy relations of
neutral pions was previously studied in \cite{agasian2001}. Using a
first order chiral perturbation theory, it was shown that
$f_{\pi^{0}}$ and $m_{\pi^{0}}$ are shifted in such a way that the
GOR relation remains valid at finite $T$ and $eB$
\cite{agasian2001}.
\par
Very strong magnetic fields are supposed to be created in the early
stages of noncentral heavy-ion collisions at the RHIC and LHC
\cite{rhicmagnetic}. Depending on the collision energies and impact
parameters, these magnetic fields are estimated to be of the order
$eB\sim 1.5m_{\pi}^{2}$, with $m_{\pi}=138$ MeV, corresponding to
$0.03$ GeV$^{2}$, and $eB\sim 15m_{\pi}^{2}$, corresponding to $0.3$
GeV$^{2}$, respectively \cite{skokov2010}.\footnote{To keep the
notation as standard as possible, we use in this paper GeV$^{2}$
instead of  (GeV)$^{2}$. Note that $eB=1$ GeV$^{2}$ corresponds to
$b\sim 1.7\times 10^{20}$ Gau\ss.} Although extremely short lived,
the strong magnetic fields created in the heavy-ion collisions
affect the properties of charged quarks in the early stage of the
collision. Consequently, the properties of pions, even neutral
pions, which are made of these magnetized quarks are affected by
background magnetic fields. Let us emphasize that neutral pions
cannot interact directly with the external magnetic fields, because
of the lack of electric charge. Our method provides a possibility to
study the effect of strong magnetic fields on neutral pions, at
least at a theoretical level. Studying the properties of neutral
pions in the final freeze-out region may provide some hints on the
prehistory of their constituent quarks in the early stages of
heavy-ion collisions.
\par
The organization of this paper is as follows: In Sec. \ref{sec2}, we
will introduce the two-flavor magnetized NJL model, and, reviewing
our analytical results from \cite{fayazbakhsh2012}, we will derive
the effective action of neutral scalar and pseudoscalar mesons,
$\sigma\sim \bar{\psi}\psi$ and $\pi^{a}\sim
\bar{\psi}i\gamma_{5}\tau^{a}\psi$, in a derivative expansion up to
the second derivative.\footnote{Here, $\tau^{a}, a=1,2,3$ are Pauli
matrices. Moreover, $\pi^{0}\equiv \pi_{3}$ and $\pi^{\pm}\equiv
\frac{1}{\sqrt{2}}(\pi_{1}\mp i\pi_{2})$.} We will show that it
includes nontrivial form factors. To study the dynamics of pions, we
will mainly focus on the kinetic part of the effective action that
implies a nontrivial ``anisotropic'' energy dispersion relation for
neutral pions,
\begin{eqnarray}\label{int1}
E_{\pi^{0}}^{2}=u_{\pi^{0}}^{(i)2}q_{i}^{2}+m_{\pi^{0}}^{2}.
\end{eqnarray}
Here, $u_{\pi^{0}}^{(i)}, i=1,2,3$ are the directional refraction
indices and $m_{\pi^{0}}$ is the mass of neutral pions. We then
present our numerical results for the $T$ dependence of
$u_{\pi^{0}}^{(i)}, i=1,2,3$ and $m_{\pi^{0}}$ for fixed magnetic
fields $eB=0.03, 0.2, 0.3$ GeV$^{2}$ and zero chemical potential.
\par
Comparing this case with the case of zero temperature and/or
magnetic field, we arrive at following conclusions: Starting with an
isospin symmetric theory at zero temperature and in a
magnetic-field-free vacuum, the refraction indices for all pions are
given by
\begin{eqnarray}\label{int2}
u_{\pi_{a}}^{(i)}=1, \qquad\forall i,a=1,2,3.
\end{eqnarray}
Thus, according to (\ref{int1}), the energy dispersion relation of
neutral pions in field-free vacuum turns out to be the well-known
``isotropic'' energy-dispersion relation
\begin{eqnarray}\label{int3}
E_{\pi^{a}}^{2}={\mathbf{q}}^{2}+m_{\pi^{a}}^{2}, \qquad\forall
a=1,2,3.
\end{eqnarray}
Hence, in the chiral limit,\footnote{The chiral limit is
characterized by $m_{0}\to 0$, where $m_{0}$ is the bare mass of up
and down quarks. This in turn implies $m_{\pi_{a}}\to 0$.} and in
field-free vacuum, massless pions propagate at the speed of light.
\par
At finite temperature, and $eB=0$, however, the relativistic
invariance of the theory is broken by the medium, but the isospin
symmetry is still preserved. In this case a single refraction index
is defined for all pions by
\begin{eqnarray}\label{int4}
u_{\pi_{a}}^{(1)}=u_{\pi_{a}}^{(2)}=u_{\pi_{a}}^{(3)}=u,\qquad\forall
a=1,2,3.
\end{eqnarray}
The energy dispersion relation (\ref{int1}) is therefore given by $
E_{\pi_{a}}^{2}=u^{2}{\mathbf{q}}^{2}+m_{\pi_{a}}^{2}$, for all
$a=1,2,3$. This relation appears e.g. in \cite{shuryak1990, son2000,
ayala2002}. According to the results in \cite{son2000, ayala2002},
the refraction index $u$, being a $T$-dependent quantity, is always
smaller than unity (i.e. $u<1$). Thus, as it turns out, in the
chiral limit, massless pions move at a speed slower than the speed
of light \cite{pisarski1996,chiral-perturbation}.
\par
In \cite{fayazbakhsh2012}, we have studied the case $T\neq 0$ and
$eB\neq 0$, and shown that the effect of the magnetic field is
twofold: First, the isospin symmetry of the theory breaks down.
Consequently, neutral and charged pions exhibit different properties
in the presence of background magnetic fields. Second, for a
magnetic field aligned in a fixed direction, a certain anisotropy
appears in the transverse and longitudinal directions with respect
to this fixed direction. In \cite{fayazbakhsh2012}, we have only
considered the case of neutral pions, and we assumed that the
magnetic field is directed in the third direction. Here, the
above-mentioned anisotropy is reflected in the directional
refraction indices as \cite{fayazbakhsh2012}
 \begin{eqnarray}\label{int5}
u_{\pi^{0}}^{(1)}=u_{\pi^{0}}^{(2)}\neq u_{\pi^{0}}^{(3)}.
\end{eqnarray}
Thus, the magnetized neutral pions satisfy the anisotropic energy
dispersion relation (\ref{int1}). Within our truncation up to second
order derivative expansion,\footnote{The contributions arising from
pion-pion interactions are not included in our computation.} the
transverse refraction indices
$u_{\pi^{0}}^{(1)}=u_{\pi^{0}}^{(2)}>1$, and the longitudinal one
$u_{\pi^{0}}^{(3)}=1$, and they do not depend on temperature and/or
magnetic fields [see Fig. \ref{fig4}(b) in Sec. \ref{sec2}]. As
concerns the chiral limit, it turns out that whereas the massless
pions propagate in the direction parallel to the magnetic field with
the speed of light, their velocity in the plane perpendicular to the
direction of the $B$ field is larger than $1$.\footnote{In our
previous paper \cite{fayazbakhsh2012}, the directional refraction
indices $u_{\pi^{0}}^{(i)}, i=1,2,3$ defined by nontrivial form
factors were computed with finite bare quark mass $m_{0}\neq 0$. We
recalculate $u_{\pi^{0}}^{(i)}, i=1,2,3$ by assuming that $m_{0}=0$.
We get again $u_{\pi^{0}}^{(1)}=u_{\pi^{0}}^{(2)}>1$ as well as
$u_{\pi^{0}}^{(3)}=1$. In this paper, since we are interested in the
low energy theorems satisfied by massive pions, we do not attempt to
go through the discussion of superluminal pions (for more
discussions see \cite{superluminal}).} The natural question arising
at this stage is how these results are reflected in the other
properties of neutral and massive pions, in particular, in their low
energy relations. In Secs. \ref{sec3}-\ref{sec4} of the present
paper, we have tried to answer this question by computing the
``directional'' weak decay constant of neutral pions using a
\emph{modified} PCAC relation, that leads naturally to
\textit{modified} GT and GOR relations.
\par
In Sec. \ref{sec3}, we prove the GT and GOR relations in two
different cases: In Sec. \ref{subsec3-A}, using the main arguments
presented in \cite{buballa2004}, we will first assume that the pions
satisfy the ordinary isotropic energy dispersion relation
(\ref{int3}). We will then combine the ordinary PCAC relation
\begin{eqnarray}\label{int6}
\langle 0|J_{\mu,5}^{a}(0)|\pi^{b}(q)\rangle=f_{\pi}q_{\mu}\delta^{ab},
\end{eqnarray}
and the Feynman integral corresponding to the one-pion-to-vacuum
matrix element, and then derive the ordinary GT as well as GOR
relations
\begin{eqnarray}\label{int7}
g_{qq\pi}f_{\pi}=m+{\cal{O}}(m_{0}^{2}),
\end{eqnarray}
as well as
\begin{eqnarray}\label{int8}
m_{\pi}^{2}f_{\pi}^{2}=\frac{m_{0}\sigma_{0}}{2G}+{\cal{O}}(m_{0}^{2}).
\end{eqnarray}
Here, $J_{\mu,5}^{a}\equiv
\bar{\psi}\gamma_{\mu}\gamma_{5}\lambda^{a}\psi$, with
$\lambda^{a}=\tau^{a}/2$, is the axial vector current, $g_{qq\pi}$
is the quark-pion coupling constant, $m=m_{0}+\sigma_{0}$ is the
constituent quark mass, including the bare quark mass $m_{0}$, and
the chiral condensate $\sigma_{0}=-2G\langle\bar{\psi}\psi\rangle$,
with $G$ being the NJL coupling constant. In Sec. \ref{subsec3-B},
we will then generalize the method presented in  Sec.
\ref{subsec3-A} to the case of pions satisfying anisotropic energy
dispersion relation (\ref{int1}). To derive the modified GT and GOR
relations in this case, we will appropriately modify the PCAC
relation as
\begin{eqnarray}\label{int9}
\langle 0|J_{\mu,5}^{a}(0)|\pi^{b}(q)\rangle=f_{b}q_{\mu}u_{\pi_{b}}^{(\mu)2}\delta^{ab},
\end{eqnarray}
for all $\mu=0,\cdots,3$ and $a,b=1,2,3$. Here, the four-vector
$u_{\pi_{b}}^{(\mu)}\equiv (1,u_{\pi_{b}}^{(i)})$, with
$u_{\pi_{b}}^{(i)}, i=1,2,3$ the directional refraction indices
appearing in (\ref{int1}). In (\ref{int9}), $f_{b}$ is first an
unknown dimensionful proportionality factor, which will then be
shown to be equal to the constituent quark mass $m$. Following the
same method as in Sec. \ref{subsec3-A}, we will derive the modified
GT and GOR relations for all $\mu=0,\cdots, 3$ and $a=1,2,3$ as
\begin{eqnarray}\label{int10}
g_{qq\pi_{a}}^{(\mu)}f_{\pi_{a}}^{(\mu)}=m+{\cal{O}}(m_{0}^{2}),
\end{eqnarray}
and
\begin{eqnarray}\label{int11}
m_{\pi_{a}}^{2}f_{\pi_{a}}^{(\mu)2}=u_{\pi_{a}}^{(\mu)2}\frac{m_{0}\sigma_{0}}{2G}+{\cal{O}}(m_{0}^{2}),
\end{eqnarray}
which, in contrast to (\ref{int7}) and (\ref{int8}), include the
``directional'' quark-pion coupling $g_{qq\pi^{0}}^{(\mu)}$ and weak
decay constants of neutral pions $f_{\pi_{a}}^{(\mu)}$, as well as
the directional refraction indices $u_{\pi_{a}}^{(\mu)}$. Note that
for $\mu=0$ (temporal direction), the GOR relation (\ref{int11})
leads to the well-known GOR relation
\begin{eqnarray}\label{int12}
m_{\pi_{a}}^{2}f_{\pi_{a}}^{(0)2}=\frac{m_{0}\sigma_{0}}{2G}+{\cal{O}}(m_{0}^{2}),
\end{eqnarray}
arising at finite temperature and zero magnetic field
\cite{pisarski1996, chiral-perturbation}.\footnote{The GOR relation
$m_{\pi}^{2}=\frac{2m_{0}\langle\bar{\psi}\psi\rangle}{(\mbox{Re}f_{\pi}^{t})^{2}}$
in finite temperature QCD appearing in \cite{pisarski1996,
chiral-perturbation} turns out to be equivalent to (\ref{int11}) in
the NJL model.} As aforementioned, the case of a nonvanishing
magnetic field can be viewed as a special case, where neutral pions
satisfy the anisotropic energy dispersion relation (\ref{int1}) with
directional refraction indices given in (\ref{int5}). In Sec.
\ref{sec4}, we will explicitly determine $g_{qq\pi^{0}}^{(\mu)}$ and
$f_{\pi_{a}}^{(\mu)}$, first analytically (see Sec. \ref{subsec4-A})
and then numerically (see Sec. \ref{subsec4-B}). As it turns out, $
g_{qq\pi^{0}}^{(0)}=g_{qq\pi^{0}}^{(3)}\neq
g_{qq\pi^{0}}^{(1)}=g_{qq\pi^{0}}^{(2)} $, as well as $
f_{\pi^{0}}^{(0)}=f_{\pi^{0}}^{(3)}\neq
f_{\pi^{0}}^{(1)}=f_{\pi^{0}}^{(2)} $. In studying the $T$
dependence of these quantities, we arrive at $
g_{qq\pi^{0}}^{(0)}>g_{qq\pi^{0}}^{(1)} \qquad\mbox{and}\qquad
f_{\pi^{0}}^{(0)}<f_{\pi^{0}}^{(1)}$. Setting $a=3$ in the GT and
GOR relations (\ref{int10}) and (\ref{int11}), we will use these
relations for neutral pions to prove numerically that $f_{3}$,
appearing in (\ref{int9}), is given by the constituent quark mass
$m$. Using the numerical data for $m_{\pi^{0}}^{2}$,
$f_{\pi^{0}}^{(\mu)}$, $u_{\pi^{0}}^{(\mu)}$ and $m$, we will
compare numerically the left- and right hand side (r.h.s.) of the
relation
\begin{eqnarray}\label{int13}
m_{\pi_{a}}^{2}f_{\pi_{a}}^{(\mu)2}=u_{\pi_{a}}^{(\mu)2}\frac{m_{0}m}{2G},
\end{eqnarray}
which, comparing to the original GOR relation (\ref{int11}),
includes the contributions of ${\cal{O}}(m_{0}^{2})$. As it turns
out, this relation seems to be exact for small $T$ and $eB$. A
slight deviation from this relation occurs at temperature above the
chiral transition temperature. In Sec. \ref{sec5}, we will summarize
our results. A number of useful relations, which are used to derive
the analytical results in Sec. \ref{sec4}, are presented in the
Appendix.
\section{The Model}\label{sec2}
\setcounter{equation}{0}
\subsection{Effective action of a two-flavor magnetized NJL model in a derivative
expansion}\label{subsec2-A}
\par\noindent
Let us start with the Lagrangian density of a two-flavor gauged NJL model
\begin{eqnarray}\label{NA1a}
{\cal{L}}&=&\bar{\psi}(x)\left(i\gamma^{\mu}D_{\mu}-m_{0}\right)\psi(x)+G~\{[\bar{\psi}(x)\psi(x)]^2\nonumber\\
&&+
[\bar{\psi}(x)i\gamma_5\vec{\tau}\psi(x)]^2\}-\frac{1}{4}F^{\mu\nu}F_{\mu\nu}.
\end{eqnarray}
Here, $\psi_{\alpha,f}^{c},\bar{\psi}_{\alpha,f}^{c}$ are the
fermionic fields, carrying Dirac $\alpha\in (1,\cdots,4)$, flavor
$f\in (1,2)=(u,d)$ and color $c\in (1,2,3)=(r,g,b)$ indices.
Choosing the bare quark mass $m_{0}\equiv m_{u}=m_{d}$, the above
Lagrangian density is invariant under $SU(2)$ isospin symmetry. The
global $SU_{L}(2_{f})\times SU_{R}(2_{f})$ chiral and $SU(3_{c})$
color symmetries are guaranteed in the chiral limit $m_{0}\to 0$.
Fixing the external $U(1)$ gauge field,
$A_{\mu}^{\mbox{\tiny{ext.}}}$, appearing in the covariant
derivative $D_{\mu}\equiv
\partial_{\mu}+ieQA_{\mu}^{\mbox{\tiny{ext.}}}$, as
$A_{\mu}^{\mbox{\tiny{ext.}}}=\left(0,0,Bx_{1},0\right)$, a constant
magnetic field is produced, which is directed in the third
direction, $\mathbf{B}=B\mathbf{e}_{3}$. Here,
$Q=\mbox{diag}\left(2/3,-1/3\right)$ is the charge matrix of the up
and down quarks. Using $A_{\mu}^{\mbox{\tiny{ext.}}}$, the
electromagnetic field strength tensor$F_{\mu\nu}\equiv
\partial_{[\mu}A_{\nu]}^{\mbox{\tiny{ext.}}}$ is a constant tensor,
with the only nonvanishing component $F_{12}=-F_{21}=B$. Defining
the auxiliary meson fields $\sigma$ and $\vec{\pi}$ as
\begin{eqnarray}\label{NA2a}
\sigma(x)&=&-2G\bar{\psi}(x)\psi(x),\nonumber\\
\vec{\pi}(x)&=&-2G\bar{\psi}(x)i\gamma_5\vec{\tau}\psi(x),
\end{eqnarray}
where $\vec{\tau}=(\tau_{1},\tau_{2}.\tau_{3})$ are the Pauli
matrices, the Lagrangian density (\ref{NA1a}) is equivalently given
by the semibosonized Lagrangian as
\begin{eqnarray}\label{NA3a}
{\cal{L}}_{sb}&=&\bar{\psi}(x)\left(i\gamma^{\mu}D_{\mu}-m_{0}\right)\psi(x)-\bar{\psi}\left(\sigma+i\gamma_5\vec{\tau}\cdot\vec{\pi}\right)\psi\nonumber\\
&&-\frac{(\sigma^2+\vec{\pi}^2)}{4G}-\frac{B^2}{2}.
\end{eqnarray}
The effective action corresponding to (\ref{NA3a}), $\Gamma_{\mbox{\tiny{eff}}}$, is then determined by integrating
out the fermionic fields $\psi$
and $\bar{\psi}$,
\begin{eqnarray}\label{NA4a}
e^{i\Gamma_{\mbox{\tiny{eff}}}[\Phi(x)]}=\int{\cal{D}}\psi{\cal{D}}\bar{\psi}\exp\left(i\int
d^{4}x~{\cal{L}}_{sb}\right).
\end{eqnarray}
As a functional of $x$-dependent meson fields
\begin{eqnarray}\label{NA5a}
\Phi(x)=\left(\varphi_{0},\varphi_{1},\varphi_{2},\varphi_{3}\right)\equiv
\left(\sigma,\pi_{1}, \pi_{2},\pi_{3}\right),
\end{eqnarray}
the effective action $\Gamma_{\mbox{\tiny{eff}}}$ consists of two
parts, $\Gamma_{\mbox{\tiny{eff}}}=\Gamma_{\mbox{\tiny{eff}}}^{(0)}
+\Gamma_{\mbox{\tiny{eff}}}^{(1)}$. The tree-level part is given by
\begin{eqnarray}\label{NA6a}
\Gamma_{\mbox{\tiny{eff}}}^{(0)}[\Phi(x)]=-\int
d^{4}x\left(\frac{\sigma^{2}+\vec{\pi}^{2}}{4G}+\frac{B^{2}}{2}\right),
\end{eqnarray}
and the one-loop part reads
\begin{eqnarray}\label{NA7a}
\Gamma_{\mbox{\tiny{eff}}}^{(1)}[\Phi(x)]=-i
{\mbox{tr}}_{\{sfcx\}}\ln(i{S_{Q}^{-1}[\Phi(x)]}).
\end{eqnarray}
Here, the trace operation includes a trace over discrete degrees of
freedom, spinor $s$, flavor $f$ and color $c$, as well as an
integration over the continuous coordinate $x$. The inverse fermion
propagator appearing in (\ref{NA7a}) is given by
\begin{eqnarray}\label{NA8a}
iS^{-1}_{Q}[\Phi(x)]\equiv
i\gamma^{\mu}D_{\mu}-\big[m(x)+i\gamma^{5}\vec{\tau}\cdot\vec{\pi}(x)\big],
\end{eqnarray}
with $m(x)\equiv m_{0}+\sigma(x)$. In what follows, we use the
method used in \cite{fayazbakhsh2012} to evaluate the effective
action $\Gamma_{\mbox{\tiny{eff}}}[\Phi(x)]$ in a derivative
expansion up to second order. To do this, let us assume a constant
field configuration $\Phi_{0}\equiv
\left(\sigma_{0},\mathbf{0}\right)$ that breaks the
$SU_{L}(2_{f})\times SU_{R}(2_f)$ chiral symmetry of the original
action in the chiral limit. We then expand $\Phi(x)$ around this
constant configuration,
\begin{eqnarray}\label{NA9a}
\Phi(x)=\Phi_{0}+\bar{\Phi}(x).
\end{eqnarray}
Plugging (\ref{NA9a}) into the effective action $\Gamma_{\mbox{\tiny{eff}}}[\Phi(x)]$, we arrive first at
\begin{eqnarray}\label{NA10a}
\lefteqn{\hspace{-0.5cm}\Gamma_{\mbox{\tiny{eff}}}[\Phi]=\Gamma_{\mbox{\tiny{eff}}}[\Phi_{0}]-\frac{1}{2}\int
d^{d}x\
{\cal{M}}^{2}_{ij}[\Phi_{0}]\bar{\varphi}_{i}(x)\bar{\varphi}_{j}(x)}\nonumber\\
&&\hspace{-0.5cm}+\frac{1}{2}\int d^{d}x\
\chi_{ij}^{\mu\nu}[\Phi_{0}]\partial_{\mu}\bar{\varphi}_{i}(x)\partial_{\nu}\bar{\varphi}_{j}(x)+\cdots,
\end{eqnarray}
where the summation over $i,j=0,\cdots,3$ is skipped. Here, the squared mass matrix ${\cal{M}}_{ij}^{2}$ and the
kinetic matrix $\chi_{ij}^{\mu\nu}$ are given by
\begin{eqnarray}
\hspace{-0.5cm}{\cal{M}}_{ij}^{2}[\Phi_{0}]&\equiv&-\int
d^{d}z\frac{\delta^{2}\Gamma_{\mbox{\tiny{eff}}}}{\delta\varphi_{i}(0)
\delta\varphi_{j}(z)}\bigg|_{\Phi_{0}},\label{NA11a}\\
\hspace{-0.5cm}\chi_{ij}^{\mu\nu}[\Phi_{0}]&\equiv& -\frac{1}{2}\int
d^{d}z
z^{\mu}z^{\nu}\frac{\delta^{2}\Gamma_{\mbox{\tiny{eff}}}}{\delta\varphi_{i}(0)\delta\varphi_{j}(z)}\bigg|_{\Phi_{0}}.\label{NA12a}
\end{eqnarray}
Choosing, at this stage, the kinetic matrix $\chi_{ij}^{\mu\nu}$, as
in \cite{miransky1995,fayazbakhsh2012}, in the form
\begin{eqnarray}\label{NA13a}
\chi_{ij}^{\mu\nu}[\Phi]=(F_{1}^{\mu\nu})_{ij}+2F_{2}^{\mu\nu}\frac{\varphi_{i}\varphi_{j}}{\Phi^{2}},\hspace{0.5cm}
\forall i,j=0,\cdots,3,\nonumber\\
\end{eqnarray}
with $\Phi^{2}=\sigma^{2}+\vec{\pi}^{2}$, and plugging (\ref{NA13a})
into (\ref{NA10a}), the kinetic part of the effective action,
$\Gamma_{\mbox{\tiny{eff}}}^{k}[\Phi]$, will then be given by
$\Gamma_{\mbox{\tiny{eff}}}^{k}[\Phi]=\int
d^{d}x~{\cal{L}}_{k}[\Phi]$, where ${\cal{L}}_{k}[\Phi]$ is the part
of the effective Lagrangian including only two derivatives,
\begin{eqnarray}\label{NA14a}
{\cal{L}}_{k}=\frac{1}{2}(F_{1}^{\mu\nu})_{ij}\partial_{\mu}\varphi_{i}\partial_{\nu}\varphi_{j}
+\frac{F_{2}^{\mu\nu}}{\Phi^{2}}\left(\varphi_{i}\partial_{\mu}\varphi_{i}\right)
\left(\varphi_{j}\partial_{\nu}\varphi_{j}\right).\nonumber\hspace{-0.5cm}\\
\end{eqnarray}
Here, according to (\ref{NA5a}), the fields $\varphi_{i},
i=1,\cdots,3$ are identified by the meson fields $\sigma$ and
$\vec{\pi}$. Following the standard algebraic manipulations
presented in \cite{fayazbakhsh2012}, the effective action of the
two-flavor NJL model including $\sigma,\vec{\pi}$ mesons, which is
valid in a truncation of the effective action up to two derivatives,
reads
\begin{eqnarray}\label{NA15a}
\lefteqn{\hspace{-0.6cm}\Gamma_{\mbox{\tiny{eff}}}[\sigma,\vec{\pi}]=\Gamma_{\mbox{\tiny{eff}}}[\sigma_{0}]}\nonumber\\
&&\hspace{-1cm}-\frac{1}{2}\int
d^{d}x~\bar{\sigma}(x)\left(M_{\sigma}^{2}+{\cal{G}}^{\mu\mu}\partial_{\mu}^{2}\right)\bar{\sigma}(x)\nonumber\\
&&\hspace{-1cm}-\frac{1}{2}\sum\limits_{\ell=1}^{3}\int
d^{d}x~\bar{\pi}_{\ell}(x)\left(M_{\vec{\pi}}^{2}+
{\cal{F}}^{\mu\mu}\partial_{\mu}^{2}\right)_{\ell\ell}\bar{\pi}_{\ell}(x).
\end{eqnarray}
For a constant field configuration $\sigma_{0}$,
$\Gamma_{\mbox{\tiny{eff}}}[\sigma_{0}]$ is given by
$\Gamma_{\mbox{\tiny{eff}}}[\sigma_{0}]
=-{\cal{V}}\Omega_{\mbox{\tiny{eff}}}$, where ${\cal{V}}$ denotes
the four-dimensional space-time volume and
$\Omega_{\mbox{\tiny{eff}}}$ is the effective (thermodynamic)
potential. In \cite{fayazbakhsh2012}, the thermodynamic potential of
this model is derived at finite temperature $T$, finite chemical
potential $\mu$, and in the presence of constant magnetic field $B$.
It reads
\begin{widetext}
\begin{eqnarray}\label{NA16a}
\lefteqn{\hspace{-0.8cm}\Omega_{\mbox{\tiny{eff}}}(m;T,\mu,eB)=\frac{\sigma_{0}^{2}}{4G}+
\frac{B^{2}}{2}-\frac{3}{2\pi^{2}}\sum\limits_{q\in\{\frac{2}{3},-\frac{1}{3}\}}|q_{f}eB|^{2}
\left\{\zeta'\left(-1,x_{q_{f}}\right)+\frac{x_{q_{f}}^{2}}{4}+\frac{x_{q_{f}}}{2}(1-x_{q_{f}})\ln
x_{q_{f}}\right\}
}\nonumber\\
&&
+\frac{3}{4\pi^{2}}\left\{m^{4}\ln\left(\frac{\Lambda+\sqrt{\Lambda^{2}+m^{2}}}{m}\right)-\Lambda(2\Lambda^{2}+m^{2})\sqrt{\Lambda^{2}+m^{2}}\right\}\nonumber\\
&&-3T\sum\limits_{q_{f}\in\{\frac{2}{3},-\frac{1}{3}\}}|q_{f}
eB|\sum^{+\infty}_{p=0}\alpha_{p}\int_{-\infty}^{+\infty}\frac{dp_{3}}{4\pi^{2}}\left
\{\ln\left(1+e^{-\beta(E_{q}+\mu)}\right)+\ln\left(1+e^{-\beta(E_{q}-\mu)}\right)\right\}.
\end{eqnarray}
\end{widetext}
In a mean field approximation, the constant field configuration
$\Phi_{0}=(\sigma_{0},\mathbf{0})$ is supposed to minimize the above
thermodynamic potential $\Omega_{\mbox{\tiny{eff}}}$. In
(\ref{NA16a}), $m=m_{0}+\sigma_{0}$, $x_{q_{f}}\equiv
\frac{m^{2}}{2|q_{f}eB|}$, $\Lambda$ is an appropriate ultraviolet
(UV) momentum cutoff and $\beta\equiv T^{-1}$. Moreover,
$\zeta'(-1,x_{q})\equiv \frac{d\zeta(s,x_{q})}{ds}\big|_{s=1}$,
where $\zeta(s,a)\equiv \sum_{p=0}^{\infty}(a+p)^{-s}$ is the
Riemann-Hurwitz $\zeta$-function, and $\alpha_{p}\equiv
2-\delta_{p,0}$ is the spin degeneracy factor. In (\ref{NA16a}),
$E_{q}$ is the energy of charged fermions in a constant magnetic
field, which is given by
 \begin{eqnarray}\label{NA17a}
E_{q}\equiv\sqrt{\bar{\mathbf{p}}_{q}^{2}+m^{2}}=\sqrt{2|q_{f}eB|p+p_{3}^{2}+m^{2}}.
\end{eqnarray}
It arises from the solution of the Dirac equation in the presence of
a constant magnetic field using the Ritus eigenfunction method
\cite{ritus}. Here, $\bar{p}_{q}$ is the Ritus four-momentum defined
by
\begin{eqnarray}\label{NA18a}
\bar{p}_{q}=(p_{0},0,-\mbox{sgn}(q_{f} eB)\sqrt{2|q_{f} eB|p},
p_{3}),
\end{eqnarray}
where $p=0,1,2,\cdots$ labels the Landau levels.
Using the standard definition (\ref{NA11a}), the squared mass matrices of $\sigma$ and $\vec{\pi}$ mesons,
$M_{\sigma}^{2}$ and $(M_{\vec{\pi}}^{2})_{\ell m}$, appearing in (\ref{NA15a}), are defined by
\begin{eqnarray}\label{NA19a}
M_{\sigma}^{2}&\equiv&-\int
d^{4}z\frac{\delta^{2}\Gamma_{\mbox{\tiny{eff}}}}{\delta\sigma(0)\delta\sigma(z)}\bigg|_{(\sigma_{0},{\mathbf{0}})},
\nonumber\\
\hspace{-0.5cm}(M_{\vec{\pi}}^{2})_{\ell m}&\equiv&-\int
d^{4}z\frac{\delta^{2}\Gamma_{\mbox{\tiny{eff}}}}{\delta\pi_{\ell}(0)\delta\pi_{m}(z)}\bigg|_{(\sigma_{0},{\mathbf{0}})},
\end{eqnarray}
$\forall \ell, m=1,2,3$. The nontrivial form factors ${\cal{G}}^{\mu\nu}$ and ${\cal{F}}^{\mu\nu}_{\ell m}$,
appearing in (\ref{NA15a}), are combinations of the form factors $(F_{\alpha}^{\mu\nu})_{ij}, \alpha=1,2,$ and
$i,j=0,\cdots,3$, appearing in (\ref{NA13a}). They are defined by ${\cal{G}}^{\mu\nu}\equiv
\big[(F_{1}^{\mu\nu})_{00}+2F_{2}^{\mu\nu}\big]$ and ${\cal{F}}^{\mu\nu}_{\ell m}\equiv
\frac{1}{2}\big[(F_{1}^{\mu\nu})_{\ell m}+(F_{1}^{\mu\nu})_{m\ell}\big]$. They can be determined from the kinetic
part of the effective action $\Gamma_{\mbox{\tiny{eff}}}^{k}$ using
\begin{eqnarray}\label{NA20a}
{\cal{G}}^{\mu\nu}&\equiv&-\frac{1}{2}\int
d^{4}z
z^{\mu}z^{\nu}\frac{\delta^{2}\Gamma_{\mbox{\tiny{eff}}}^{k}}{\delta\sigma(0)\delta\sigma(z)}\bigg|_{(\sigma_{0},
{\mathbf{0}})},\nonumber\\
{\cal{F}}^{\mu\nu}_{\ell m}&\equiv&-\frac{1}{2}\int d^{4}z
z^{\mu}z^{\nu}\frac{\delta^{2}\Gamma_{\mbox{\tiny{eff}}}^{k}}{\delta\pi_{\ell}(0)\delta\pi_{m}(z)}\bigg|_{
(\sigma_{0},{\mathbf{0}})}\hspace{-0.7cm},
\end{eqnarray}
$\forall \ell,m=1,2,3$. Note that, according to our results in
\cite{fayazbakhsh2012}, the pion mass squared matrix
$(M^{2}_{\vec{\pi}})_{\ell m}$ and the form factors
${\cal{G}}^{\mu\nu}$ and ${\cal{F}}^{\mu\nu}_{\ell m}$ have the
following properties:
\begin{eqnarray}\label{NA21a}
\hspace{-0.3cm}(M^{2}_{\vec{\pi}})_{\ell m}=-(M^{2}_{\vec{\pi}})_{m\ell},~~~\mbox{and}~~~
 {\cal{F}}^{\mu\nu}_{\ell m}=- {\cal{F}}^{\mu\nu}_{m\ell},
\end{eqnarray}
$\forall\ell\neq m$. Moreover, $\forall \ell,m=1,2,3$,
${\cal{F}}^{\mu\nu}_{\ell m}={\cal{F}}^{\mu\mu}_{\ell m}g^{\mu\nu}$.
The latter property is also shared by ${\cal{G}}^{\mu\nu}$.
\par
In the present paper, as in \cite{fayazbakhsh2012}, we are mainly
interested in the properties of neutral $\sigma$ and $\pi^{0}$
mesons in hot and magnetized quark matter. We therefore identify the
$\pi_{3}$ meson with the neutral meson $\pi^{0}$. To simplify the
notations in the rest of this paper, we will denote
$(M_{\vec{\pi}}^{2})_{33}$ from (\ref{NA19a}) by $M_{\pi^{0}}^{2}$
and ${\cal{F}}^{\mu\nu}_{33}$ from (\ref{NA20a}) by
${\cal{F}}^{\mu\nu}$. Plugging the effective action
$\Gamma_{\mbox{\tiny{eff}}}$ from (\ref{NA6a}) and (\ref{NA7a}) in
(\ref{NA19a}) and (\ref{NA20a}), the squared mass matrices of
neutral mesons, $M_{\sigma}^{2}$ and $M_{\pi^{0}}^{2}$, as well as
the form factors ${\cal{G}}^{\mu\nu}$ and ${\cal{F}}^{\mu\nu}$ at
zero $(T,\mu)$ and in the presence of a background $B$ field, are
given by
\begin{eqnarray}\label{NA22a}
M_{\sigma}^{2}&=&\frac{1}{2G}-i\int
d^{4}z\mbox{tr}_{sfc}\big[S_{Q}(z,0)S_{Q}(0,z)\big],\nonumber\\
\hspace{-0.5cm}M_{\pi^{0}}^{2}&=&\frac{1}{2G}+i\int
d^{4}z\mbox{tr}_{sfc}\big[S_{Q}(z,0)\tau_{3}\gamma^{5}S_{Q}(0,z)\gamma^{5}\tau_{3}\big],\nonumber\\
\end{eqnarray}
and
\begin{eqnarray}\label{NA23a}
{\cal{G}}^{\mu\nu}&=&-\frac{i}{2}\int
d^{4}z z^{\mu}z^{\nu}\mbox{tr}_{sfc}\big[S_{Q}(z,0)S_{Q}(0,z)\big],\nonumber\\
{\cal{F}}^{\mu\nu}&=&\frac{i}{2}\int
d^{4}z z^{\mu}z^{\nu}\mbox{tr}_{sfc}\big[S_{Q}(z,0)\tau_{3}\gamma^{5}S_{Q}(0,z)\gamma^{5}\tau_{3}\big].\nonumber\\
\end{eqnarray}
Here, $S_{Q}(x,y)$ is the Ritus fermion propagator
\cite{fayazbakhsh2012, fukushima2009},
\begin{eqnarray}\label{NA24a}
S_{Q}(x,y)=i\sum_{p=0}^{\infty}\hspace{-0.5cm}\int{\cal{D}}\tilde{p}~e^{-i\tilde{p}\cdot
(x-y)}P_{p}(x_{1})D_{Q}^{-1}(\bar{p})~P_{p}(y_{1}),\nonumber\\
\end{eqnarray}
with ${\cal{D}}\tilde{p}\equiv
\frac{dp_{0}dp_{2}dp_{3}}{(2\pi)^{3}}$,
$\tilde{p}=(p_{0},0,p_{2},p_{3})$, and $\bar{p}$ is the Ritus
momentum,
$$ \bar{p}=(p_{0},0,-s_{Q}\sqrt{2|QeB|p},p_{3}),$$
with $s_{Q}\equiv \mbox{sgn}(QeB)$.\footnote{The notation
${\cal{D}}\tilde{p}$ appearing in (\ref{NA24a}), is used in a number
of our previous papers (see e.g.,
\cite{fayazbakhsh2012,taghinavaz2012} to denote the Ritus measure of
integration
${\cal{D}}\tilde{p}=\frac{dp_{0}dp_{2}dp_{3}}{(2\pi)^{3}}$. The
symbol ${\cal{D}}$ is not to be confused with ${\cal{D}}$ used
normally in the measure of the path integrals in quantum field
theory, e.g., ${\cal{D}}$ appearing in ${\cal{D}}\psi$ or
${\cal{D}}\bar{\psi}$ in (\ref{NA4a}).} Having in mind that $Q$ is a
diagonal matrix with the entries $q_{f}=\{2/3,-1/3\}$, we will use
$s\equiv \mbox{sgn}(q_{f}eB)$ for the elements of this $2\times 2$
matrix $s_{Q}$. In (\ref{NA24a}),
\begin{eqnarray}\label{NA25a}
\hspace{-0.8cm}P_{p}(x_{1})&\equiv&\frac{1}{2}[f_{p}^{+s}(x_{1})+\Pi_{p}f_{p}^{-s}(x_{1})]
\nonumber\\
&&\hspace{-0.2cm}+\frac{is_{Q}}{2}[f_{p}^{+s}(x_{1})-\Pi_{p}f_{p}^{-s}(x_{1})]
\gamma^{1}\gamma^{2}.
\end{eqnarray}
Here, $\Pi_{p}\equiv 1-\delta_{p,0}$ considers the spin degeneracy
in the lowest Landau level with $p=0$.  Moreover, $f_{p}^{\pm
s}(x_{1})$ are defined by
\begin{eqnarray}\label{NA26a}
\begin{array}{rclcrcl}
f_{p}^{+s}(x_{1})&\equiv&\phi_{p}\left(x_{1}-s_{Q}p_{2}\ell_{B}^{2}\right),&&
p&=&0,1,2,\cdots,\nonumber\\
f_{p}^{-s}(x_{1})&\equiv&\phi_{p-1}\left(x_{1}-s_{Q}p_{2}\ell_{B}^{2}\right),&&
p&=&1,2,3,\cdots,
\end{array}
\hspace{-0.4cm}\nonumber\\
\end{eqnarray}
where $\phi_{p}(x)$ is a function of Hermite polynomials $H_{p}(x)$
in the form
\begin{eqnarray}\label{NA27a}
\phi_{p}(x)\equiv a_{p}\exp\left(-\frac{x^{2}}{2\ell_{B}^{2}}\right)H_{p}\left(\frac{x}{\ell_{B}}\right).
\end{eqnarray}
Here, $a_{p}\equiv (2^{p}p!\sqrt{\pi}\ell_{B})^{-1/2}$ is the
normalization factor and $\ell_{B}\equiv |QeB|^{-1/2}$ is the
magnetic length. In (\ref{NA24a}),
$D_{Q}(\bar{p})\equiv\gamma\cdot\bar{p}_{Q}-m$, with $\bar{p}_{Q}$
the Ritus four-momentum from (\ref{NA18a}). Note that since
$Q=\mbox{diag}\left(2/3,-1/3\right)$ is a $2\times 2$ matrix in the
flavor space, building the trace in the flavor space is equivalent
to evaluating the sum over $q_{f}\in\{2/3,-1/3\}$.
\par
In \cite{fayazbakhsh2012}, the squared mass matrices of neutral
bosons $(M_{\sigma}^{2},M_{\pi^{0}}^{2})$ as well as the nontrivial
form factors $({\cal{G}}^{\mu\nu}, {\cal{F}}^{\mu\nu})$, appearing
in (\ref{NA15a}), are determined numerically at finite $(T,\mu)$ and
$eB$. To introduce $T$ and $\mu$, we used the standard replacements
\begin{eqnarray}\label{NA28a}
p_{0}=i\omega_{n}-\mu,~~\mbox{and}~~\int\frac{dp_{0}}{2\pi}\to
iT\sum_{n},
\end{eqnarray}
where $\omega_{n}\equiv (2n+1)\pi T$ are the fermionic Matsubara frequencies. As it turns out, at finite temperature
and zero magnetic field, the meson form factors satisfy
\begin{eqnarray}\label{NA29a}
{\cal{G}}^{00}&=&-{\cal{G}}^{11}=-{\cal{G}}^{22}=-{\cal{G}}^{33},\nonumber\\
{\cal{F}}^{00}&=&-{\cal{F}}^{11}=-{\cal{F}}^{22}=-{\cal{F}}^{33}.
\end{eqnarray}
For nonvanishing magnetic fields, however, they exhibit a certain
anisotropy arising from the explicit breaking of Lorentz invariance
by the background magnetic field directed in the third direction,
\begin{eqnarray}\label{NA30a}
{\cal{G}}^{00}=-{\cal{G}}^{33}\neq {\cal{G}}^{11}={\cal{G}}^{22},\nonumber\\
{\cal{F}}^{00}=-{\cal{F}}^{33}\neq {\cal{F}}^{11}={\cal{F}}^{22}.
\end{eqnarray}
\subsection{Pole and screening masses of neutral mesons and their directional refraction indices:\\ A review of our
previous numerical results}\label{subsec2-B}
\par
\begin{figure}[hbt]
\includegraphics[width=8cm,height=5.5cm]{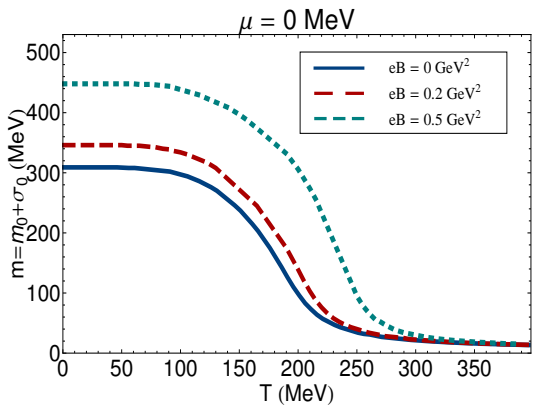}
\caption{The $T$ dependence of constituent quark mass
$m=m_{0}+\sigma_{0}$ for fixed $eB=0,0.2, 0.5$ GeV$^{2}$ and fixed
$\mu=0$. Here, $m_{0}\simeq 5$ MeV is the bare quark mass and
$\sigma_{0}$ is the chiral condensate.}\label{fig1}
\end{figure}

\noindent
The effective action $\Gamma_{\mbox{\tiny{eff}}}[\sigma,\vec{\pi}]$ from (\ref{NA15a}) can be used to determine the
energy dispersion relations of neutral $\sigma$ and $\pi^{0}$ mesons, which are generically denoted by $M$,
\begin{eqnarray}\label{NA31a}
E_{M}^{2}=u_{M}^{(i)2}q_{i}^{2}+m_{M}^{2},~~~ M\in\{\sigma,\pi^{0}\}.
\end{eqnarray}
Here, $u_{M}^{(i)}, i=1,2,3$ are the directional refraction indices
in the spatial $i=1,2,3$ directions. They are defined by
\begin{eqnarray}\label{NA32a}
u_{\sigma}^{(i)}\equiv\bigg|\frac{\mbox{Re}
{\cal{G}}^{ii}}{\mbox{Re} {\cal{G}}^{00}}\bigg|^{1/2},\qquad
u_{\pi^0}^{(i)}\equiv\bigg|\frac{\mbox{Re} {\cal{F}}^{ii}}{\mbox{Re}
{\cal{F}}^{00}}\bigg|^{1/2}.
\end{eqnarray}
Moreover, $m_{M}$ is the neutral mesons' pole mass given by
\begin{eqnarray}\label{NA33a}
m_{\sigma}\equiv\bigg|\frac{\mbox{Re} M_{\sigma}^{2}}{\mbox{Re} {\cal{G}}^{00}}\bigg|^{1/2},\qquad
m_{\pi^0}\equiv\bigg|\frac{\mbox{Re} M_{\pi^0}^{2}}{\mbox{Re} {\cal{F}}^{00}}\bigg|^{1/2}.
\end{eqnarray}
In the above relations, the squared mass matrices of neutral mesons
$(M_{\sigma}^{2}, M_{\pi^{0}}^{2})$, as well as form factors
${\cal{G}}^{\mu\nu}$
 and ${\cal{F}}^{\mu\nu}$, are defined in (\ref{NA19a}) as well as (\ref{NA20a}).
Combining the refraction indices (\ref{NA32a}) and the pole masses
(\ref{NA33a}), the screening masses of neutral mesons are defined by
\begin{eqnarray}\label{NA34a}
\hspace{-0.5cm}m_{M}^{(i)}=\frac{m_{M}}{u_{M}^{(i)}}, ~\mbox{for}~ M\in\{\sigma,\pi^{0}\},~\mbox{and}~i=1,2,3.
\end{eqnarray}
In \cite{fayazbakhsh2012}, we used (\ref{NA32a})-(\ref{NA34a}) and
determined the $T$ dependence of the neutral mesons' pole and
screening masses as well as their refraction indices for vanishing
and nonvanishing $\mu$ and $eB$. To do this, we first determined the
$T$ dependence of $\sigma_{0}$, the minima of the thermodynamic
potential $\Omega_{\mbox{\tiny{eff}}}$ from (\ref{NA16a}). In Fig.
\ref{fig1}, the $T$ dependence of the constituent quark mass
$m=m_{0}+\sigma_{0}$ is plotted for fixed $\mu=0$ MeV and
$eB=0,0.2,0.5$ GeV$^{2}$. We observe that, for a fixed temperature,
the constituent quark mass increases with increasing $eB$. Moreover,
it turns out that the transition from the broken chiral symmetry
phase, with $m\neq 0$, to the normal phase, with $m\simeq
m_{0}\approx 0$, is a smooth crossover, and for stronger magnetic
fields the transition to the normal phase occurs at higher
temperatures. All these effects are related to the phenomenon of
magnetic catalysis \cite{klimenko1992, miransky1995}, according to
which the constant magnetic field enhances the production of the
chiral condensate $\sigma_{0}\sim \langle\bar{\psi}\psi\rangle$, and
therefore catalyzes the dynamical chiral symmetry breaking.
\par\vspace{0.3cm}
\par
\begin{figure}[hbt]
\includegraphics[width=8cm,height=5.5cm]{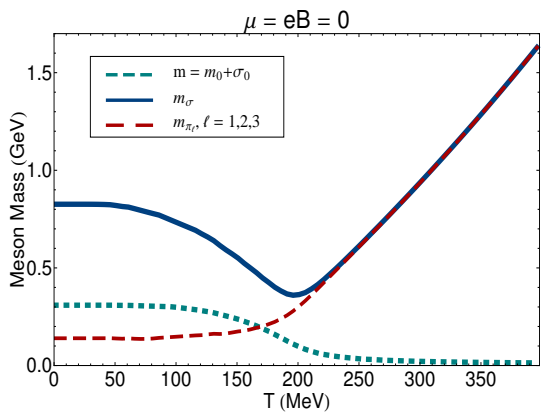}
\caption{The $T$ dependence of $\sigma$ and $\vec{\pi}$ meson masses
is demonstrated at $\mu=0$ and for a vanishing magnetic field (solid
line for $m_{\sigma}$ and dashed line for $m_{\vec{\pi}}$).
Comparing these curves with the $T$ dependence of the constituent
quark mass $m=m_{0}+\sigma_{0}$ (dotted line) shows that a mass
degeneracy between $m_{\sigma}$ and $m_{\vec{\pi}}$ occurs in the
chirally restored phase $T>220$ MeV.}\label{fig2}
\end{figure}

\noindent
To determine $m_{\sigma}$ and $m_{\pi^{0}}$, we further developed in
\cite{fayazbakhsh2012} the necessary technique to analytically
evaluate the integrals (\ref{NA22a}) and (\ref{NA23a}) up to an
integration over the $p_{3}$ momentum and a summation over Landau
levels and eventually performed them numerically. In Fig.
\ref{fig2}, the $T$ dependence of $m_{\sigma}$ and $m_{\pi_{\ell}},
\ell=1,2,3$ is plotted for vanishing $\mu$ and $eB$. The free
parameters of the theory, $m_{0}, \Lambda$ and $G$, are chosen so
that they reproduce the pion mass $m_{\pi}=137.7$ MeV and
$m_{\sigma}\simeq 824.3$ MeV, for vanishing $(T,\mu)$ and $eB$ (see
\cite{fayazbakhsh2012} and Sec. \ref{sec4} for more details). As it
turns out, for $eB=0$, the pion masses are degenerate, i.e.,
$m_{\pi_{1}}=m_{\pi_{2}}=m_{\pi_{3}}$. In the chirally broken phase,
the temperature dependence of pions, $m_{\vec{\pi}}$, is very weak.
This is because, in this phase, the pions play the role of
pseudo-Goldstone modes. In the symmetry restored phase, however,
pions are no longer bound states but only $q\bar{q}$ resonances
\cite{buballa2012}. In other words, they are expected to decay into
(free) quarks and antiquarks. This can only happen when $m_{\pi}$
increases with the temperature so that $m_{\pi}\simeq 2m$ (see below
for the definition of Mott temperature at which $m_{\pi}= 2m$). This
is why, for temperatures larger than the crossover temperature,
$m_{\pi}$ increases with increasing $T$.\footnote{In the present
paper, as in \cite{fayazbakhsh2012}, we mainly focus on the masses
of neutral $\sigma$ and $\pi^{0}$ mesons. The $(T,\mu,eB)$
dependence of $m_{\pi^{0}}$ is studied in detail in
\cite{fayazbakhsh2012}.}
\par
\begin{figure}[hbt]
\includegraphics[width=7.7cm,height=4.5cm]{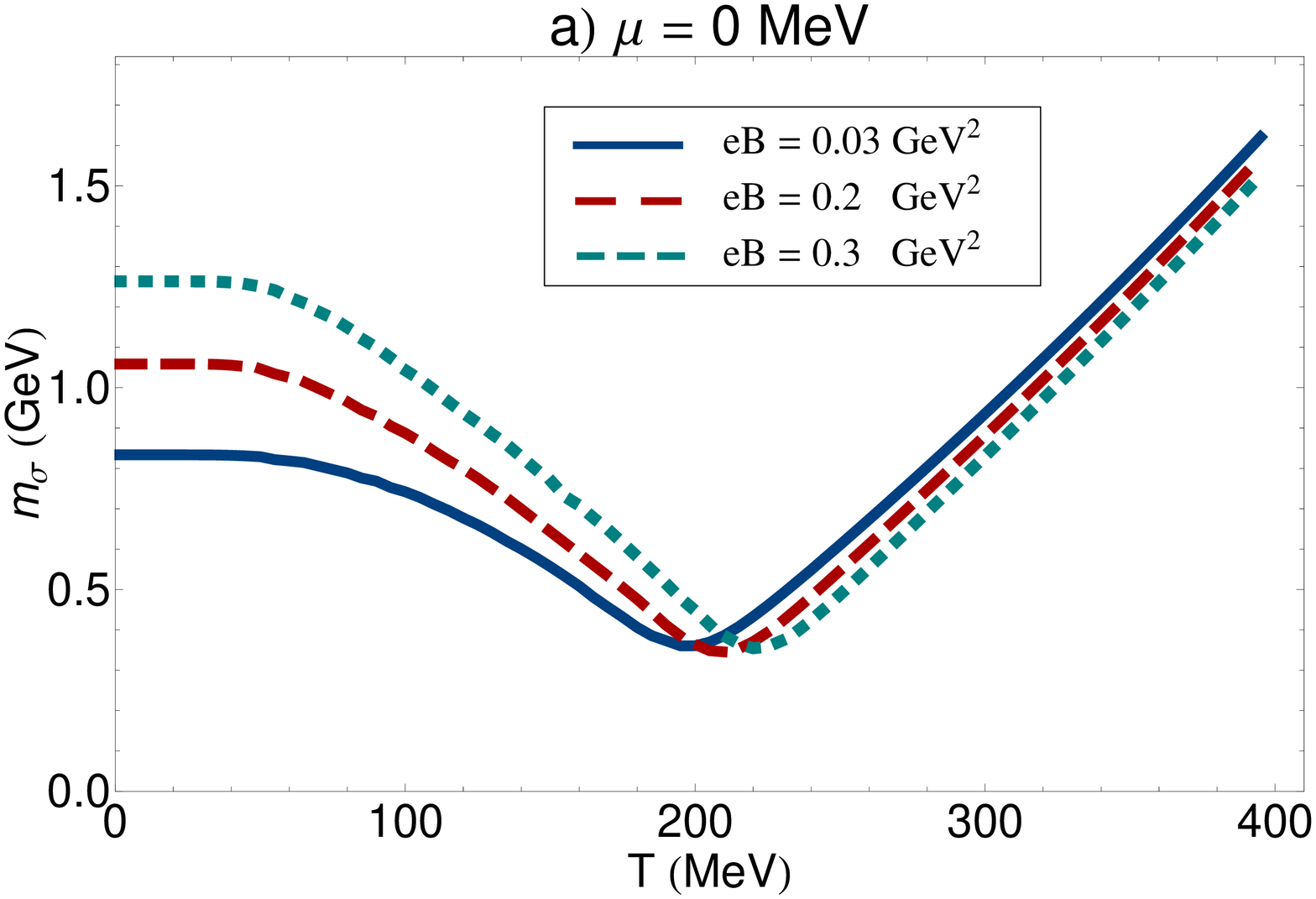}
\includegraphics[width=7.7cm,height=4.5cm]{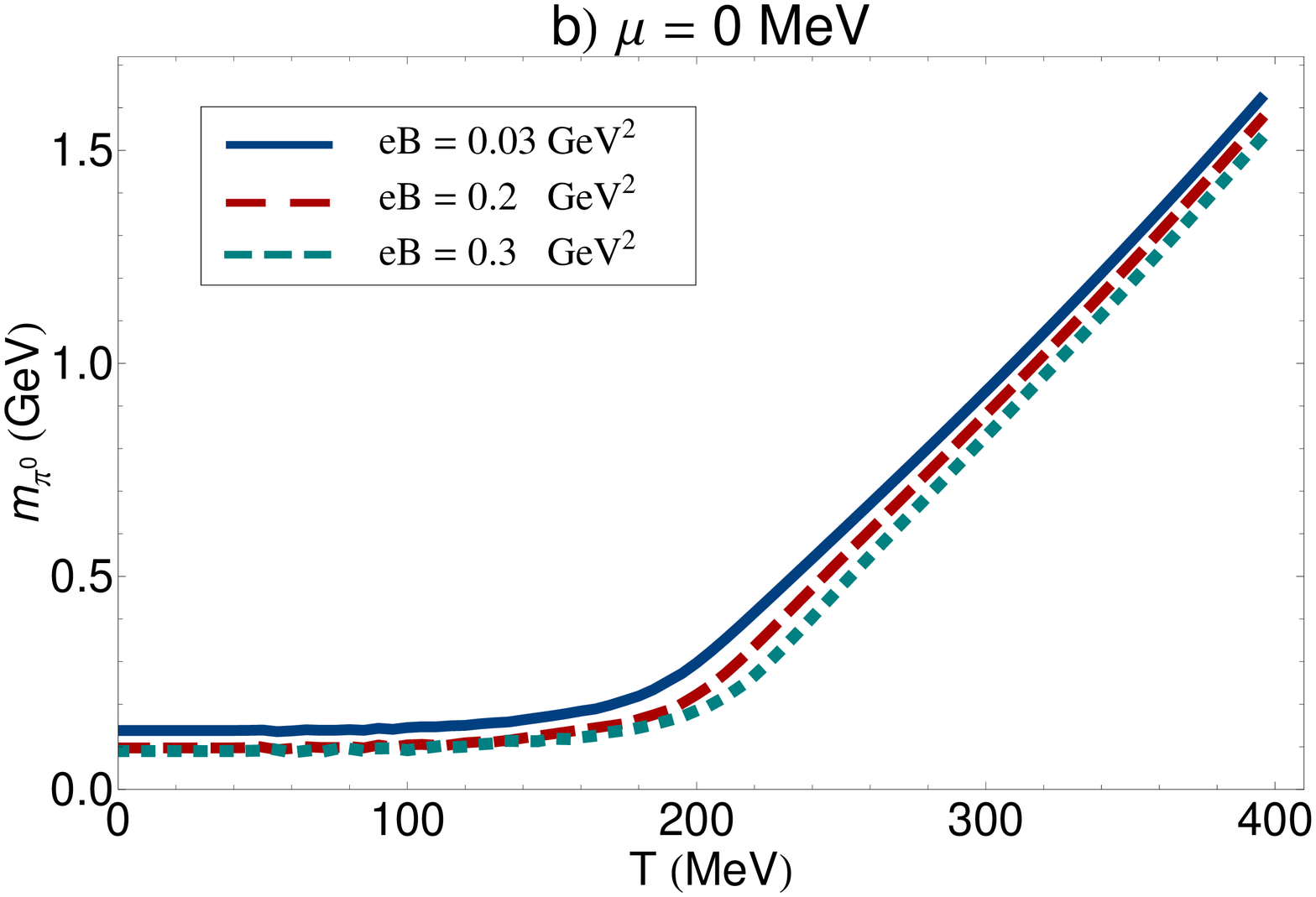}
\caption{The $T$ dependence of $m_{\sigma}$ [panel (a)] and
$m_{\pi^{0}}$ [panel (b)] is plotted for $eB=0.03,0.2,0.3$ GeV$^{2}$
and $\mu=0$ MeV.}\label{fig3}
\end{figure}

\noindent
For nonvanishing $eB$, as it turns out, the pion masses are not
degenerate anymore. We have $m_{\pi^{0}}\neq
m_{\pi^{+}}=m_{\pi^{-}}$. Here, the global $SU(2_{f})$ symmetry
breaks to a global $U(1)$ symmetry, where up and down quarks,
because of their opposite electric charges, rotate with opposite
angles. Thus, in the $U(1)$ symmetry broken phase and in the chiral
limit, whereas charged pions become massive, the neutral pion is the
only possible massless Goldstone mode \cite{shushpanov1997}. As we
have shown in \cite{fayazbakhsh2012}, for $eB\neq 0$, the $T$
dependence of the neutral pion mass exhibits the same behavior as
$m_{\vec{\pi}}(T)$ for $eB=0$ (see also Figs. \ref{fig2} and
\ref{fig3}).
\par
As concerns the $\sigma$-meson mass, in the symmetry broken phase,
$m_{\sigma}$ is large ($m^{2}_{\sigma}= 4 m^{2}+m_{\pi}^{2}$ for
$eB=0$) and drops when approaching the crossover temperature
\cite{buballa2012}. In the crossover region, the $\sigma$ meson
dissociates into two pions (see below). For $eB=0$, a possible
explanation for this effect is provided in \cite{klevansky1994}.
Here, it is shown that the $\sigma$ dissociation in the crossover
region is due to the occurrence of an $s$-channel pole in the
scattering amplitude of the $\pi+\pi\to\pi+\pi$ scattering, in a
process where a $\sigma$ meson is coupled to the external pions via
quark triangles (see e.g., \cite{klevansky1994, buballa2012} for
more details). For $eB \neq 0$, according to our results, the $T$
dependence of the $\sigma$-meson mass is the same as $m_{\sigma}(T)$
for $eB=0$ [see Fig. \ref{fig3}(a) and, in particular, Fig. 9(a)-(c)
in \cite{fayazbakhsh2012} for more details]. Whether this behavior
for $eB\neq 0$ is also related to the appearance of a certain pole
in the $\pi\pi$-scattering amplitude is an open question, which
shall be investigated in the future. In the chirally restored phase
$m_{\sigma}$ and $m_{\vec{\pi}}$, become degenerate and increase
with increasing $T$. These results are in agreement with the results
for $eB=0$ and $T\neq 0$ recently presented in \cite{buballa2012},
where a possible definition of the crossover temperature is also
provided. The latter is given by two characteristic temperatures,
the $\sigma$-dissociation temperature $T_{\mbox{\tiny{diss.}}}$,
defined by
\begin{eqnarray}\label{NA35a}
m_{\sigma}(T_{\mbox{\tiny{diss.}}})=2m_{\pi}(T_{\mbox{\tiny{diss.}}}),
\end{eqnarray}
and the Mott temperature $T_{\mbox{\tiny{Mott}}}$, defined by
\begin{eqnarray}\label{NA36a}
m_{\pi}(T_{\mbox{\tiny{Mott}}})=2m(T_{\mbox{\tiny{Mott}}}).
\end{eqnarray}
These temperatures were originally introduced in
\cite{klevansky1994}. According to \cite{buballa2012,klevansky1994},
for $T<T_{\mbox{\tiny{diss.}}}$, the $\sigma$ meson can decay into
two pions, and for $T>T_{\mbox{\tiny{Mott}}}$, the pion decays into
a constituent quark and antiquark; i.e., it is no longer a bound
state \cite{buballa2012}. In the chiral limit, both temperatures are
equal to the critical temperature of the chiral phase transition
$T_{c}$. In Table \ref{tab1}, we have listed
$T_{\mbox{\tiny{diss.}}}$ as well as $T_{\mbox{\tiny{Mott}}}$ for
$eB=0,0.03,0.2, 0.3,0.5$ GeV$^{2}$ and for our set of parameters.
\begin{table}[htb]
\begin{tabular}{ccccc}
          \hline\hline
$eB$ in GeV$^{2}$&$\qquad$&$T_{\mbox{\tiny{diss.}}}$ in MeV &$\qquad$&$T_{\mbox{\tiny{Mott}}}$ in MeV\\
\hline
0&$\qquad$&175&$\qquad$&190\\
0.03&$\qquad$&175&$\qquad$&190\\
0.2&$\qquad$&193&$\qquad$&205\\
0.3&$\qquad$&206&$\qquad$&210\\
0.5&$\qquad$&250&$\qquad$&255\\ \hline\hline
\end{tabular}
\caption{The $\sigma$ dissociation and Mott temperatures,
$T_{\mbox{\tiny{diss.}}}$ and $T_{\mbox{\tiny{Mott}}}$, defined in
(\ref{NA35a}) and (\ref{NA36a}), for vanishing and nonvanishing $eB$
and for our set of parameters, $m_{0}=0.005$ GeV, $\Lambda=0.6643$
GeV and $G=4.668$ GeV$^{-2}$.}\label{tab1}
\end{table}
\par\noindent
\par\noindent
Comparing $T_{\mbox{\tiny{diss.}}}$ and $T_{\mbox{\tiny{Mott}}}$ for
different nonvanishing $eB$ in Table \ref{tab1}, it seems that the
crossover region is shifted to higher temperatures for increasing
$eB$. In the chiral limit, where
$T_{\mbox{\tiny{diss.}}}=T_{\mbox{\tiny{Mott}}}=T_{c}$, as mentioned
above, this would mean that the critical temperature increases with
increasing $eB$. This phenomenon is related to the magnetic
catalysis of dynamical chiral symmetry breaking in the presence of
external magnetic fields. Recent lattice results \cite{bali2011}
show, however, a certain discrepancy with this conclusion. As it is
shown in \cite{bali2011}, for certain $eB$, the critical temperature
decreases with increasing $eB$. This shows that the $m$ dependence
of $T_{c}$ is indeed not trivial. There are a number of attempts in
the literature to resolve this disagreement
\cite{inversemagnetic,fukushima2012}, and to explain the so-called
inverse magnetic catalysis \cite{rebhan2010}. Note, that, to the
best of our knowledge, inverse magnetic catalysis was first observed
in \cite{inagaki2003} within a two-flavor NJL model. In one of our
previous papers \cite{fayazbakhsh2010}, we have also plotted the
$T$-$eB$ phase diagram for different fixed chemical potential $\mu$.
As it is shown in Fig. 14(a) and (b) of \cite{fayazbakhsh2010} (see
also Fig. 2 in \cite{fayazbakhsh2012}), there are some regions in
$eB$ in which $T_{c}$ decreases with increasing $eB$ and the inverse
magnetic catalysis occurs.
\par
\par
In Fig. \ref{fig3}, the $T$ dependence of $m_{\sigma}$ [Fig. 3(a)]
and $m_{\pi^{0}}$ [Fig. 3(b)] are plotted for fixed $\mu$ and
$eB=0.03, 0.2,0.3$ GeV$^{2}$. At a fixed temperature below (above)
the crossover region, the $\sigma$-meson mass increases (decreases)
with increasing $eB$. In contrast, $m_{\pi^{0}}$ decreases with
increasing $eB$ in both symmetry broken and restored phases. The
qualitative behavior of the $T$ dependence of $m_{\sigma}$ and
$m_{\pi^{0}}$ for $\mu=0$ remains unchanged for zero and nonzero
$eB$ (see Fig. 9 in \cite{fayazbakhsh2012}, where the $T$ dependence
of $m_{\sigma}$ and $m_{\pi^{0}}$ is compared). In particular,
$m_{\sigma}$ and $m_{\pi^{0}}$ are degenerate in the symmetry
restored phase $T>220$ MeV.
\par\noindent
\begin{figure}[hbt]
\includegraphics[width=7.7cm,height=4.5cm]{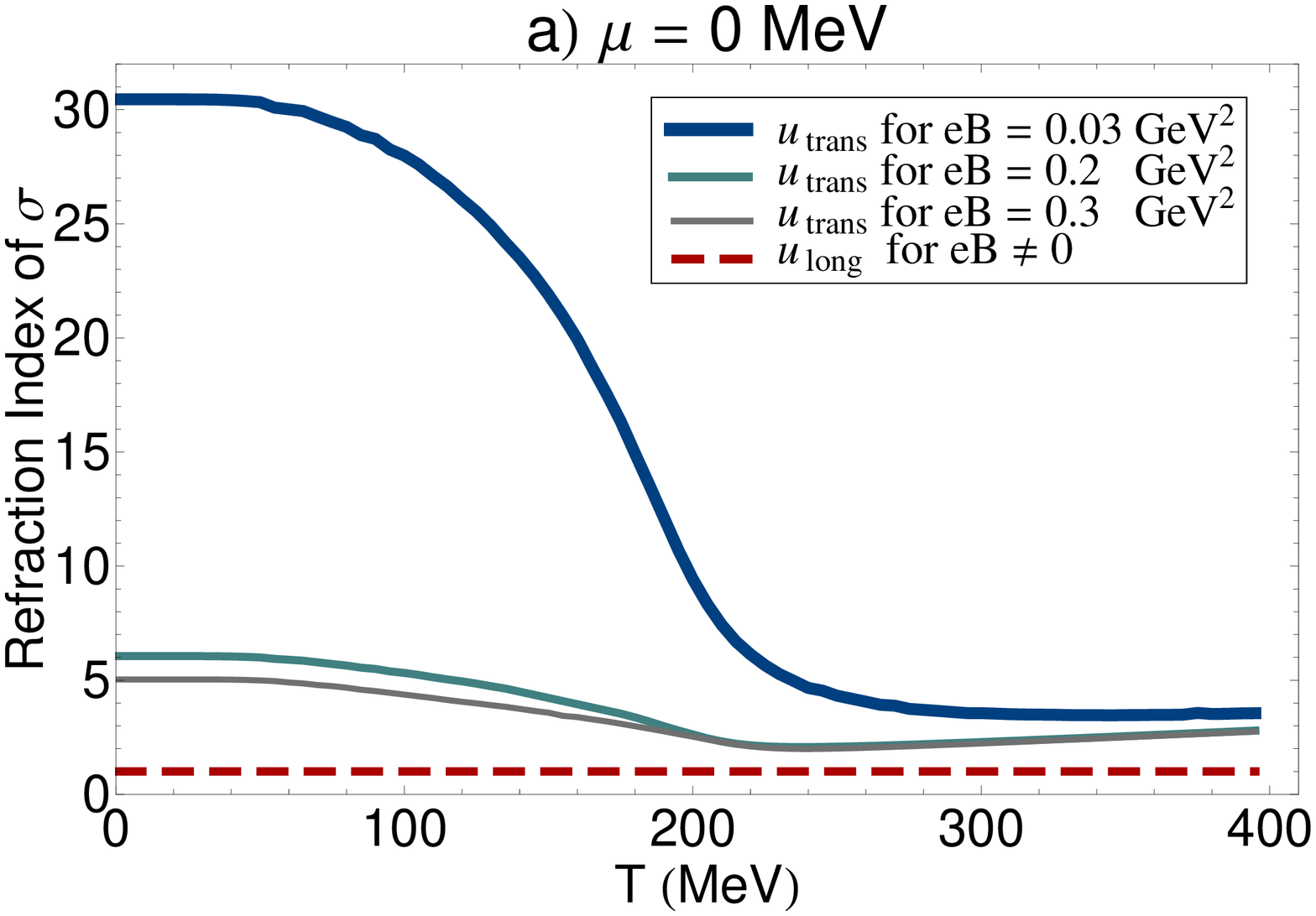}
\includegraphics[width=7.7cm,height=4.5cm]{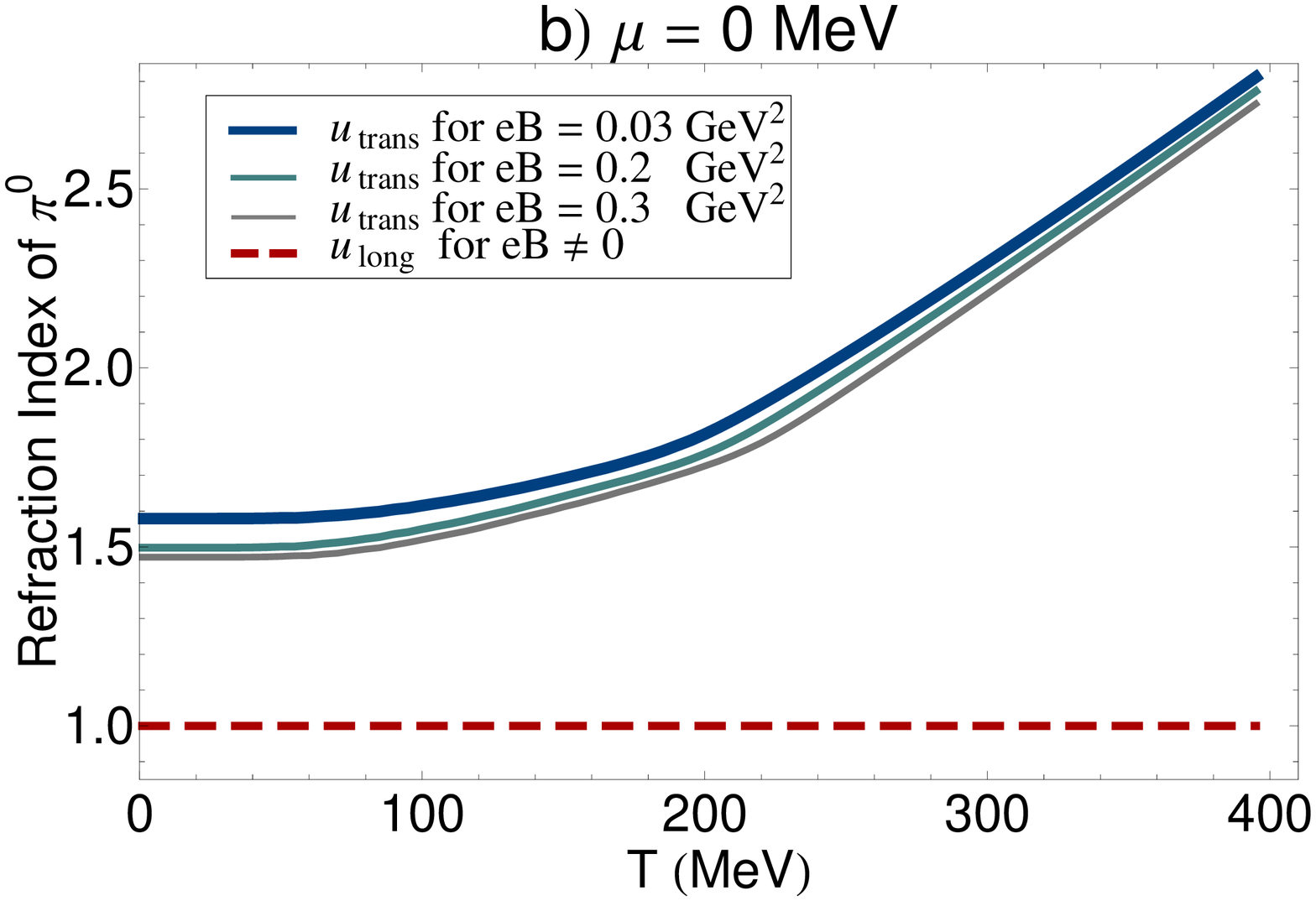}
\caption{The $T$ dependence of the transverse and longitudinal
refraction indices of $\sigma$ [panel (a)] and $\pi^{0}$ mesons
[panel (b)] is plotted for various $eB$. The longitudinal refraction
index of neutral mesons is equal to unity and independent of $T$
(red dashed lines). In contrast, the transverse refraction index of
neutral mesons is larger than unity and decreases with increasing
$eB$. For a $B$ field directed in the third direction,
$u_{\mbox{\tiny{trans}}}\equiv u_{M}^{(1)}=u_{M}^{(2)}$ and
$u_{\mbox{\tiny{long}}}\equiv u_{M}^{(3)}$ for
$M\in\{\sigma,\pi^{0}\}$.}\label{fig4}
\end{figure}
\par
Let us finally consider the directional refraction indices of
neutral mesons, $u_{M}^{(i)}, M\in \{\sigma,\pi^{0}\}$, that appear
in the energy dispersion relation (\ref{NA31a}). They are defined in
(\ref{NA32a}). In what follows, we will briefly review the results
appearing first in our previous paper \cite{fayazbakhsh2012} for
$eB\neq 0$ and $T\neq 0$. In Fig. \ref{fig4}, the $T$ dependence of
$u_{M}^{(i)}$ is plotted for fixed $\mu=0$ and nonvanishing
$eB=0.03,0.2,0.3$ GeV$^{2}$. For $eB\neq 0$, the refraction indices
of $\sigma$ and $\pi^{0}$ mesons in the longitudinal direction with
respect to the direction of the external magnetic field turn out to
be equal to unity, independent of $T$ and $\mu$ (see the red dashed
lines in Figs. \ref{fig4}(a) and (b)]. In the transverse directions,
however, the refraction indices of neutral mesons,
$u_{\sigma}^{\mbox{\tiny{trans}}}$ and
$u_{\pi^{0}}^{\mbox{\tiny{trans}}}$, are $T$ dependent and always
larger than unity. It seems that, in the directions perpendicular to
the direction of the external magnetic field, the neutral mesons
move with a speed larger than the speed of light $c$. But, this is
simply not true. Let us emphasize that, according to the dispersion
relation (\ref{NA31a}), this conclusion is only valid for
\textit{massless} mesons, where directional group velocities
coincide with $u_{M}^{(i)}$ defined in (\ref{NA32a}). Our mesons
are, however, massive,\footnote{We do not neglect $m_{0}$ in our
computations. The original theory is therefore nonchiral.
Consequently, the neutral pion is not a massless Goldstone mode of
the theory, even in the chirally broken phase.} and the fact that
the transverse refraction index is larger than unity does not
naturally lead to superluminal propagation. Having this in mind, the
natural question is what are the consequences of
$u_{\sigma}^{\mbox{\tiny{trans}}}>1$ and
$u_{\pi^{0}}^{\mbox{\tiny{trans}}}>1$? One of our main motivations
in this paper, is to answer this interesting question. In the next
section, we will show that directional refraction indices play an
important role in the low energy relations of neutral and massive
pions, once we have a medium at $T\neq 0$ and/or $eB\neq 0$. But,
before coming to this, let us consider again the plots in Fig.
\ref{fig4}. As it turns out, for a fixed temperature,
$u_{M}^{\mbox{\tiny{trans}}}, M\in \{\sigma,\pi^{0}\}$ decrease with
increasing $eB$. This effect is strongly related to the dynamics of
fermions in an external magnetic field. As it is stated in
\cite{miransky1995}, for very large magnetic fields, the motion of
charged fermions is restricted in directions perpendicular to the
magnetic field. A dimensional reduction from $3+1$ to $1+1$
dimensions occurs in the presence of very strong magnetic fields.
Naively, one would expect that the properties of $\sigma$ and
$\pi^{0}$ mesons are not affected by external magnetic fields,
simply because they are neutral. But, as is also argued in our
previous paper \cite{fayazbakhsh2012}, the way we have introduced
the background magnetic field in the original Lagrangian
(\ref{NA1a}) and the semibosonized Lagrangian (\ref{NA3a}) shows
that the effects of the external magnetic field on neutral mesons
arise mainly from its interaction with the constituent quarks of
these mesonic bound states. According to this argument, the behavior
of $u_{M}^{\mbox{\tiny{trans}}}, M\in \{\sigma,\pi^{0}\}$ for fixed
$T$ and with increasing $eB$ is a realization of the
well-established dimensional reduction, appearing mainly in the
lowest Landau level (LLL). Note that by increasing the magnetic
field strength, the effect of higher Landau levels can be neglected,
and the dynamics of fermions is solely determined by LLL, where the
above-mentioned dimensional reduction occurs. This is the reason why
at a fixed temperature $u_{M}^{\mbox{\tiny{trans}}}, M\in
\{\sigma,\pi^{0}\}$ decrease with increasing $eB$. The same property
occurs at zero temperature, where the transverse velocity of neutral
pions is determined within an effective chiral model in the LLL
approximation \cite{fukushima2012}.
\par
Coming back to the plots of Fig. 4, it turns out that for fixed
$eB$, $u_{\sigma}^{\mbox{\tiny{trans}}}$ decreases with increasing
$T$, while $u_{\pi^{0}}^{\mbox{\tiny{trans}}}$ increases with
increasing $T$. The same behavior is also observed for the $T$
dependence of the mass of neutral mesons $m_{\sigma}$ and
$m_{\pi^{0}}$ for a fixed $eB$ (see Fig. \ref{fig2}). In this paper,
we do not intend to describe the nontrivial relation between
$u_{M}^{\mbox{\tiny{trans}}}$ and $m_{M}$, with $M\in
\{\sigma,\pi^{0}\}$. All we want to do is use the neutral pions'
directional refraction indices, and study the low energy GT and GOR
relations satisfied by these neutral and massive pions. To do this,
we will first modify, in the next section, the PCAC relation and
prove the GT and GOR relations for $T\neq 0$ and/or $eB\neq 0$ in a
model-independent way. In Sec. \ref{sec4}, we will then use the
numerical data from the present section to prove the GT and GOR
relations in a two-flavor magnetized NJL model at finite $T$. We
will show, in particular, the role played by pions' directional
refraction indices in satisfying these relations.

\section{Directional weak decay constant of pions in random phase approximation}\label{sec3}
\setcounter{equation}{0}
\par\noindent
It is known that finite temperature and/or constant magnetic fields
explicitly break the relativistic invariance. One of the main
consequences of this explicit Lorentz symmetry breaking is that the
massive free particles satisfy nontrivial anisotropic energy
dispersion relations, including nontrivial directional energy
refraction indices. As concerns the massive neutral pions, the
anisotropic energy dispersion relation is given by (\ref{NA31a})
with $M=\pi^{0}$. Whereas at $T=eB=0$ all refraction indices are
equal to unity, at $T\neq 0$ and $eB=0$, a single refraction index
is defined by
$u=u_{\pi^{0}}^{(1)}=u_{\pi^{0}}^{(2)}=u_{\pi^{0}}^{(3)}$, and at
$T\neq 0$ and $eB\neq 0$, we have
$u_{\pi^{0}}^{(1)}=u_{\pi^{0}}^{(2)}\neq u_{\pi^{0}}^{(3)}$.
\par
It is the purpose of this paper to determine the weak decay constant
of neutral pions, $f_{\pi}$, at $T\neq 0$ and $eB\neq 0$. To do
this, we will generalize in this section the method presented in
\cite{buballa2004, buballa2000}, where $f_{\pi}$ is computed at zero
$T$ and $eB$ using a multiflavor NJL model. To do this, we will
first review in Sec. \ref{subsec3-A} the method presented in
\cite{buballa2000, buballa2004}, by assuming that pions satisfy
ordinary isotropic energy dispersion relations
$E_{\pi}^{2}=q_{i}^{2}+m_{\pi}^{2}$. Combining the usual PCAC
relation and the Feynman integral corresponding to a
one-pion-to-vacuum matrix element in a random phase approximation
(RPA), we arrive at the main relation leading to $f_{\pi}$ at zero
$(T,\mu,eB)$. As a by-product the quark-pion coupling constant
$g_{qq\pi}$ will also be defined in terms of a quark-antiquark
polarization loop, in the pion channel. In \cite{buballa2004}, it is
argued that in the chiral limit $m_{0}\to 0$, $f_{\pi}$ and
$g_{qq\pi}$ satisfy the GT and GOR relations. We will review the
method presented in \cite{buballa2000}, where these relations are
proved in the leading and next-to-leading order $1/N_{c}$ expansion.
\par
In Sec. \ref{subsec3-B}, we will then generalize the arguments in
\cite{buballa2000, buballa2004} for free pions satisfying the
anisotropic energy dispersion relation,
$E_{\pi_{\ell}}^{2}=u_{\pi_{\ell}}^{(i)2}q_{i}^{2}+m_{\pi_{\ell}}^{2}$.
To do this, we will first define the directional quark-pion coupling
constant $g_{qq\pi_{\ell}}^{(\mu)}, \mu=0,\cdots,3,$ in terms of
nontrivial form factors ${\cal{F}}_{\ell\ell}^{\mu\nu}$ and
refraction indices $u_{\pi_{\ell}}^{(\mu)}= (1,
u_{\pi_{\ell}}^{(i)})$. We will then introduce a modified PCAC
relation including ${\cal{F}}_{\ell\ell}^{\mu\nu}$, the directional
refraction indices $u_{\pi_{\ell}}^{(\mu)}$ and a dimensionful
proportionality factor $f_{b}$. The directional decay constant of
pions, $f_{\pi_{\ell}}^{(\mu)}$, is then defined by combining
${\cal{F}}_{\ell\ell}^{\mu\nu}$ and $f_{b}$. Following the same
method as in Sec. \ref{subsec3-A}, we will eventually arrive at the
main relation leading to $f_{\pi_{\ell}}^{(\mu)}$. We will show that
$f_{\pi_{\ell}}^{(\mu)}$ and $g_{qq\pi_{\ell}}^{(\mu)}$ satisfy
modified GT and GOR relations.
\subsection{Quark-pion coupling constant and pion decay constant from an isotropic
energy dispersion relation}\label{subsec3-A}
\begin{figure}[t]
\includegraphics[width=16cm,height=20cm]{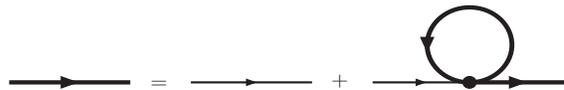}
\par\vspace{-16.5cm}
\caption{The Schwinger-Dyson gap equation for the quark propagator
in the four-fermi NJL model in the Hartree approximation. The bare
(dressed) propagator is denoted by thin (thick) lines.}\label{fig5}
\end{figure}
\par\noindent
Let us start with the Lagrangian density (\ref{NA1a}) of a
two-flavor NJL model with $A_{\mu}^{\mbox{\tiny{ext.}}}=0$. As we
have mentioned in the previous section, for sufficiently strong $G$,
this model exhibits, in the chiral limit $m_{0}\to 0$, a dynamical
mass generation. The resulting mass gap is determined via minimizing
the effective (thermodynamic) potential (free energy), in a mean
field approximation. Equivalently, it is given by solving the
corresponding Schwinger-Dyson (SD) equation for the quark
propagator. In the Hartree approximation, the constituent quark mass
$m$ is given by
\begin{eqnarray}\label{NE1c}
m=m_{0}+\Sigma_{H},
\end{eqnarray}
where $m_{0}$ is the bare quark mass and $\Sigma_{H}$ is the
self-energy of quarks within the four-Fermi NJL model. In the lowest
order perturbative expansion, Eq. (\ref{NE1c}) reduces to
$m=m_{0}+\sigma_{0}$, where
$\sigma_{0}=-2G\langle\bar{\psi}\psi\rangle$ is the quark
condensate. This is also consistent with the diagrammatic
representation of the SD equation, presented in Fig. \ref{fig5}. The
SD relation leads to
\begin{eqnarray}\label{NE2c}
\hspace{-0.5cm}m=m_{0}+\sigma_{0}\equiv
m_{0}+2iG\int\frac{d^{4}p}{(2\pi)^{4}}\mbox{tr}_{sfc} S(p).
\end{eqnarray}
Here, $S(p)=(\gamma\cdot p-m)^{-1}$ is the dressed quark propagator.
The nontrivial solution of the above integral equation leads to a
mass gap $\Delta E=2m$ in the quark spectrum.
\par
\begin{figure}[b]
\includegraphics[width=14.2cm,height=19cm]{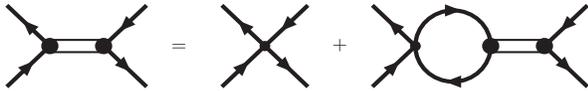}
\par\vspace{-15cm}
\caption{The Bethe-Salpeter equation for quark-antiquark scattering
in RPA \cite{buballa2004}. The double lines denote the meson
propagator $-(q^{2}-m_{M}^{2})^{-1}$ appearing in (\ref{NE5c}), and
the solid lines indicate the dressed quark propagators
$S(p)=(\gamma\cdot p-m)^{-1}$, including the constituent quark mass
$m$. Fat vertices denote the quark-meson coupling constant defined
in (\ref{NE7c}).}\label{fig6}
\end{figure}
\par
\begin{figure}[t]
\includegraphics[width=16cm,height=20cm]{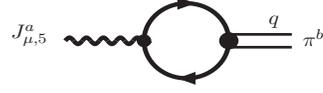}
\par\vspace{-16cm}
\caption{One-pion-to-vacuum matrix element, $\langle 0|J_{\mu,
5}^{a}(0)|\pi^{b}(q)\rangle$, in RPA, leading to pion decay
constant. Here, $J_{\mu,5}^{a}\equiv
\bar{\psi}\gamma_{\mu}\gamma_{5}\lambda^{a}\psi$ is the axial vector
current, with $\lambda^{a}\equiv \tau^{a}/2$. The solid lines are
the dressed quark propagator, $S(p)=(\gamma\cdot p-m)^{-1}$,
including the constituent quark mass $m$. The quark-pion coupling
$g_{qq\pi_{a}}$ (fat vertex) is given in (\ref{NE7c}).}\label{fig7}
\end{figure}
\par
Similarly, mesons are described via a Bethe-Salpeter (BS) equation. In Fig. \ref{fig6}, the BS equation for
quark-antiquark scattering is demonstrated in RPA. In this approximation, the quark-antiquark scattering matrix
\begin{eqnarray}\label{NE3c}
T_{M}(q^{2})=\frac{2G}{1-2G\Pi_{M}(q^{2})},
\end{eqnarray}
 is given in terms of the quark-antiquark polarization $\Pi_{M}(q^{2})$ in the $M\in\{\sigma, \pi\}$ channel
\begin{eqnarray}\label{NE4c}
\Pi_{M}(q^{2})=i\int d^{4}x~\mbox{tr}_{sfc}\big[\Gamma_{M}S(x)\Gamma_{M}S(-x)\big]~e^{iq\cdot x}.\nonumber\\
\end{eqnarray}
Here, $\Gamma_{\sigma}=1$ and
$\Gamma_{\pi_{a}}=i\gamma_{5}\tau_{a}$. The scattering matrix
$T_{M}(q^{2})$ can equivalently be written as an effective meson
exchange between the external quark-antiquark legs [see the
left-hand side (l.h.s.) of the relation appearing in Fig.
\ref{fig6}],
\begin{eqnarray}\label{NE5c}
T_{M}(q^{2})=-\frac{g^{2}_{qqM}}{q^{2}-m_{M}^{2}},
\end{eqnarray}
where $g_{qqM}$ is the quark-meson coupling constant and $m_{M}$ is
the meson pole mass. Note that the denominator
 in (\ref{NE5c}) reflects the isotropic meson energy dispersion relation 
\begin{eqnarray}\label{NE6c}
E_{M}^{2}=q_{i}^{2}+m_{M}^{2}, ~~\mbox{for}~~M\in\{\sigma,
\vec{\pi}\}.
\end{eqnarray}
Equating $T_{M}(q^{2})$ from (\ref{NE3c}) with (\ref{NE5c}) leads to
the definition of the quark-meson coupling constant in terms of
$\Pi_{M}(q^{2})$,
\begin{eqnarray}\label{NE7c}
g_{qqM}^{-2}=\frac{d\Pi_{M}}{dq^{2}}\bigg|_{q^{2}=m_{M}^{2}},
\end{eqnarray}
as well as
\begin{eqnarray}\label{NE8c}
1-2G\Pi_{M}(q^{2}=m_{M}^{2})=0,
\end{eqnarray}
which is equivalently given by \cite{buballa2000}
\begin{eqnarray}\label{NE9c}
2G\Pi_{M}(q^{2}=0)=\frac{\sigma_{0}}{m}\to 1,~~\mbox{for}~~m_{0}\to 0.
\end{eqnarray}
As it is argued in \cite{buballa2000}, relations (\ref{NE8c}) and
(\ref{NE9c}) are consistent with Hartree approximation (\ref{NE1c})
as well as the RPA scheme. To determine the pion decay constant
$f_{\pi}$, we consider the one-pion-to-vacuum matrix element
$\langle 0|J_{\mu, 5}^{a}(0)|\pi^{b}(q)\rangle$, which satisfies the
usual PCAC relation\footnote{For $eB=0$, because of the Lorentz and
isospin symmetry of the original Lagrangian, there is no difference
between the properties of different species of pions $\pi_{\ell},
\ell=1,2,3$. In so far, we will skip the indices $\ell=1,2,3$ on
$\pi_{\ell}$ as long as these symmetries hold.}
\begin{eqnarray}\label{NE10c}
\langle 0|J_{\mu, 5}^{a}(0)|\pi^{b}(q)\rangle=f_{\pi}q_{\mu}\delta^{ab}.
\end{eqnarray}
Here, $J_{\mu,5}^{a}\equiv
\bar{\psi}\gamma_{\mu}\gamma_{5}\lambda^{a}\psi$ with
$\lambda^{a}\equiv \tau^{a}/2$ is the axial vector current of an
axial $SU(2_{f})$ global transformation. The matrix element
appearing on the r.h.s. of (\ref{NE10c}) is demonstrated
diagrammatically in Fig. \ref{fig7}. Note that in RPA, the quark
propagator appearing in the quark-antiquark loop in Fig. \ref{fig7}
is the dressed quark propagator $S(p)=(\gamma\cdot p-m)^{-1}$,
including the constituent quark mass $m$. Combining the PCAC
relation (\ref{NE10c}) and the Feynman integral corresponding to the
one-pion-to-vacuum amplitude demonstrated in Fig. \ref{fig7}, we
arrive at
\begin{eqnarray}\label{NE11c}
f_{\pi}q_{\mu}\delta^{ab}=g_{qq\pi}\Upsilon_{\mu}^{ab}(q),
\end{eqnarray}
with
\begin{eqnarray}\label{NE12c}
\Upsilon_{\mu}^{ab}(q)\equiv i\int d^{4}x~\mbox{tr}_{sfc}\big[\gamma_{\mu}\gamma_{5}\lambda^{a}S(x)
\gamma_{5}\tau^{b}S(-x)\big]e^{iq\cdot x}.\hspace{-0.3cm}\nonumber\\
\end{eqnarray}
Contracting (\ref{NE11c}) with $q^{\mu}$ and using the energy
dispersion relation (pion on-mass-shell condition)
$q^{2}=m_{\pi}^{2}$, we arrive finally at $f_{\pi}$ in terms of
$g_{qq\pi}$ and $\Upsilon_{\mu}^{ab}(q)$. As it is shown in
\cite{buballa2000}, $f_{\pi}$ satisfies the GT relation
\begin{eqnarray}\label{NE13c}
g_{qq\pi}f_{\pi}=m+{\cal{O}}(m_{0}^{2}),
\end{eqnarray}
and the GOR relation
\begin{eqnarray}\label{NE14c}
m_{\pi}^{2}f_{\pi}^{2}=\frac{m_{0}\sigma_{0}}{2G}+{\cal{O}}(m_{0}^{2}),
\end{eqnarray}
up to second order in $m_{0}$. Let us emphasize again that the
isotropic energy dispersion relation (\ref{NE6c}) plays an important
role in proving (\ref{NE13c}) and (\ref{NE14c}) (see
\cite{buballa2000} for more details). In the next section, we will
show how the above relations shall be modified, when we start from a
nontrivial anisotropic energy dispersion relation
$q_{0}^{2}=u_{\pi_{\ell}}^{(i)2}q_{i}^{2}+m_{\pi_{\ell}}^{2}$ for
free $\pi_{\ell}, \ell=1,2,3$ mesons. In Sec. \ref{sec4}, we will
use the results arising from this general consideration to determine
the directional quark-pion coupling constant and decay constant for
neutral pions in a hot and magnetized medium.
\subsection{Directional quark-pion coupling constant and pion decay constant
from an anisotropic energy dispersion
relation}\label{subsec3-B}\par\noindent
Let us consider the Lagrangian density
\begin{eqnarray}\label{NE15c}
\hspace{-0.5cm}{\cal{L}}_{\pi_{\ell}}=-\frac{1}{2}\Pi_{\ell}(x)
\left(m_{\pi_{\ell}}^{2}+g^{\mu\mu}u_{\pi_{\ell}}^{(\mu)2}\partial_{\mu}^{2}\right)\Pi_{\ell}(x),
\end{eqnarray}
for each pion species $\pi_{\ell}, \ell=1,2,3$. The above Lagrangian
density is comparable to the Lagrangian of pion fields appearing in
the effective action (\ref{NA15a}) for free mesons and leads to a
nontrivial anisotropic energy dispersion relation
\begin{eqnarray}\label{NE16c}
E_{\pi_{\ell}}^{2}=u_{\pi_{\ell}}^{(i)2}q_{i}^{2}+m_{\pi_{\ell}}^{2},\qquad \forall \ell=1,2,3,
\end{eqnarray}
similar to (\ref{NA31a}). In (\ref{NE15c}), we have introduced the
``normalized'' pion field $\Pi_{\ell}(x)\equiv
|{\cal{F}}_{\ell\ell}^{00}|^{1/2}\pi_{\ell}(x)$, where the form
factors ${\cal{F}}_{\ell\ell}^{00}$ appear in the effective action
(\ref{NA15a}). Moreover, in analogy to (\ref{NA32a}) and
(\ref{NA33a}), the mass $m_{\pi_{\ell}}$ and a four-vector for the
refraction index, $u_{\pi_{\ell}}^{(\mu)}$, for each pion species
$\ell=1,2,3$, are defined by the pion mass squared matrix
$M_{\pi_{\ell}}^{2}$ and form factors
${\cal{F}}_{\ell\ell}^{\mu\nu}$,
\begin{eqnarray}\label{NE17c}
m_{\pi_{\ell}}\equiv\bigg|\frac{M_{\pi_{\ell}}^{2}}{{\cal{F}}^{00}_{\ell\ell}}\bigg|^{1/2},\qquad
u^{(\mu)}_{\pi_{\ell}}\equiv (1,u^{(i)}_{\pi_{\ell}})=\bigg|\frac{{\cal{F}}_{\ell\ell}^{\mu\mu}}
{{\cal{F}}_{\ell\ell}^{00}}\bigg|^{1/2}.\nonumber\\
\end{eqnarray}
Using ${\cal{L}}_{\pi_{\ell}}$ from (\ref{NE15c}), the pion
propagator $D_{\pi_{\ell}}(x,y) \equiv\langle
0|T\left(\pi_{\ell}(x)\pi_{\ell}(y)\right)|0\rangle$ in the momentum
space reads
\begin{eqnarray}\label{NE18c}
\widetilde{D}_{\pi_{\ell}}(q)=-\frac{|{\cal{F}}_{\ell
\ell}^{00}|^{-1}}{q_{0}^{2}-
u_{\pi_{\ell}}^{(i)2}q_{i}^{2}-m_{\pi_{\ell}}^{2}}.
\end{eqnarray}
The metric $g=\mbox{diag}(1,-1,-1,-1)$. In contrast to (\ref{NE5c}),
the denominator reflects the anisotropic pion on-mass-shell
condition
\begin{eqnarray}\label{NE19c}
q_{0}^{2}-u_{\pi_{\ell}}^{(i)2}q_{i}^{2}=m_{\pi_{\ell}}^{2},\qquad \forall \ell=1,2,3,
\end{eqnarray}
including the directional refraction indices $u_{\pi_{\ell}}^{(i)}$.
The factor $|{\cal{F}}_{\ell\ell}^{00}|^{-1}$ in (\ref{NE18c})
arises from the normalization of pion fields, as the pion propagator
can equivalently be defined by $D_{\Pi_{\ell}}(x,y)\equiv\langle
0|T\left(\Pi_{\ell}(x)\Pi_{\ell}(y)\right)|0\rangle$ with
$D_{\Pi_{\ell}}=|{\cal{F}}_{\ell\ell}^{00}|D_{\pi_{\ell}}$. Using
(\ref{NE18c}), the quark-antiquark scattering amplitude
$T_{M}(q^{2})$ from (\ref{NE5c}) with $M=\pi_{\ell}$ is then
modified as
\begin{eqnarray}\label{NE20c}
T_{\pi_{\ell}}(q^{2})=-\frac{g_{qq\pi_{\ell}}^{2}|{\cal{F}}_{\ell\ell}^{00}|^{-1}}
{g^{\mu\mu}u_{\pi_{\ell}}^{(\mu)2}q_{\mu}^{2}-m_{\pi_{\ell}}^{2}}.
\end{eqnarray}
Combining the ``bare'' quark-pion coupling constant
$g_{qq\pi_{\ell}}$ and the normalization parameter (form factor)
$|{\cal{F}}_{\ell\ell}^{00}|$ appearing also in the definition of
normalized pion fields, a ``normalized'' coupling constant
$g_{qq\Pi_{\ell}}$ is defined as
\begin{eqnarray}\label{NE21c}
g_{qq\Pi_{\ell}}\equiv
g_{qq\pi_{\ell}}|{\cal{F}}_{\ell\ell}^{00}|^{-1/2}.
\end{eqnarray}
Equating, at this stage, $T_{\pi_{\ell}}(q^{2})$ from (\ref{NE20c})
and the quark-antiquark scattering amplitude from (\ref{NE3c}) with
$M=\pi_{\ell}$, we arrive first at
\begin{eqnarray}\label{NE22c}
\hspace{-0.5cm}g^{\mu\mu}\frac{d\Pi_{\pi_{\ell}}(q^{2})}{d
q_{\mu}^{2}}\bigg|_{\tilde{q}_{\pi_{\ell}}=(m_{\pi_{\ell}},\mathbf{0})}=u_{\pi_{\ell}}^{(\mu)2}|{\cal{F}}_{\ell\ell}^{00}|g_{qq\pi_{\ell}}^{-2},
\end{eqnarray}
$\forall \mu=0,\cdots,3$ and $\forall \ell=1,2,3$. Note that in
contrast to (\ref{NE7c}), the expression on the l.h.s. of
(\ref{NE22c}) is to be evaluated in the rest frame of $\ell$th pion
species, i.e. in
$\tilde{q}_{\pi_{\ell}}=(m_{\pi_{\ell}},{\mathbf{0}})$. Defining, at
this stage, a directional quark-pion coupling constant
$g_{qq\pi_{\ell}}^{(\mu)}$ by
\begin{eqnarray}\label{NE23c}
\hspace{-0.5cm}
\big(g_{qq\pi_{\ell}}^{(\mu)}(\tilde{q}_{\pi_{\ell}})\big)^{-2}\equiv
g^{\mu\mu} \frac{d\Pi_{\pi_{\ell}}(q^{2})}{d
q_{\mu}^{2}}\bigg|_{\tilde{q}_{\pi_{\ell}}=(m_{\pi_{\ell}},
\mathbf{0})},
\end{eqnarray}
plugging  this definition into the l.h.s. of (\ref{NE22c}), and
using (\ref{NE17c}), we get
\begin{eqnarray}\label{NE24c}
g^{(\mu)}_{qq\pi_{\ell}}=g_{qq\pi_{\ell}}|{\cal{F}}_{\ell\ell}^{\mu\mu}|^{-1/2}.
\end{eqnarray}
Similarly, the directional quark-sigma meson coupling constant is defined by
\begin{eqnarray}\label{NE25c}
\big(g_{qq\sigma}^{(\mu)}(\tilde{q}_{\sigma})\big)^{-2}\equiv g^{\mu\mu}
\frac{d\Pi_{\sigma}(q^{2})}{d q_{\mu}^{2}}\bigg|_{\tilde{q}_{\sigma}=(m_{\sigma}, \mathbf{0})}.
\end{eqnarray}
It satisfies
\begin{eqnarray}\label{NE26c}
g^{(\mu)}_{qq\sigma}=g_{qq\sigma}|{\cal{G}}^{\mu\mu}|^{-1/2}.
\end{eqnarray}
As it turns out, the directional coupling constant
$g_{qq\pi_{\ell}}^{(\mu)}$ from (\ref{NE23c}) plays a crucial role
in the definition of the directional weak decay constant of pions.
This is determined by introducing the modified PCAC relation, which
shall replace (\ref{NE10c}), whenever pions satisfy the anisotropic
energy dispersion relation (\ref{NE16c}),
\begin{eqnarray}\label{NE27c}
\hspace{-0.3cm}\langle 0|J_{\mu,5}^{\ell}(0)|\Pi^{m}(q)\rangle=f_{\ell}|{\cal{F}}_{\ell\ell}^{00}|^{1/2}
u_{\pi_{\ell}}^{(\mu)2}q_{\mu}\delta^{\ell m},
\end{eqnarray}
$\forall \mu=0,\cdots,3$ and $\forall \ell,m=1,2,3$. Here,
$f_{\ell}$ is an unknown dimensionful constant, which depends on
$(T,\mu,eB)$, whenever they are nonvanishing. Later, we will show
that, for neutral pions in the chiral limit $m_{0}\to 0$,
$f_{3}=\sigma_{0}$, where $\sigma_{0}$ is the chiral condensate. For
nonvanishing $m_{0}$, we get $f_{3}=m$, where $m=m_{0}+\sigma_{0}$.
Note that in the chiral limit, where pions are (massless) Goldstone
bosons, Eq. (\ref{NE27c}) leads to
\begin{eqnarray}\label{NE28c}
\hspace{-0.7cm}0=\langle 0|\partial^{\mu}J_{\mu,5}^{\ell}(0)|\Pi^{m}(q)\rangle=
f_{\ell}|{\cal{F}}_{\ell\ell}^{00}|^{1/2}m_{\pi_{\ell}}^{2}\delta^{\ell m},
\end{eqnarray}
in accordance with the Goldstone theorem. Here, the pion
on-mass-shell condition (\ref{NE19c}) and the axial $SU(2_{f})$
invariance, $\partial^{\mu}J_{\mu,5}^{\ell}(x)=0$, are used. Using
the Feynman diagram appearing in Fig. \ref{fig7} for normalized pion
fields, $\Pi_{\ell}$, the one-pion-to-vacuum matrix element $\langle
0|J_{\mu,5}^{\ell}(0)|\Pi^{m}(q)\rangle$ on the l.h.s. of
(\ref{NE27c}) is given by
\begin{eqnarray}\label{NE29c}
\langle 0|J_{\mu,5}^{\ell}(0)|\Pi^{m}(q)\rangle=g_{qq\Pi_{\ell}}\Upsilon_{\mu}^{\ell m}(q),
\end{eqnarray}
where $g_{qq\Pi_{\ell}}$ and $\Upsilon^{\ell m}(q)$ are given in
(\ref{NE21c}) and (\ref{NE12c}), respectively. Relation
(\ref{NE29c}) leads, upon using (\ref{NE21c}) and (\ref{NE24c}), as
well as the modified PCAC relation (\ref{NE27c}), to
\begin{eqnarray}\label{NE30c}
f_{\pi_{\ell}}^{(\mu)}q_{\mu}\delta^{\ell m}=g^{(\mu)}_{qq\pi_{\ell}}\Upsilon_{\mu}^{\ell m}(q),
\end{eqnarray}
$\forall \mu=0,\cdots,3$ and $\forall \ell,m=1,2,3$.  Here, the
directional pion decay constant $f_{\pi_{\ell}}^{(\mu)}$ for the
$\ell$th pion species is defined by
\begin{eqnarray}\label{NE31c}
f_{\pi_{\ell}}^{(\mu)}\equiv f_{\ell}|{\cal{F}}_{\ell\ell}^{\mu\mu}|^{1/2}.
\end{eqnarray}
In the next section, we will use (\ref{NE30c}) to determine
$f_{\pi^{0}}^{(\mu)}$ for $\mu=0,\cdots,3$.
\par
Similar to the previous
section, the directional quark-pion coupling and decay constants
satisfy the modified GT and GOR relations,
\begin{eqnarray}\label{NE32c}
f_{\pi_{\ell}}^{(\mu)}g_{qq\pi_{\ell}}^{(\mu)}=m+{\cal{O}}(m_{0}^{2}),
\end{eqnarray}
and
\begin{eqnarray}\label{NE33c}
m_{\pi_{\ell}}^{2}f_{\pi_{\ell}}^{(\mu)2}=u_{\pi_{\ell}}^{(\mu)2}\frac{m_{0}\sigma_{0}}{2G}+{\cal{O}}(m_{0}^{2}),
\end{eqnarray}
where $u_{\pi_{\ell}}^{(\mu)}$ is defined in (\ref{NE17c}). To prove
these relations, we generalize the method described in
\cite{buballa2000} for the case when pions satisfy nontrivial,
anisotropic energy dispersion relation (\ref{NE16c}).
\par
As concerns (\ref{NE32c}), let us consider the modified PCAC
relation (\ref{NE27c}) and combine it with (\ref{NE29c}).
Differentiating the l.h.s. of (\ref{NE27c}) and using
\cite{buballa2004}
\begin{eqnarray}\label{NE34c}
\sum_{\mu=0}^{3}q^{\mu}\Upsilon_{\mu}^{\ell\ell}(q)=m[\Pi_{\pi_{\ell}}(q^{2})-\Pi_{\pi_{\ell}}(0)],
\end{eqnarray}
with the quark-antiquark polarization, $\Pi_{\pi_{\ell}}(q^{2})$
from (\ref{NE4c}), and $\Upsilon_{\mu}^{\ell
m}=\Upsilon_{\mu}^{\ell\ell}\delta^{\ell m}$ from (\ref{NE12c}), we
first get
\begin{eqnarray}\label{NE35c}
f_{\ell}|{\cal{F}}_{\ell\ell}^{00}|m_{\pi_{\ell}}^{2}=mg_{qq\pi_{\ell}}[\Pi_{\pi_{\ell}}(q^{2})-\Pi_{\pi_{\ell}}(0)].
\end{eqnarray}
Expanding $\Pi_{\pi_{\ell}}(q^{2})$ on the r.h.s. of the above
relation around $q=0$, and evaluating the resulting expression at
$\tilde{q}_{\pi_{\ell}}$, we obtain
\begin{eqnarray}\label{NE36c}
\hspace{-0.6cm}f_{\ell}|{\cal{F}}_{\ell\ell}^{00}|=mg_{qq\pi_{\ell}}
\frac{d\Pi_{\pi_{\ell}}}{dq_{0}^{2}}\bigg|_{\tilde{q}_{\pi_{\ell}}=(m_{\pi_{\ell}},{\mathbf{0}})}
+{\cal{O}}(m_{\pi_{\ell}}^{4}),
\end{eqnarray}
which leads, upon using (\ref{NE22c}) for $\mu=0$, to
\begin{eqnarray}\label{NE37c}
g_{qq\pi_{\ell}}f_{\ell}=m+{\cal{O}}(m_{0}^{2}).
\end{eqnarray}
Plugging (\ref{NE24c}) as well as (\ref{NE31c}) into (\ref{NE37c}),
we arrive finally at the GT relation (\ref{NE32c}), including
directional quark-pion coupling and pion decay constants. To show
the GOR relation (\ref{NE33c}), we use (\ref{NE8c}) in a slightly
modified form,
\begin{eqnarray}\label{NE38c}
1-2G\Pi_{\pi_{\ell}}(\tilde{q}_{\pi_{\ell}}=(m_{\pi_{\ell}},{\mathbf{0}}))=0.
\end{eqnarray}
This relation arises, similarly to (\ref{NE8c}), by equating
(\ref{NE20c}) and (\ref{NE3c}) with $M=\pi_{\ell}$. Expanding  the
l.h.s. of (\ref{NE38c}) around $q^{2}=0$ and evaluating the
resulting expression at $\tilde{q}_{\pi_{\ell}}=(m_{\pi_{\ell}},
{\mathbf{0}})$, we get first
\begin{eqnarray}\label{NE39c}
0&=&1-2G\Pi_{\pi_{\ell}}(0)-2Gm_{\pi}^{2}\frac{d\Pi_{\pi_{\ell}}(q^2)}{dq_{0}^{2}}\bigg|_{\tilde{q}_{\pi_{\ell}}
=(m_{\pi_{\ell}},{\mathbf{0}})}\nonumber\\
&&+{\cal{O}}(m_{\pi_{\ell}}^{4}),
\end{eqnarray}
which yields
\begin{eqnarray}\label{NE40c}
m_{\pi_{\ell}}^{2}=\frac{g_{qq\pi_{\ell}}^{2}m_{0}}{2Gm|{\cal{F}}_{\ell\ell}^{00}|}+{\cal{O}}(m_{\pi_{\ell}}^{4}).
\end{eqnarray}
Here,  Eq. (\ref{NE9c}) and the definition of
$g_{qq\pi_{\ell}}^{(\mu)}$ from (\ref{NE23c}) are used. Plugging
further the GT relation (\ref{NE37c}) and (\ref{NE31c}), as well as
the definition of $u_{\pi_{\ell}}^{(\mu)2}$ from (\ref{NE17c}) into
the r.h.s. of (\ref{NE40c}), we arrive finally at the modified GOR
relation (\ref{NE33c}), including $f_{\pi_{\ell}}^{(\mu)}$.
\par
As it turns out, whereas (\ref{NE34c}) plays an important role in
proving (\ref{NE13c}) and (\ref{NE14c}), it is not so crucial to
derive (\ref{NE32c}) and (\ref{NE33c}). In the next section, we will
first use the above results to derive a number of analytical
expressions for the directional quark-pion coupling and decay
constants for neutral pions. Using these expressions, together with
the modified PCAC relation (\ref{NE27c}), we will then analytically
prove the GT and GOR relations (\ref{NE32c}) and (\ref{NE33c}),
without using (\ref{NE34c}). We will then verify these low energy
relations for neutral pions numerically and eventually present
numerical results for $g_{qq\pi^{0}}^{(\mu)}$ and
$f_{\pi^{0}}^{(\mu)}$.
\section{Directional quark-pion coupling and pion decay constants at finite $(T,\mu,eB)$}\label{sec4}
\subsection{Analytical results}\label{subsec4-A}
\subsubsection{Directional quark-meson couplings for neutral mesons}\label{subsec4-A-1}\par\noindent
\setcounter{equation}{0}
\par\noindent
Let us consider the quark-antiquark polarization $\Pi_{M}(q^{2})$
from (\ref{NE4c}) in the $M\in\{\sigma,\pi^{0}\}$ channel. In the
presence of an external magnetic field the fermion propagators
$S(x)$ appearing in (\ref{NE4c}) are to be replaced by Ritus
propagators, $S_{Q}(x)$ from (\ref{NA24a})-(\ref{NA27a}). Plugging
(\ref{NA24a}) into (\ref{NE4c}), we arrive first at
\begin{eqnarray}\label{ND1}
\lefteqn{\Pi_{M}(q)}\nonumber\\
&&\hspace{-0.5cm}=i\sum_{p,k=0}^{\infty}\int\ {\cal{D}}\tilde{p}\ {\cal{D}}\tilde{k}
\int d^{4}x~e^{-i(\tilde{p}-\tilde{k}-q)\cdot x}\Lambda^{M}_{pk}(\bar{p},\bar{k},x_1),\nonumber\\
\end{eqnarray}
where
\begin{eqnarray}\label{ND2}
\Lambda^{M}_{pk}(\bar{p},\bar{k},x_1)&=&\mbox{tr}_{sfc}\big[\Gamma_{M}P_{p}(x_{1})S_{Q}(\bar{p})P_{p}(0)\Gamma_{M}
\nonumber\\
&&\times K_{k}(0)S_{Q}(\bar{k})K_{k}(x_{1})\big].
\end{eqnarray}
Here, $\Gamma_{\sigma}=1$ and
$\Gamma_{\pi^{0}}=i\gamma_{5}\tau^{3}$, and
$\tau^{3}=\mbox{diag}(1,-1)$ is the third Pauli matrix. The momenta
$\tilde{p}$ and $\bar{p}$ are defined below (\ref{NA24a}). In
(\ref{ND1}), integrating first over $x_{0},x_{2},x_{3}$ and then
over $p_{0},p_{2}, p_{3}$ components, we arrive at the boundary
conditions $p_{i}=k_{i}+q_{i}$ for $i=0,2,3$. In what follows, we
will abbreviate these boundary conditions by a subscript ``b.c.''
Plugging $P_{p}(x_{1})$ from (\ref{NA25a}) into (\ref{ND2}), we
arrive first at
\begin{eqnarray}\label{ND3}
\lefteqn{\hspace{-0.5cm}\Pi_{M}(q)
=i\sum_{p,k=0}^{\infty}\int\ \frac{dk_{0}dk_{3}}{(2\pi)^{3}}
}\nonumber\\
&&\times \int dk_{2}dx_{1}~e^{-iq_{1}x_{1}}\Lambda^{M}_{pk}(\bar{p},\bar{k},x_1)\bigg|_{\mbox{\tiny{b.c.}}},
\end{eqnarray}
with
\begin{widetext}
\begin{eqnarray}\label{ND4}
\hspace{-1cm}\Lambda^{M}_{pk}(\bar{p},\bar{k},x_{1})&=&-12\sum_{q_f}\bigg\{\frac{\bar{k}.\bar{p}+\kappa_{M}m^2}{(\bar{p}^2-m^2)(\bar{k}^2-m^2)}
[A^{+(0)}_{pk}\alpha^+_{pk}+A^{-(0)}_{pk}\alpha^-_{pk}]+\frac{2\bar{p}_2\bar{k}_2}{(\bar{p}^2-m^2)
(\bar{k}^2-m^2)}A^{-(0)}_{pk}\alpha^-_{pk}\bigg\},
\end{eqnarray}
\end{widetext}
where $\kappa_{\pi^{0}}=-1$ and $\kappa_{\sigma}=1$. From the decomposition
\begin{eqnarray}\label{ND5}
P_{p}(0)K_{k}(0)&\equiv&\alpha_{pk}^{+}(p_{2},k_{2})+is\gamma_{1}\gamma_{2}\alpha^{-}(p_{2},k_{2}),\nonumber\\
K_{k}(x_{1})P_{p}(x_{1})&\equiv&A_{pk}^{+(0)}(x_{1})+is\gamma_{1}\gamma_{2}A_{pk}^{-(0)}(x_{1}),\nonumber\\
\end{eqnarray}
we have
\begin{eqnarray}\label{ND6}
\alpha_{pk}^{\pm(0)}(p_{2},k_{2})&\equiv&\frac{1}{2}\big[f_{p}^{+s}(0)f_{k}^{+s}(0)
\nonumber\\
&&\pm\Pi_{p}\Pi_{k}f_{p}^{-s}(0)f_{k}^{-s}(0)\big],
\end{eqnarray}
and
\begin{eqnarray}\label{ND7}
A_{pk}^{\pm(0)}(x_{1})&\equiv&\frac{1}{2}\big[f_{p}^{+s}(x_{1})f_{k}^{+s}(x_{1})\nonumber\\
&&\pm\Pi_{p}\Pi_{k}f_{p}^{-s}(x_{1})f_{k}^{-s}(x_{1})\big].
\end{eqnarray}
In (\ref{ND4}), the summation over $q_{f}\in\{2/3,-1/3\}$ replaces
the trace in the flavor space. Moreover, we set $N_{c}=3$.  In what
follows, we will use $\Pi_{M}(q^{2})$ from (\ref{ND3}) to derive the
directional quark-meson coupling $g_{qqM}^{(\mu)},
M\in\{\sigma,\pi^{0}\}$ in the direction longitudinal, $\mu=(0,3)$,
and transverse, $\mu=(1,2)$, with respect to the direction of the
external magnetic field. Here, the definitions (\ref{NE23c}) with
$\ell=3$ and (\ref{NE25c}), for
$\big(g_{qqM}^{(\mu)}(\tilde{q}_{M})\big)^{-2}, M\in\{\pi^{0},
\sigma\}$ in the rest frame of mesons $\tilde{q}_{M}\equiv
(m_{M},{\mathbf{0}})$, will be used.
\par
Using $\Pi_{M}(q)$ from (\ref{ND3}) and plugging
$\Lambda_{pk}^{M}(\bar{p},\bar{k},x_{1})$ from (\ref{ND4}) in
\begin{eqnarray}\label{ND8}
\lefteqn{\hspace{-0.5cm}
\big(g^{(\mu)}_{qqM}(\tilde{q}_{M})\big)^{-2}=i\sum_{p,k=0}^{\infty}\int
\frac{dk_{0}dk_{3}}{(2\pi)^{3}}\int dk_{2} dx_{1}
}\nonumber\\
&&\hspace{-0.8cm}\times g^{\mu\mu}\frac{d}{dq^{2}_{\mu}}\bigg[e^{-iq_1x_1}\Lambda^{M}_{pk}(\bar{p},
\bar{k},x_{1})\bigg]\bigg|_{\tilde{q}_{M}, \mbox{\tiny{b.c.}}},
\end{eqnarray}
which arises from (\ref{NE23c}) and (\ref{NE25c}), choosing $\mu=0$,
and eventually using
\begin{eqnarray}\label{ND9}
\hspace{-1.5cm}\int dk_{2}dx_{1}A_{pk}^{-(0)}\alpha_{pk}^{-}\bigg|_{p_{2}=k_{2}}=
\frac{1}{4}|q_{f}eB|\delta_{p,k}\delta_{k,0},
\end{eqnarray}
as well as
\begin{eqnarray}\label{ND10}
\lefteqn{\hspace{-2cm}\int dk_{2}dx_{1}\big[A_{pk}^{+(0)}\alpha_{pk}^{+}+A_{pk}^{-(0)}\alpha_{pk}^{-}\big]\bigg|_{p_{2}=k_{2}}
}\nonumber\\
&&\hspace{1cm}=\frac{1}{2}|q_{f}eB|\alpha_{k}\delta_{p,k},
\end{eqnarray}
we arrive first at
\begin{eqnarray}\label{ND11}
\lefteqn{\hspace{-0.5cm}\big(g^{(0)}_{qqM}(\tilde{q}_{M})\big)^{-2}
=6i\sum_{q_f}|q_feB|\sum_{k=0}^{\infty}\alpha_k\int_0^1 dx~x(1-x)
}\nonumber\\
&&\times\int\frac{dk_0dk_3}{(2\pi)^3}\bigg\{\frac{3}{(\bar{k}^2+x(1-x)m_{M}^{2}-m^2)^2}\nonumber\\
&&-\frac{4\Xi_{M}(x)}{(\bar{k}^2+x(1-x)m_{M}^2-m^2)^3}\bigg\},
\end{eqnarray}
with
\begin{eqnarray}\label{ND12}
\Xi_{\pi^{0}}(x)&\equiv& x(1-x)m_{\pi^{0}}^{2},\nonumber\\
\Xi_{\sigma}(x)&\equiv& x(1-x)m_{\sigma}^{2}-m^{2},
\end{eqnarray}
for a nonzero magnetic field and at $T=\mu=0$. In the above
relations $\bar{k}^{2}= {\mathbf{k}}_{\|}^{2}-2|q_{f}eB|k$ with
${\mathbf{k}}_{\|}^{2}=k_{0}^{2}-k_{3}^{2}$. To derive (\ref{ND9})
and (\ref{ND10}), we use the same technique, developed in
\cite{fayazbakhsh2012} (A number of useful relations, that can be
used to derive (\ref{ND9}) and (\ref{ND10}) are presented in the
Appendix). At finite $(T,\mu,eB)$, we obtain
\begin{eqnarray}\label{ND13}
\lefteqn{\hspace{0cm}\big(g^{(0)}_{qqM}(\tilde{q}_{M})\big)^{-2}=-6\sum_{q_f}|q_{f}eB|
\sum_{k=0}^{\infty}\alpha_{k}\int_0^1dxx(1-x)}\nonumber\\
&&\hspace{-0.5cm}\times \int\frac{dk_3}{(2\pi)^2} \bigg\{3{\cal{S}}^{(0)}_2(\omega^{M}_{k}(x))-
4\Xi_{M}(x){\cal{S}}^{(0)}_3(\omega^{M}_k(x))\bigg\},\nonumber\\
\end{eqnarray}
where $\Xi_{M}(x)$ is given in (\ref{ND12}). To introduce $(T,\mu)$,
Eq. (\ref{NA28a}) is used. In the above relations
$\omega_{k}^{M}(x), M\in\{\sigma,\pi^{0}\}$ is defined by
\begin{eqnarray}\label{ND14}
\omega_{k}^{M}(x)\equiv\big[k_{3}^{2}+2|q_{f}eB|k-x(1-x)m_{M}^{2}+m^{2}\big]^{1/2}.\nonumber\\
\end{eqnarray}
Moreover, similar to the computation presented in
\cite{fayazbakhsh2012}, we have used the functions
\begin{eqnarray}\label{ND15}
{\cal{S}}_{\ell}^{(m)}(\omega)\equiv T\sum\limits_{n=-\infty}^{+\infty}\frac{(k_{0}^{2})^{m}}{(k_{0}^{2}
-\omega^{2})^{\ell}},
\end{eqnarray}
with $\ell\geq 1, m\geq 0$.  Using
\begin{eqnarray}\label{ND16}
{\cal{S}}_{1}^{(0)}(\omega)=\frac{1}{2\omega}[1-N_{f}(\omega)],
\end{eqnarray}
 with $N_{f}(\omega)\equiv n_{f}^{+}(\omega)+n_{f}^{-}(\omega)$
and the fermion distribution functions
\begin{eqnarray}\label{ND17}
n_{f}^{\pm}(\omega)\equiv \frac{1}{e^{\beta(\omega\mp \mu)}+1},
\end{eqnarray}
the following recursion relations can be used to evaluate
${\cal{S}}_{\ell}^{(m)}(\omega)$, $\forall \ell\geq 1, m\geq 0$,
\begin{eqnarray}\label{ND18}
{\cal{S}}_{\ell}^{(0)}(\omega)&=&\frac{1}{2(\ell-1)\omega}\frac{d{\cal{S}}_{\ell-1}^{0}(\omega)}{d\omega},\qquad \ell\geq 2,\nonumber\\
{\cal{S}}_{\ell}^{(m)}(\omega)&=&{\cal{S}}_{\ell-1}^{(m-1)}(\omega)+\omega^{2}{\cal{S}}_{\ell}^{(m-1)}(\omega).
\end{eqnarray}
In Sec. \ref{subsec4-B}, we will numerically evaluate the sum over
Landau levels $k$ and the integration over $k_{3}$ momentum,
appearing in (\ref{ND13}).
\par
To determine $\big(g^{(1)}_{qqM}(\tilde{q}_{M})\big)^{-2}$, we use
(\ref{ND8}) with $\mu=1$. To compute the differentiation with
respect to $q_{1}^{2}$, we use the identity $\frac{d
f(q_{1})}{dq_{1}^{2}}|_{\tilde{q}_{M},\mbox{\tiny{b.c.}}}=\frac{1}{2}\frac{d^{2}f(q_{1})}{dq_{1}^{2}}|_{\tilde{q}_{M},
\mbox{\tiny{b.c.}}}$,
which is correct for $\tilde{q}_{M}=(m_{M},{\mathbf{0}})$. The
integration over $k_{2}$ and $x_{1}$ can be carried out using
\begin{eqnarray}\label{ND19}
\int
dk_{2}dx_{1}x_{1}A_{pk}^{\pm(0)}\alpha_{pk}^{\pm}\bigg|_{p_{2}=k_{2}}=0.
\end{eqnarray}
We arrive first at
\begin{eqnarray}\label{ND20}
\lefteqn{\hspace{-0.8cm}\big(g^{(1)}_{qqM}(\tilde{q}_{M})\big)^{-2}=\frac{i}{2}
\sum\limits_{p,k=0}^{\infty}\int \frac{dk_{0}dk_{3}}{(2\pi)^{3}}
}\nonumber\\
&&\hspace{-1cm}\times \int dk_{2}dx_{1}~e^{-iq_{1}x_{1}}x_{1}^{2}
\Lambda_{pk}^{M}(\bar{p},\bar{k},x_{1})\bigg|_{{\tilde{q}_{M}, \mbox{\tiny{b.c.}}}},
\end{eqnarray}
for $M\in\{\sigma,\pi^{0}\}$. Plugging (\ref{ND4}) into
(\ref{ND20}), and using
\begin{eqnarray}\label{ND21}
\int dk_{2}dx_{1}x_{1}^{2}\big[A_{pk}^{+(0)}\alpha_{pk}^{+}+A_{pk}^{-(0)}\alpha_{pk}^{-}\big]\bigg|_{p_{2}=k_{2}}
&=&\frac{1}{2}C_{pk}^{(1)},\nonumber\\
\int dk_{2}dx_{1}x_{1}^{2}A_{pk}^{-(0)}\alpha_{pk}^{-}\bigg|_{p_{2}=k_{2}}&=&-\frac{1}{4}C_{pk}^{(2)},\nonumber\\
\end{eqnarray}
with $C_{pk}^{(i)}, i=1,2,$ defined by
\begin{eqnarray}\label{ND22}
C_{pk}^{(1)}&=&\delta_{p,k}(4k+\delta_{k,0})-\Pi_{k}(2k-1)\delta_{p,k-1}\nonumber\\
&&-(2k+1)\delta_{p,k+1},\nonumber\\
C_{pk}^{(2)}&=&-\delta_{p,k}\delta_{k,0}+\Pi_{k}\left(2k-1-2\sqrt{k(k-1)}\right)\delta_{p,k-1}\nonumber\\
&&+\left(2k+1-2\sqrt{k(k+1)}\right)\delta_{p,k+1},
\end{eqnarray}
\begin{widetext}
\par\noindent we first arrive for $B\neq 0$ and at $T=\mu=0$, at
\begin{eqnarray}\label{ND23}
\big(g^{(1)}_{qqM}(\tilde{q}_{M})\big)^{-2}&=&-3i\sum_{q_f}\sum_{p,k=0}^{\infty}\int_0^1dx\int\frac{dk_0dk_3}{(2\pi)^3}
\bigg\{\frac{\big[{\mathbf{k}}_{\parallel}^2-x(1-x)m^2_{M}-\Delta_{pk}^{(M)}\big]}{\big[{\mathbf{k}}_{\parallel}^2+
x(1-x)m^2_{M}-\Delta_{pk}\big]^2}C_{pk}^{(1)}\nonumber\\
&&-\frac{\bar{p}_2\bar{k}_2 }{\big[{\mathbf{k}}_{\parallel}^2+x(1-x)m^2_{M}-\Delta_{pk}\big]^2}C^{(2)}_{pk}\bigg\},
\end{eqnarray}
with
\begin{eqnarray}\label{ND24}
\hspace{-0.5cm}\Delta_{pk}^{\pi^{0}}&\equiv& 2|q_{f}eB|\sqrt{pk}+m^{2},\nonumber\\
\hspace{-0.5cm}\Delta_{pk}^{\sigma}&\equiv& 2|q_{f}eB|\sqrt{pk}-m^{2},\nonumber\\
\hspace{-0.5cm}\Delta_{pk}&\equiv& 2|q_{f}eB|[(1-x)k+xp]+m^{2}.
\end{eqnarray}
At finite $(T,\mu,eB)$, we therefore have
 \begin{eqnarray}\label{ND25}
\lefteqn{\big(g^{(1)}_{qqM}(\tilde{q}_{M})\big)^{-2}=3\sum_{q_f}\sum_{k=0}^{\infty}\int_{0}^{1}
dx\int\frac{dk_3}{(2\pi)^2}
\bigg\{(4k+\delta_{k0})\big[{\cal{S}}^{(0)}_1(\omega^{M}_{k}(x))-2\Xi_{M}(x){\cal{S}}^{(0)}_{2}
(\omega^{M}_{k}(x))\big]}\nonumber\\
&&-2(2k+1)\big[{\cal{S}}^{(0)}_{1}(\omega^{M}_{k+x}(x))-2\Xi_{M}(x){\cal{S}}^{(0)}_{2}
(\omega^{M}_{k+x}(x))\big]+4|qeB|\left(k-x-2kx\right){\cal{S}}^{(0)}_{2}(\omega^{M}_{k+x}(x))\bigg\},
\end{eqnarray}
where $\Xi_{M}(x)$ and $\omega_{k}^{M}(x)$ are defined in
(\ref{ND12}) and (\ref{ND14}), respectively. In the Appendix, we
will present the method we have used to sum over $p$ in (\ref{ND23})
and to arrive at (\ref{ND25}).
\par
As concerns $\big(g^{(2)}_{qqM}(\tilde{q}_{M})\big)^{-2}$, we set
$\mu=2$ in (\ref{ND8}). Using   $\frac{d
f(q_{2})}{dq_{2}^{2}}|_{\tilde{q}_{M},\mbox{\tiny{b.c.}}}=\frac{1}{2}\frac{d^{2}f(q_{2})}
{dq_{2}^{2}}|_{\tilde{q}_{M},\mbox{\tiny{b.c.}}}$
and
\begin{eqnarray}\label{ND26}
\int dx_{1}dk_{2}\frac{d^{2}}{dq_{2}^{2}}\left(A_{pk}^{+(0)}\alpha_{pk}^{+}+A_{pk}^{-(0)}\alpha_{pk}^{-}\right)
\bigg|_{p_{2}=k_{2}}=-\frac{1}{2}C_{pk}^{(1)},\qquad
\int dx_{1}dk_{2}\frac{d^{2}}{dq_{2}^{2}}A_{pk}^{-(0)}\alpha_{pk}^{-}\bigg|_{p_{2}=k_{2}}=\frac{1}{4}C_{pk}^{(2)},
\end{eqnarray}
\end{widetext}
with $C_{pk}^{(i)}, i=1,2$ given in (\ref{ND22}), and eventually
plugging these relations into the resulting expression, it turns out
that
\begin{eqnarray}\label{ND27}
\hspace{-0.5cm}g^{(1)}_{qqM}(\tilde{q}_{M})=g^{(2)}_{qqM}(\tilde{q}_{M}),~~\mbox{for}~~M\in\{\sigma,\pi^{0}\}.
\end{eqnarray}
At finite $(T,\mu,eB)$, the inverse squared value of
$g^{(1)}_{qqM}(\tilde{q}_{M})$ is given in (\ref{ND25}). In the
Appendix, we will present the necessary relations that can be used
to derive (\ref{ND26}). To derive $g^{(3)}_{qqM}(\tilde{q}_{M})$, we
set $\mu=3$ in (\ref{ND8}). Using (\ref{ND9}) and (\ref{ND10}), we
arrive at
\begin{eqnarray}\label{ND28}
\hspace{-0.5cm}g^{(0)}_{qqM}(\tilde{q}_{M})=g^{(3)}_{qqM}(\tilde{q}_{M}),~~\mbox{for}~~M\in\{\sigma,\pi^{0}\},
\end{eqnarray}
where, at finite $(T,\mu,eB)$, the inverse squared value of
$g^{(0)}_{qqM}(\tilde{q}_{M})$ is given in (\ref{ND13}). Note that,
at this stage, in the limit of $m_{\pi}\to 0$,
$g_{qq\sigma}^{(\mu)}(\tilde{q}_{\sigma})$ and
$g_{qq\pi^{0}}^{(\mu)}(\tilde{q}_{\pi^{0}})$, arising in
(\ref{ND13}) and (\ref{ND25}), as well as (\ref{ND27}) and
(\ref{ND28}), satisfy
$g_{qq\sigma}^{(\mu)}=|{\cal{G}}^{\mu\mu}|^{-1/2}$ as well as
$g_{qq\pi^{0}}^{(\mu)}=|{\cal{F}}^{\mu\mu}_{33}|^{-1/2}$. Here, the
analytical expressions of ${\cal{G}}^{\mu\mu}$ and
${\cal{F}}^{\mu\mu}_{33}$ are explicitly presented in
\cite{fayazbakhsh2012}. This is indeed expected from (\ref{NE24c})
and (\ref{NE26c}), which are valid for arbitrary meson masses
$m_{\sigma}$ and $m_{\pi^{0}}$. This shows that $g_{qqM},
M\in\{\sigma,\pi^{0}\}$ arising in (\ref{NE24c}) and (\ref{NE26c})
are, in the limit $m_{M}\to 0$, equal to unity. In Sec.
\ref{subsec4-B}, we will evaluate numerically the $k_{3}$
integration and the summation over Landau levels $k$, which appear
in (\ref{ND13}) as well as (\ref{ND25}). We will present the $T$
dependence of directional quark-meson couplings $g_{qqM}^{(\mu)},
\mu=0,\cdots,3,$ and $M\in\{\sigma,\pi^{0}\}$ at zero chemical
potential and for $eB=0.03, 0.2,0.3$ GeV$^{2}$. We will, in
particular, verify (\ref{NE24c}) for all $T$-dependent pion masses.
Moreover, we will use the data arising from these results for
$g_{qqM}^{(\mu)}, \mu=0,\cdots,3$, to determine the $T$ dependence
of directional decay constants of neutral pions,
$f_{\pi^{0}}^{(\mu)}, \mu=0,\cdots,3$, at zero chemical potential
and for finite and fixed magnetic fields. In what follows, we will
present the analytical results for $f_{\pi^{0}}^{(\mu)},
\mu=0,\cdots,3$, using (\ref{NE30c}) with $\ell=m=3$.
\subsubsection{Directional pion decay constant for neutral pions}\label{subsec4-A-2}\par\noindent
\par\noindent
Let us consider (\ref{NE30c}) with $\ell=3$ for neutral pions. Using
the fact that $\Upsilon_{\mu}^{\ell m}(q)=\delta_{\ell
m}\Upsilon_{\mu}^{\ell\ell}, \forall \ell,m=1,2,3$, with
$\Upsilon_{\mu}^{\ell m}$ given in (\ref{NE12c}), and setting
$\ell=3$, we get
\begin{eqnarray}\label{ND29}
f_{\pi^{0}}^{(\mu)}q_{\mu}=g_{qq\pi^{0}}^{(\mu)}\Upsilon_{\mu}^{33}(q).
\end{eqnarray}
As it turns out, Eq. (\ref{ND29}) is not appropriate to determine
$f_{\pi^{0}}^{(\mu)}$ in the rest frame of neutral mesons,
especially in the spatial directions. Later we will show that this
is mainly because the spatial components of $\Upsilon_{\mu}^{33}(q)$
vanish in the rest frame of neutral mesons, i.e.
$\Upsilon_{i}^{33}(\tilde{q}_{\pi^{0}})=0$ for
$\tilde{q}_{\pi^{0}}=(m_{\pi^{0}},{\mathbf{0}})$. In what follows,
after determining $\Upsilon_{\mu}^{33}(q)$ in the presence of
external magnetic fields, we will present a method from which
$f_{\pi^{0}}^{(\mu)}$ can be determined for all $\mu=0,\cdots, 3$.
\par
We start by plugging the Ritus propagators
(\ref{NA24a})-(\ref{NA27a}) into (\ref{NE12c}), to arrive at
\begin{eqnarray}\label{ND30}
\lefteqn{\hspace{-0.5cm}
\Upsilon_{\mu}^{33}(q)=\frac{i}{2}\sum\limits_{p,k=0}^{\infty}\int{\cal{D}}\tilde{p}~{\cal{D}}\tilde{k}
}\nonumber\\
&&\hspace{-0.5cm}\times\int d^{4}x~e^{-i(\tilde{p}-\tilde{k}-q)\cdot x}\Lambda_{pk}^{(\mu,3)}\left(\bar{p},
\bar{k},x_{1}\right),
\end{eqnarray}
with
\begin{eqnarray}\label{ND31}
\lefteqn{\hspace{-1cm}\Lambda^{(\mu,3)}_{pk}(\bar{p},\bar{k},x_1)=\mbox{tr}_{sfc}
\big[\gamma^{\mu}P_p(x_{1})S_{Q}(\bar{p})P_p(0)}\nonumber\\
&&\times i\gamma_5\tau^{3}K_k(0)S_{Q}(\bar{k})K_k(x)i\gamma_5\tau^{3}\big].
\end{eqnarray}
Integrating over $x_{0},x_{2},x_{3}$, and eventually over
$p_{0},p_{2},p_{3}$ components, we arrive at the same boundary
conditions $p_{i}=k_{i}+q_{i}, i=0,2,3$, as described in the
paragraph following (\ref{ND2}). Plugging $P_{p}(x_{1})$ from
(\ref{NA25a}) into (\ref{ND31}), and using the decomposition
$P_{p}(0)K_{k}(0)$, as given in (\ref{ND5}), and
\begin{eqnarray}\label{ND32}
\lefteqn{\hspace{-0.5cm}K_{k}(x_{1})\gamma^{\mu}P_{p}(x_{1})
}\nonumber\\
&&\hspace{-0.5cm}=\gamma^{\mu}\big[(A_{pk}^{+(\mu)}(x_{1})+is\gamma_{1}\gamma_{2}A_{pk}^{-(\mu)}(x_{1})\big],
\end{eqnarray}
with
\begin{eqnarray}\label{ND33}
\lefteqn{A_{pk}^{\pm(0)}=A_{pk}^{\pm(3)}}\nonumber\\
&&=\frac{1}{2}\big[f_{p}^{+s}(x_{1})f_{k}^{+s}(x_{1})\pm \Pi_{p}\Pi_{k}f_{p}^{-s}(x_{1})f_{k}^{-s}(x_{1})\big],
\nonumber\\
\lefteqn{A_{pk}^{\pm(1)}=A_{pk}^{\pm(2)}}\nonumber\\
&&=\frac{1}{2}\big[\Pi_{k}f_{p}^{+s}(x_{1})f_{k}^{-s}(x_{1})\pm \Pi_{p}f_{p}^{-s}(x_{1})f_{k}^{+s}(x_{1})\big],
\nonumber\\
\end{eqnarray}
we arrive at
\begin{eqnarray}\label{ND34}
\lefteqn{\hspace{-0.5cm}
\Upsilon_{\mu}^{33}(q)=\frac{i}{2}\sum\limits_{p,k=0}^{\infty}\int\frac{dk_{0}dk_{3}}{(2\pi)^{3}}
}\nonumber\\
&&\hspace{-0.5cm}\times \int dk_{2}dx_{1}~e^{-iq_{1}\cdot x_{1}}\Lambda_{pk}^{(\mu,3)}
\left(\bar{p},\bar{k},x_{1}\right)\bigg|_{\mbox{\tiny{b.c.}}},
\end{eqnarray}
with
\begin{widetext}
\begin{eqnarray}\label{ND35}
\Lambda^{(\mu,3)}_{pk}(\bar{p},\bar{k},x_1)=12m\sum_{q_f}\bigg\{
\frac{(\bar{p}-\bar{k})_{\mu}{\cal{N}}_{pk}^{(\mu)} -isg^{\mu
1}\big[\bar{k}_{2}{\cal{M}}_{pk}^{+(\mu)}+\bar{p}_{2}{\cal{M}}_{pk}^{-(\mu)}\big]+2g^{\mu
2} \bar{p}_{2}{\cal{P}}_{pk}^{(\mu)}
}{(\bar{p}^2-m^2)(\bar{k}^2-m^2)}\bigg\},
\end{eqnarray}
and
\begin{eqnarray}\label{ND36}
{\cal{N}}_{pk}^{(\mu)}\equiv
A^{+(\mu)}_{pk}\alpha^+_{pk}+A^{-(\mu)}_{pk}\alpha^-_{pk},\qquad
{\cal{M}}_{pk}^{\pm(\mu)}\equiv A_{pk}^{+(\mu)}\alpha_{pk}^{-}\pm
A_{pk}^{-(\mu)}\alpha_{pk}^{+},\qquad
{\cal{P}}_{pk}^{(\mu)}=\alpha^{-}_{pk}A^{-(\mu)}_{pk}.
\end{eqnarray}
Using (\ref{ND10}) as well as
\begin{eqnarray}\label{ND37}
\int dk_{2}dx_{1}A_{pk}^{\pm(1)}\alpha_{pk}^{\pm}\bigg|_{p_{2}=k_{2}}=0,\qquad
\int dk_{2}dx_{1}A_{pk}^{\pm(1)}\alpha_{pk}^{\mp}\bigg|_{p_{2}=k_{2}}=0,
\end{eqnarray}
\end{widetext}
and performing the integration over $k_{2}$ and $x_{1}$ in (\ref{ND34}) for $q=\tilde{q}_{\pi^{0}}$, we get
\begin{eqnarray}\label{ND38}
\Upsilon_{i}^{33}(\tilde{q}_{\pi^{0}})=0,\qquad\mbox{for}\qquad i=1,2,3.
\end{eqnarray}
Thus, in order to determine $f_{\pi^{0}}^{(\mu)}$ from (\ref{ND29}),
especially in the spatial directions, we have to expand the r.h.s.
of (\ref{ND29}) around $\tilde{q}_{\pi^{0}}$, and then evaluate both
sides at $\tilde{q}_{\pi^{0}}$. Keeping in mind that
$g_{qq\pi^{0}}^{(\mu)}$, appearing on the r.h.s. of (\ref{ND29}) and
defined in (\ref{NE23c}), depends, in general, on $q$, we obtain
\begin{eqnarray}
f_{\pi^{0}}^{(0)}&=&m_{\pi^{0}}^{-1}g_{qq\pi^{0}}^{(0)}(\tilde{q}_{\pi^{0}})\Upsilon_{0}^{33}(\tilde{q}_{\pi^{0}}),
\label{ND39}
\\
f_{\pi^{0}}^{(i)}&=&g_{qq\pi^{0}}(\tilde{q}_{\pi^{0}})\frac{d}{dq_{i}}\Upsilon_{i}^{33}(\tilde{q}_{\pi^{0}}),
\label{ND40}
\end{eqnarray}
for $i=1,2,3$. In what follows, Eqs. (\ref{ND39}) and (\ref{ND40})
will be used to determine the analytical expressions for
$f_{\pi^{0}}^{(\mu)}, \mu=0,\cdots, 3$.
\par
Setting $\mu=0$ in (\ref{ND34}) with $\Lambda_{pk}^{(\mu,3)}$ given
in  (\ref{ND35}), and using (\ref{ND10}) to evaluate the integration
over $k_{2}$ and $x_{1}$, $f_{\pi^{0}}^{(0)}$ from (\ref{ND39})
reads
 \begin{eqnarray}\label{ND41}
\lefteqn{f^{(0)}_{\pi^{0}}=3img^{(0)}_{qq\pi^{0}}(\tilde{q}_{\pi^{0}})\sum_{k=0}^{\infty}\alpha_k\sum_{q_f}|q_feB|}
\nonumber\\
&&\times \int_{0}^{1}dx\int\frac{dk_0dk_3}{(2\pi)^3} \frac{1}{\big[\bar{k}^2+x(1-x)m_{\pi^0}^2-m^2\big]^2},\nonumber\\
\end{eqnarray}
for nonzero $eB$ and vanishing $T$ and $\mu$, as well as
\begin{eqnarray}\label{ND42}
\lefteqn{\hspace{-1.5cm}
f^{(0)}_{\pi^0}=-3m\ g^{(0)}_{qq\pi^0}\sum_{q_f}|q_feB|\sum_{k=0}^{\infty}\alpha_k}\nonumber\\
&&\hspace{-0.5cm}\times \int_0^1 dx\int\frac{dk_3}{(2\pi)^2} S^{(0)}_2(\omega^{\pi^0}_k(x)),
\end{eqnarray}
at finite $(T,\mu,eB)$. Here,
${\cal{S}}_{2}^{(0)}(\omega_{k}^{\pi^{0}}(x))$ with
$\omega_{k}^{\pi^{0}}(x)$ from (\ref{ND14}) is defined in
(\ref{ND15}) and can be evaluated using the recursion relations in
(\ref{ND18}).
\par
Plugging $i=1$ into (\ref{ND40}), we arrive first at
\begin{eqnarray}\label{ND43}
\lefteqn{\hspace{-0cm}f_{\pi^{0}}^{(1)}=
6img_{qq\pi^{0}}^{(1)}(\tilde{q}_{\pi^{0}})\sum_{p,k=0}^{\infty}\sum_{q_{f}}s\int\frac{dk_{0}dk_{3}}{(2\pi)^{3}}
}\nonumber\\
&&\hspace{-0.5cm}\times\int
dk_{2}dx_{1}\frac{[\bar{k}_{2}{x_{1}\cal{M}}_{pk}^{+(1)}+
\bar{p}_{2}x_{1}{\cal{M}}_{pk}^{-(1)}]}{(\bar{p}^{2}-m^{2})(\bar{k}^{2}-m^{2})}\bigg|_{\tilde{q}_{\pi^{0}},
\mbox{\tiny{b.c.}}},\nonumber\\
\end{eqnarray}
where ${\cal{M}}_{pk}^{\pm(1)}$ are defined in (\ref{ND36}) with
$\mu=1$. The integrations over $k_{2}$ and $x_{1}$ can be performed
using
\begin{eqnarray}\label{ND44}
\lefteqn{\hspace{-1cm}
\int dk_{2}dx_{1}x_{1}A_{pk}^{\pm(1)}\alpha_{pk}^{\mp}=-\frac{\sqrt{|q_{f}eB|}}{4\sqrt{2}}
}\nonumber\\
&&\times \big[\Pi_{k}\delta_{p,k-1}(\sqrt{k}\mp\sqrt{k-1})\nonumber\\
&&\pm\Pi_{k+1}\delta_{p,k+1}(\sqrt{k+1}
\mp\sqrt{k})\big].
\end{eqnarray}
After plugging (\ref{ND44}) into  (\ref{ND43}) and summing over $p$,
by making use of the method described in the  Appendix, we arrive,
after some work, at
\begin{eqnarray}\label{ND45}
\lefteqn{f_{\pi^{0}}^{(1)}=6img_{qq\pi^{0}}^{(1)}(\tilde{q}_{\pi^{0}})\sum_{k=0}^{\infty}\sum_{q_{f}}|q_{f}eB|\int\frac{dk_{0}dk_{3}}{(2\pi)^{3}}
}\nonumber\\
&&\times\frac{1}{[{\mathbf{k}}_{\|}^{2}+x(1-x)m_{\pi^{0}}^{2}-m^{2}-2|q_{f}eB|(k+x)]^{2}},\nonumber\\
\end{eqnarray}
for nonvanishing $eB$ and at zero $(T,\mu)$. At finite $(T,\mu,eB)$, we therefore get
\begin{eqnarray}\label{ND46}
\lefteqn{\hspace{-0.5cm}f_{\pi^{0}}^{(1)}=-6mg_{qq\pi^{0}}^{(1)}(\tilde{q}_{\pi^{0}})\sum\limits_{k=0}^{\infty}\sum_{q_{f}}|q_{f}eB|}\nonumber\\
&&\times\int_{0}^{1}
dx\int\frac{dk_{3}}{(2\pi)^{2}}{\cal{S}}_{2}^{(0)}(\omega_{k+x}^{\pi^{0}}(x)).
\end{eqnarray}
To determine $f_{\pi^{0}}^{(2)}$, we use (\ref{ND40}) with $i=2$.
Plugging (\ref{ND34})-(\ref{ND35}) into the resulting expression,
using
\begin{eqnarray}\label{ND47}
\int dk_{2}dx_{1}\alpha_{pk}^{\pm}\frac{d}{dk_{2}}A_{pk}^{\pm(1)}\bigg|_{p_{2}=k_{2}}=0,
\end{eqnarray}
as well as
\begin{eqnarray}\label{ND48}
\lefteqn{\hspace{-0.2cm}\int dk_{2}dx_{1}A_{pk}^{\pm(1)}\frac{d}{dk_{2}}
\alpha_{pk}^{\pm}\bigg|_{p_{2}=k_{2}}=\frac{s|q_{f}eB|}{4\sqrt{2}}
}\nonumber\\
&&\hspace{-0.5cm}\times \big[ \Pi_{k}\delta_{p,k-1}
(\sqrt{k}\pm\sqrt{k-1})\mp\delta_{p,k+1}(\sqrt{k+1}\pm\sqrt{k})\big],\nonumber\\
\end{eqnarray}
and eventually summing over $p$, using the method presented in the
Appendix, we arrive, after some algebraic computations, at
\begin{eqnarray}\label{ND49}
f_{\pi^{0}}^{(2)}=f_{\pi^{0}}^{(1)},
\end{eqnarray}
with $f_{\pi^{0}}^{(1)}$ given in (\ref{ND45}) for nonvanishing $eB$
and at zero $(T,\mu)$, and in (\ref{ND46}) for finite $(T,\mu,eB)$.
Finally, setting $i=3$ in (\ref{ND40}), and using (\ref{ND10}) to
perform the integration over $k_{2}$ and $x_{1}$, keeping in mind
that, according to (\ref{ND33}), $A_{pk}^{\pm(3)}=A_{pk}^{\pm(0)}$,
we arrive easily at
\begin{eqnarray}\label{ND50}
f_{\pi^{0}}^{(3)}=f_{\pi^{0}}^{(0)},
\end{eqnarray}
with $f_{\pi^{0}}^{(0)}$ given in (\ref{ND41}) for nonvanishing $eB$
and at zero $(T,\mu)$, and in (\ref{ND42}) for finite $(T,\mu,eB)$.
Using (\ref{ND42}) and (\ref{ND46}) as well as (\ref{ND49}) and
(\ref{ND50}) for $f_{\pi^{0}}^{(\mu)}$, it can easily be checked
that, in the limit $m_{\pi^{0}}\to 0$, $f_{\pi^{0}}^{(\mu)}$
satisfies $f_{\pi^{0}}^{(\mu)}=m|{\cal{F}}_{33}^{\mu\mu}|^{1/2}$,
with ${\cal{F}}_{33}^{\mu\mu}$ given explicitly in
\cite{fayazbakhsh2012}.  Comparing this relation with the definition
of $f_{\pi^{0}}^{(\mu)}$ from (\ref{NE31c}) with $\ell=3$, it turns
out that the dimensionful constant $f_{3}$, appearing originally in
the Ansatz (\ref{NE27c}), is, in the limit of vanishing
$m_{\pi^{0}}$, equal to the constituent mass $m$. In Sec.
\ref{subsec4-B}, after numerically computing $f_{\pi^{0}}^{(\mu)}$
from (\ref{ND42}) and (\ref{ND46}), we will determine the
coefficient $f_{3}$ using (\ref{NE31c}) and compare its $T$
dependence for fixed $\mu$ and $eB$ with the $T$ dependence of the
constituent mass $m$ for arbitrary $T$-dependent $m_{\pi^{0}}$. But,
before doing this, let us analytically verify the modified GT
relation (\ref{NE32c}). In Sec. \ref{subsec3-B}, we have used
(\ref{NE36c}) to prove (\ref{NE32c}). For zero magnetic fields,  the
explicit form of $\Upsilon_{\mu}^{\ell m}$ as a function of the
fermion propagator $S(p)$ is to be used to verify (\ref{NE35c}). In
the presence of magnetic fields, however, where nontrivial Ritus
propagators $S_{Q}(p)$  from (\ref{NA24a}) are to be used to
determine $\Upsilon_{\mu}^{\ell m}$, relation (\ref{NE35c}) cannot
be directly proved. To show the GT relation for neutral magnetized
pions, we use, instead, the analytical results of
$\Upsilon_{0}^{(33)}$ and $(g_{qq\pi^{0}}^{(0)})^{-2}$ and verify
first the following relation:
\begin{eqnarray}\label{ND51}
\Upsilon_{0}^{(33)}(\tilde{q}_{\pi^{0}})=mm_{\pi^{0}}(g_{qq\pi^{0}}^{(0)})^{-2}+{\cal{O}}(m_{\pi^{0}}^{2}).
\end{eqnarray}
Combining further (\ref{NE27c}) and (\ref{NE29c}) for $\ell=m=3$, and evaluating the resulting expression
\begin{eqnarray*}
g_{qq\Pi^{0}}\Upsilon_{\mu}^{33}(q)=f_{3}|{\cal{F}}_{33}^{00}|^{1/2}u_{\pi^{0}}^{(\mu)2}q_{\mu},
\end{eqnarray*}
on the pion mass shell, i.e., for
$q=\tilde{q}_{\pi^{0}}=(m_{\pi^{0}},{\mathbf{0}})$, we obtain
\begin{eqnarray}\label{ND52}
g_{qq\pi^{0}}\Upsilon_{0}^{(33)}(\tilde{q}_{\pi^{0}})=f_{3}|{\cal{F}}_{33}^{00}|m_{\pi^{0}},
\end{eqnarray}
which leads, together with (\ref{NE24c}) and (\ref{ND51}), to
\begin{eqnarray}\label{ND53}
g_{qq\pi^{0}}f_{3}=m+{\cal{O}}(m_{\pi^{0}}),
\end{eqnarray}
 and eventually, upon using (\ref{NE24c}) and (\ref{NE31c}), to the GT relation
\begin{eqnarray}\label{ND54}
g_{qq\pi^{0}}^{(\mu)}f_{\pi^{0}}^{(\mu)}=m+{\cal{O}}(m_{\pi^{0}}),
\end{eqnarray}
for neutral pions. Note that (\ref{ND53}) is the same as
(\ref{NE37c}) with $\ell=3$. In Sec. \ref{subsec3-B}, we have shown
how the GOR relation (\ref{NE33c}) for neutral pions,
\begin{eqnarray}\label{ND55}
m_{\pi^{0}}^{2}f_{\pi^{0}}^{(\mu)2}=u_{\pi^{0}}^{(\mu)2}\frac{m_{0}\sigma_{0}}{2G}+{\cal{O}}(m_{0}^{2}),
\end{eqnarray}
can  be derived from (\ref{ND53}) and (\ref{NE38c}). As concerns the
proof of (\ref{ND51}), we use (\ref{ND34}) and (\ref{ND35}) with
$\mu=0$. Using (\ref{ND10}) to perform the integration over $k_{3}$
and $x_{1}$, we arrive at
\begin{eqnarray}\label{ND56}
\Upsilon_{0}^{(33)}(\tilde{q}_{\pi^{0}})&=&3imm_{\pi^{0}}\sum\limits_{k=0}^{\infty}\alpha_{k}\sum_{q_{f}}\int\frac{dk_{0}dk_{3}}{(2\pi)^{3}}\nonumber\\
&&\times\frac{1}{[(k_{0}+m_{\pi^{0}})^{2}-k_{3}^{2}-2|q_{f}eB|k-m^{2}]}\nonumber\\
&&\times \frac{1}{[{\mathbf{k}}_{\|}^{2}-2|q_{f}eB|-m^{2}]}.
\end{eqnarray}
Expanding the r.h.s. of this relation in orders of $m_{\pi^{0}}$, we
get
\begin{eqnarray}\label{ND57}
\Upsilon_{0}^{(33)}(\tilde{q}_{\pi^{0}})&=&-3mm_{\pi^{0}}
\sum\limits_{k=0}^{\infty}\alpha_{k}\sum\limits_{q_{f}}
\int\frac{dk_{3}}{(2\pi)^{2}}~{\cal{S}}_{2}^{(0)}(\omega_{k})\nonumber\\
&&+{\cal{O}}(m_{\pi^{0}}^{2}),
\end{eqnarray}
at finite $(T,\mu,eB)$.   Here, ${\cal{S}}_{2}^{(0)}(\omega_{k})$ is
defined in (\ref{ND15}) and
$\omega_{k}^{2}=k_{3}^{2}+m^{2}+2|q_{f}eB|k$. On the other hand, let
us consider $(g_{qqM}^{(0)}(\tilde{q}_{M}))^{-2}$ from (\ref{ND13}).
For $M=\pi^{0}$, it consists of two terms. Neglecting the second
term proportional to $m_{\pi^{0}}^{2}$, expanding
${\cal{S}}_{2}^{(0)}(\omega_{k}^{\pi^{0}}(x))$, appearing in the
first term around $m_{\pi^{0}}=0$, and eventually integrating the
resulting expression over the Feynman parameter $x$, we get
\begin{eqnarray}\label{ND58}
(g_{qq\pi^{0}}^{(0)}(\tilde{q}_{\pi^{0}}))^{-2}&=&-3\sum\limits_{k=0}^{\infty}\alpha_{k}\sum\limits_{q_{f}}\int\frac{dk_{3}}{(2\pi)^{2}}~{\cal{S}}_{2}^{(0)}(\omega_{k})\nonumber\\
&&+{\cal{O}}(m_{\pi^{0}}^{2}),
\end{eqnarray}
which, together with (\ref{ND57}), leads to (\ref{ND51}). This
proves the GT and GOR relations in the limit $m_{\pi^{0}}\to 0$. In
what follows, we will numerically verify, among others, these
relations for arbitrary pion mass at finite $(T,\mu,eB)$.
\subsection{Numerical results}\label{subsec4-B}\par\noindent
\begin{figure}[hbt]
\includegraphics[width=7.7cm,height=4.5cm]{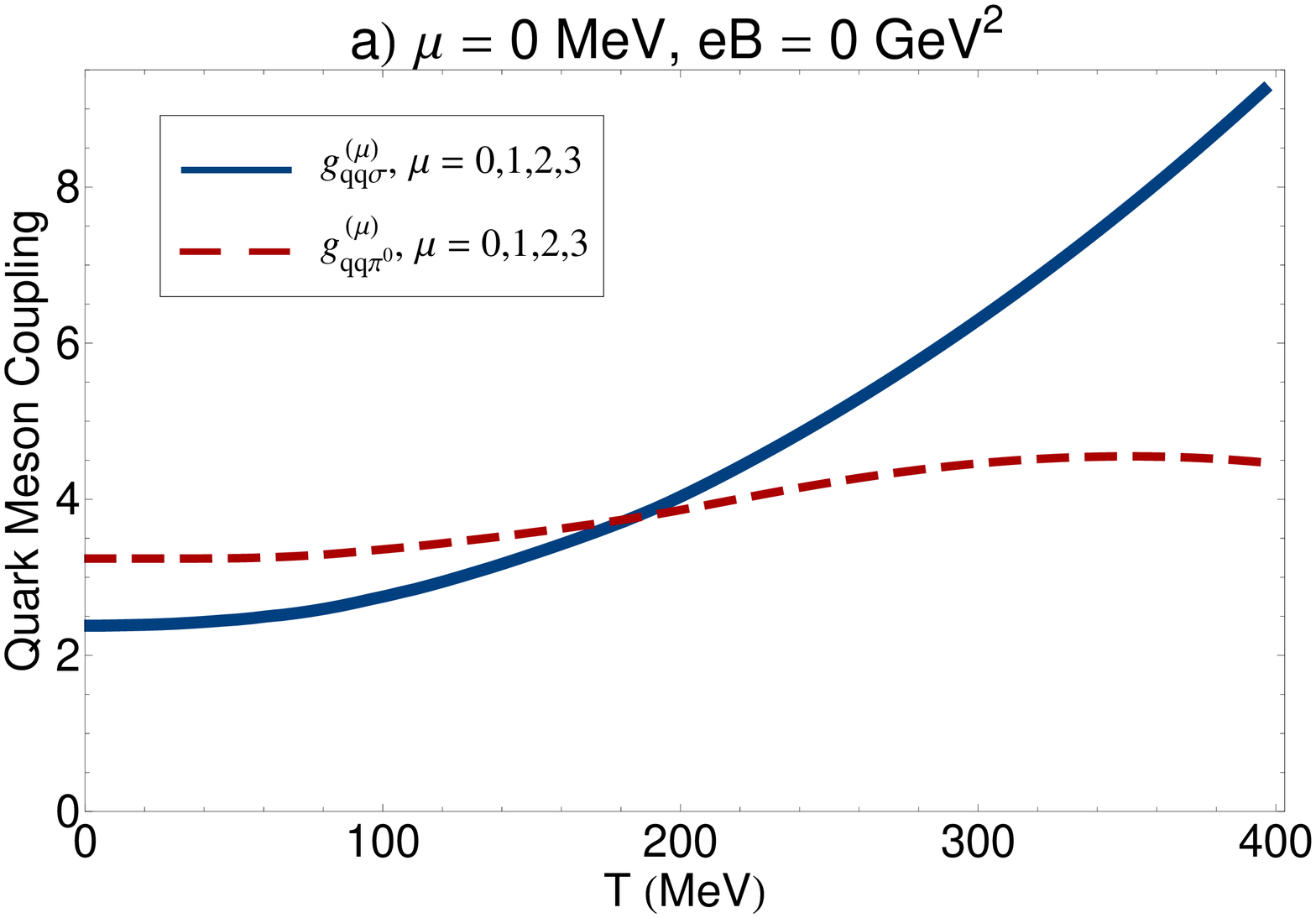} 
\includegraphics[width=7.7cm,height=4.5cm]{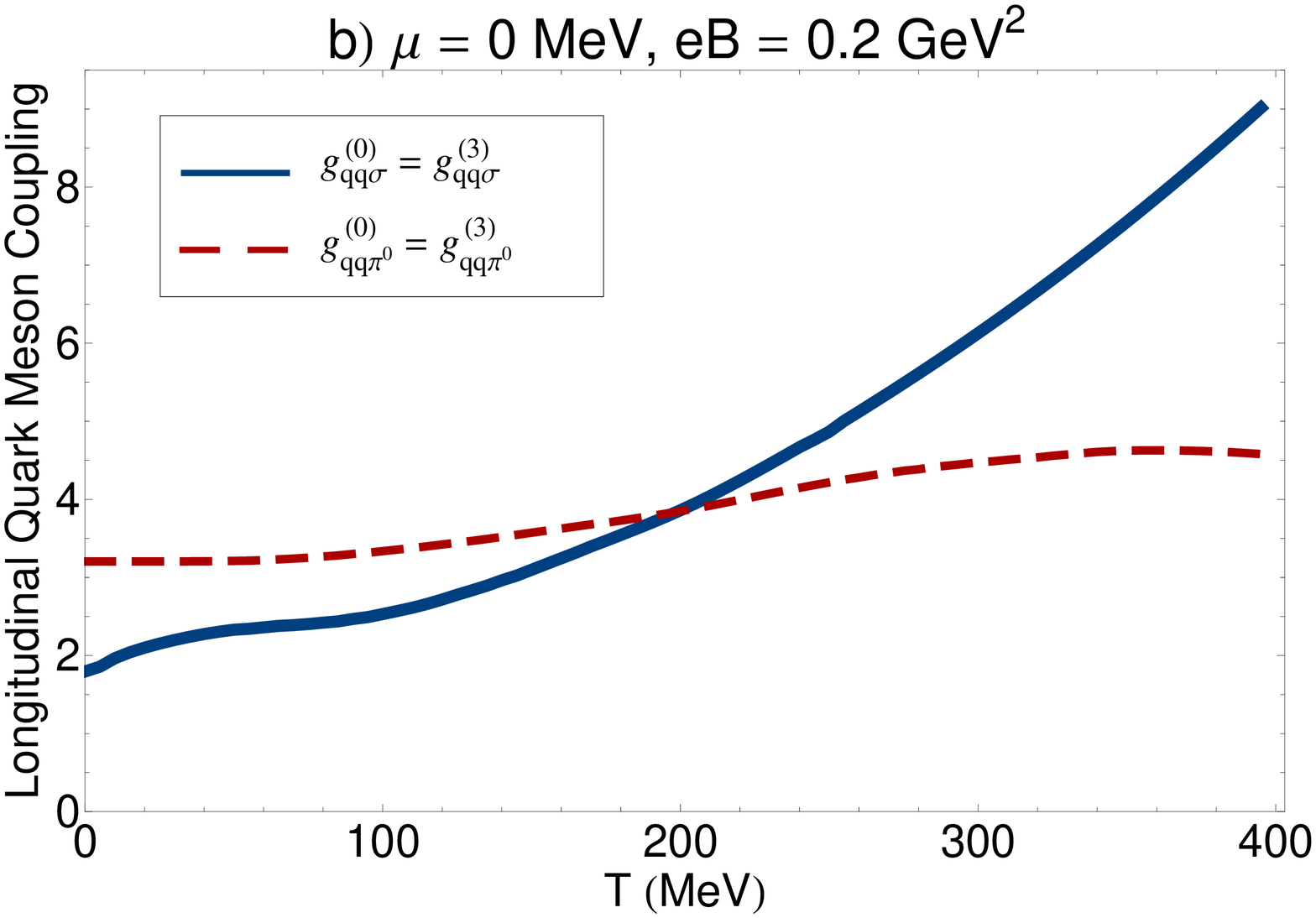} 
\includegraphics[width=7.7cm,height=4.5cm]{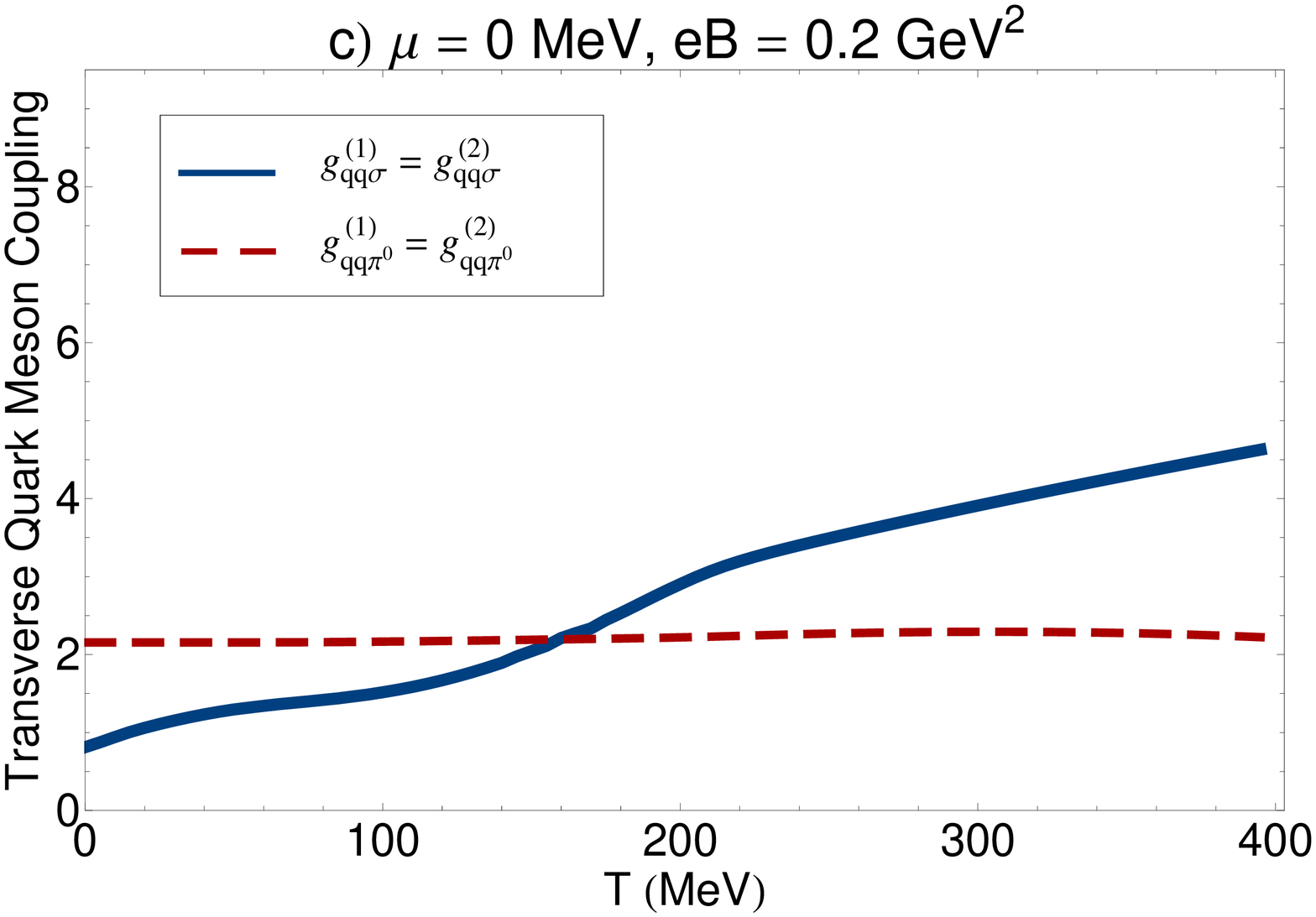}
\caption{The $T$ dependence of directional quark-meson couplings is
demonstrated for $eB=0$ [panel (a)] and $eB=0.2$ GeV$^{2}$ [panels
(b) and (c)]. Solid blue curves are directional quark-sigma
couplings $g_{qq\sigma}^{(\mu)}$, and the dashed red curves are the
directional quark-pion couplings $g_{qq\pi^{0}}^{(\mu)}$. In
contrast to $eB\neq 0$, for $eB=0$, there is no difference between
the quark-meson couplings in longitudinal and transverse directions
with respect to the direction of the external magnetic field
$eB$.}\label{fig8a}
\end{figure}
\begin{figure}[b]
\includegraphics[width=7.7cm,height=4.5cm]{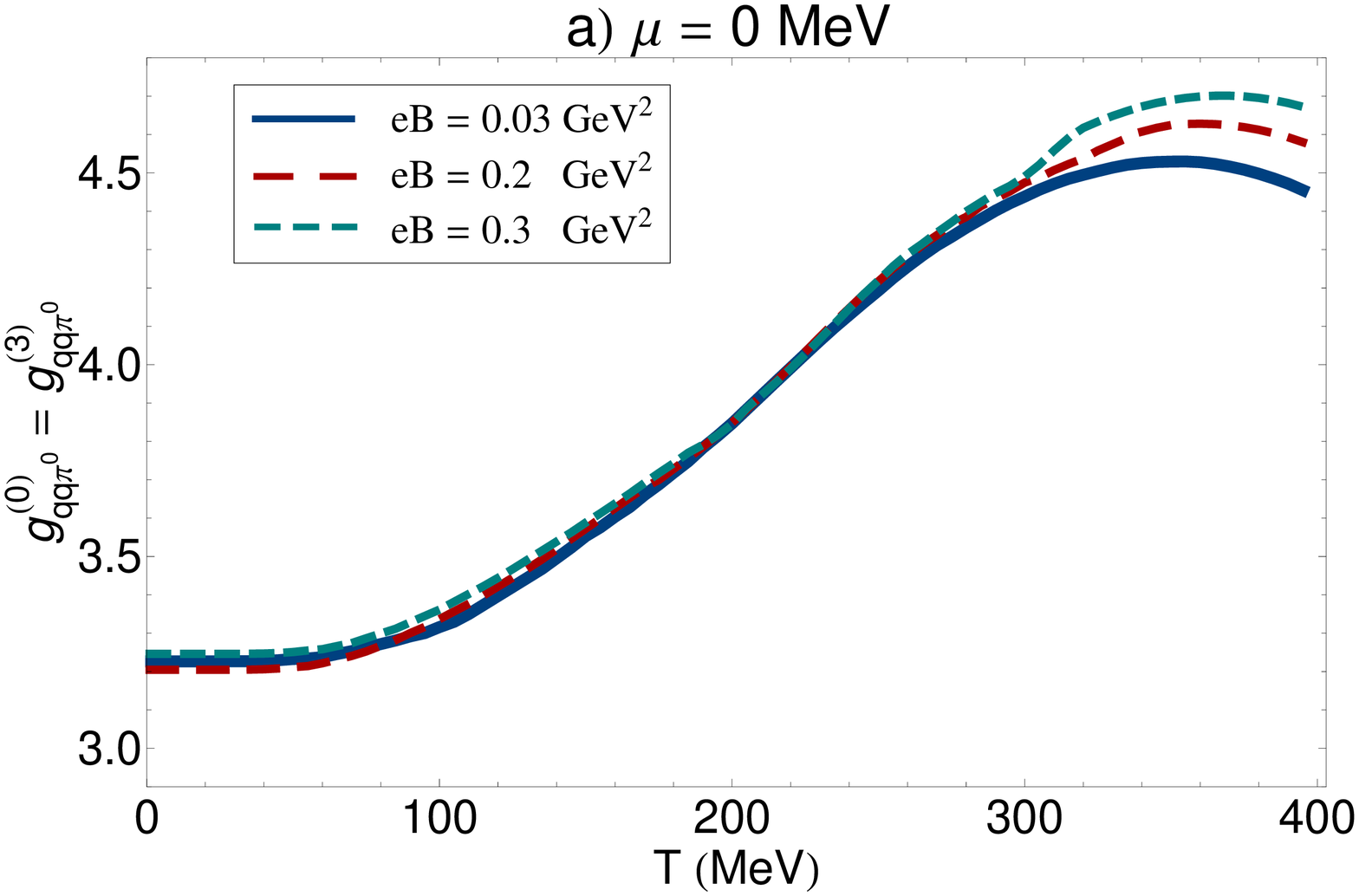}
\par
\includegraphics[width=7.7cm,height=4.5cm]{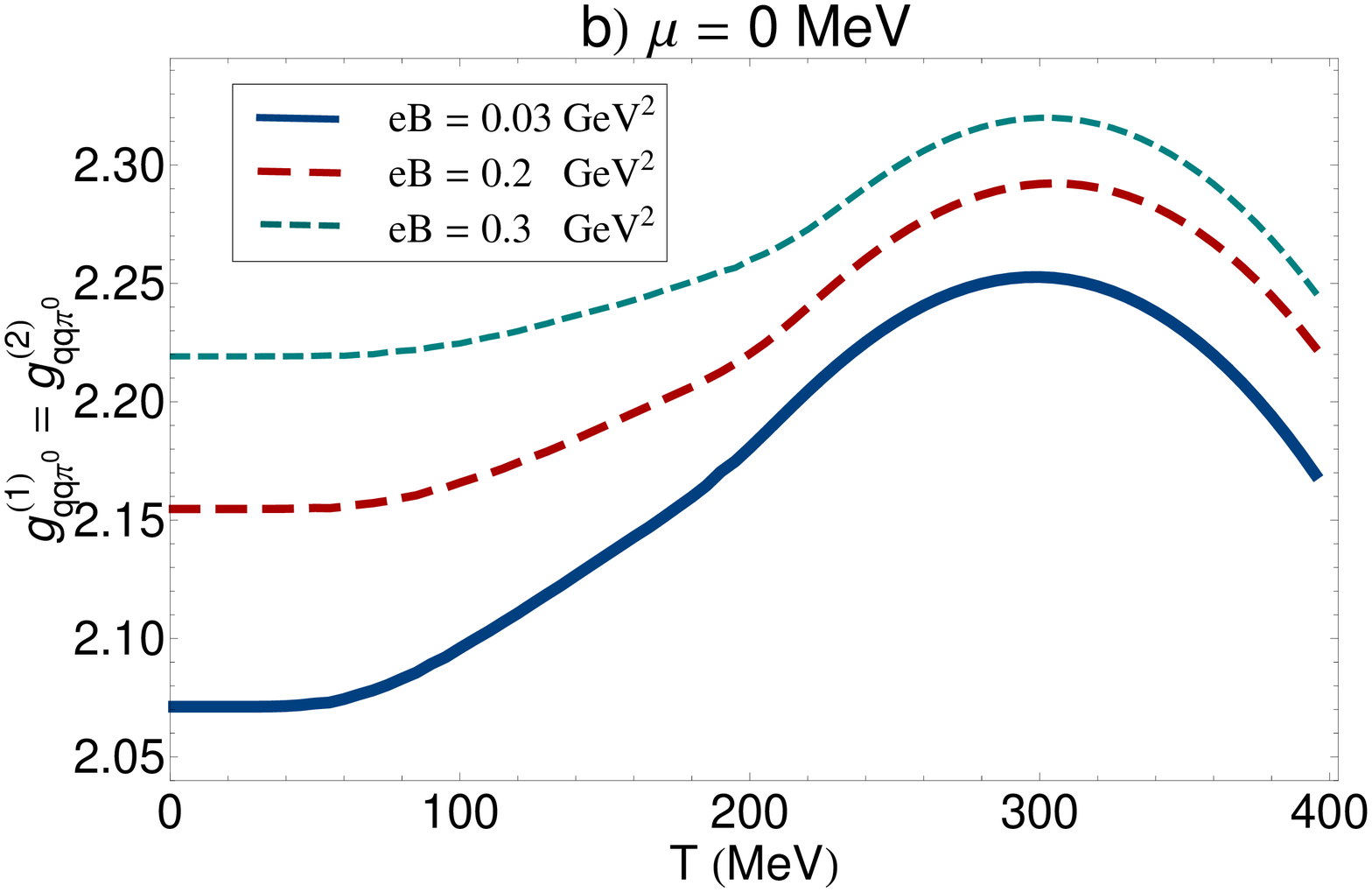}
\caption{The $T$ dependence of directional quark-pion couplings
$g_{qq\pi^{0}}^{(\mu)}$ are demonstrated for vanishing chemical
potential and $eB=0.03, 0.2,0.3$ GeV$^{2}$. In panels (a) and (b),
the longitudinal coupling $g_{qq\pi^{0}}^{(0)}=g_{qq\pi^{0}}^{(3)}$
and transverse coupling $g_{qq\pi^{0}}^{(1)}=g_{qq\pi^{0}}^{(2)}$
are plotted, respectively. The couplings show different behaviors by
increasing the strength of the magnetic field $eB$ at fixed $T$, and
similar behaviors before and after the transition temperature
$T_{c}\sim 200$ MeV.}\label{fig9a}
\end{figure}
\begin{figure}[b]
\includegraphics[width=7.7cm,height=4.5cm]{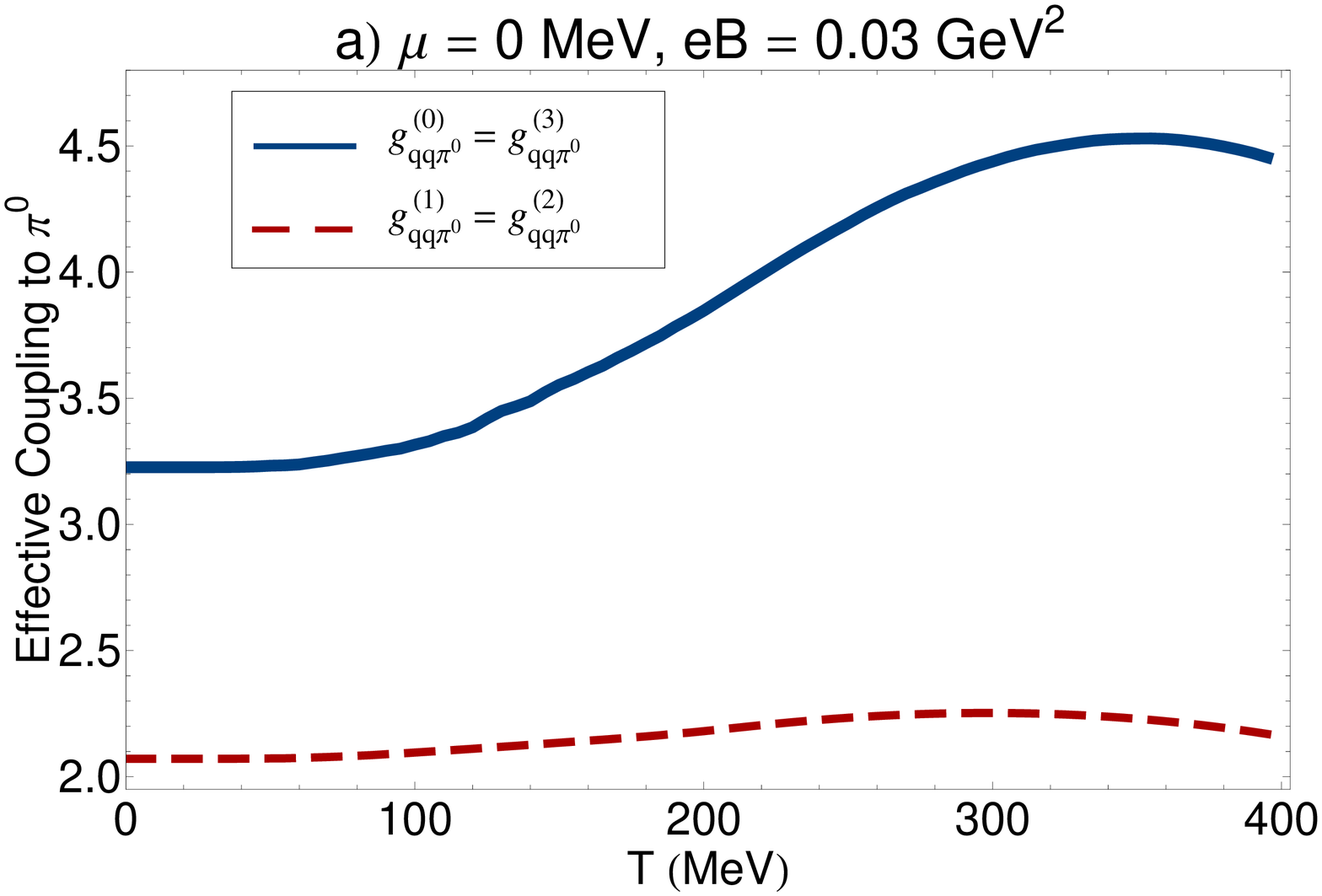}
\\
\includegraphics[width=7.7cm,height=4.5cm]{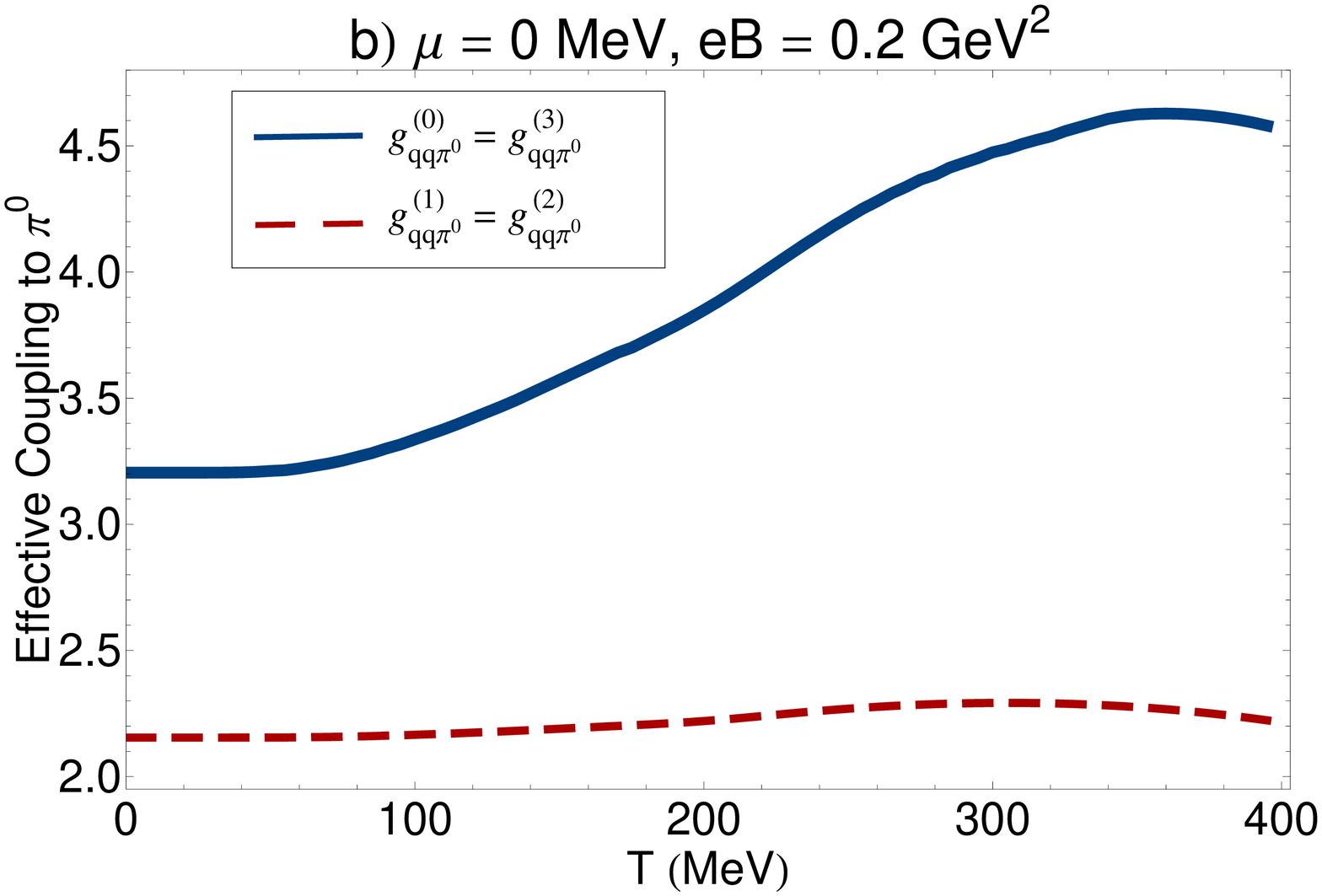}
\\
\includegraphics[width=7.7cm,height=4.5cm]{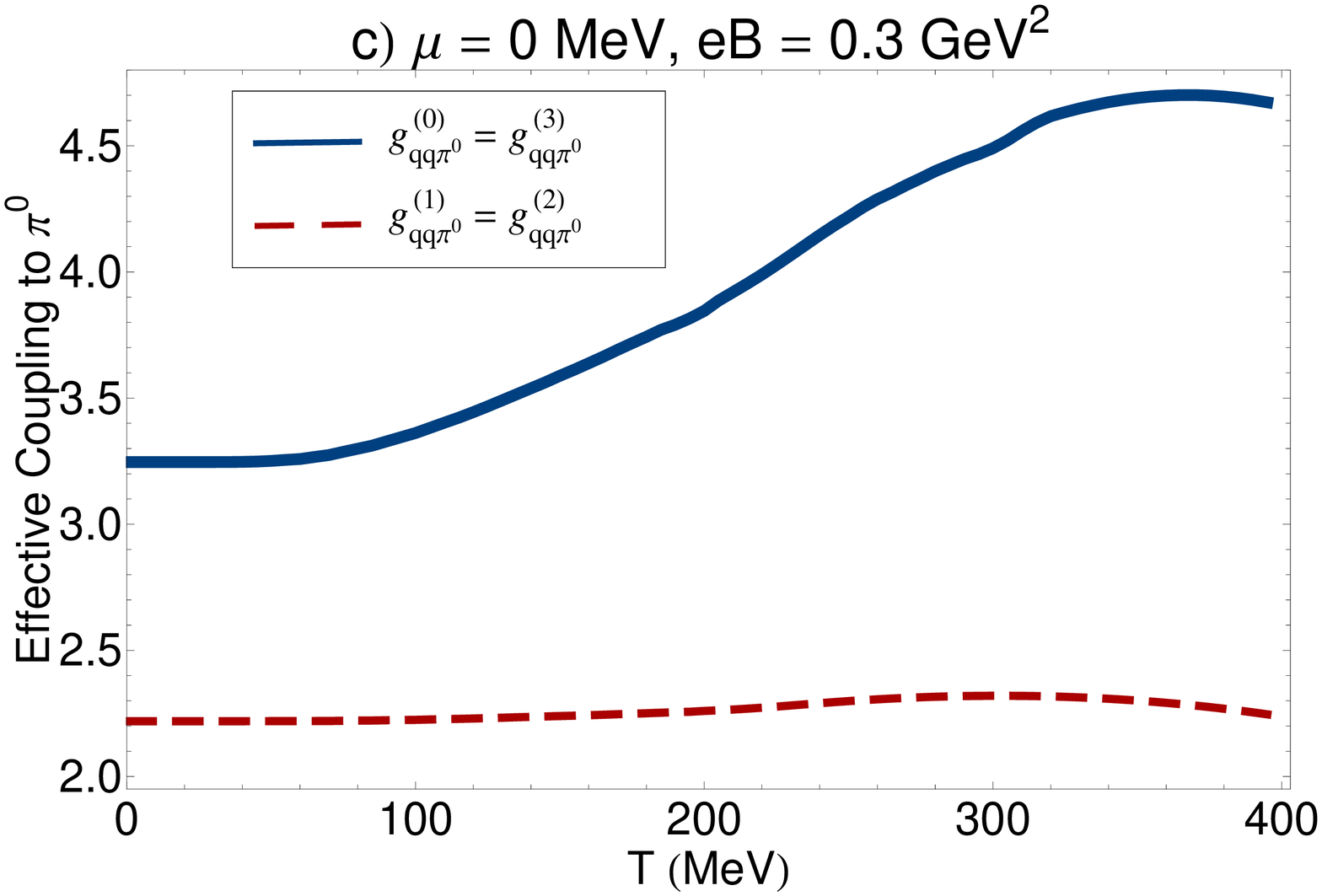}
\caption{Longitudinal (blue solid curves) and transverse (red dashed
curves) quark-pion couplings,
$g_{qq\pi^{0}}^{(0)}=g_{qq\pi^{0}}^{(3)}$  and
$g_{qq\pi^{0}}^{(1)}=g_{qq\pi^{0}}^{(2)}$, are compared in the
interval $T\in[0,400]$ MeV for vanishing chemical potential and
various $eB=0.03$ GeV$^{2}$ [panel (a)], $eB=0.2$ GeV$^{2}$ [panel
(b)] and $eB=0.3$ GeV$^{2}$ [panel (c)]. As it turns out, the
longitudinal coupling is always greater than the transverse
coupling.}\label{fig10a}
\end{figure}
\begin{figure*}[t]
\includegraphics[width=5.5cm,height=4cm]{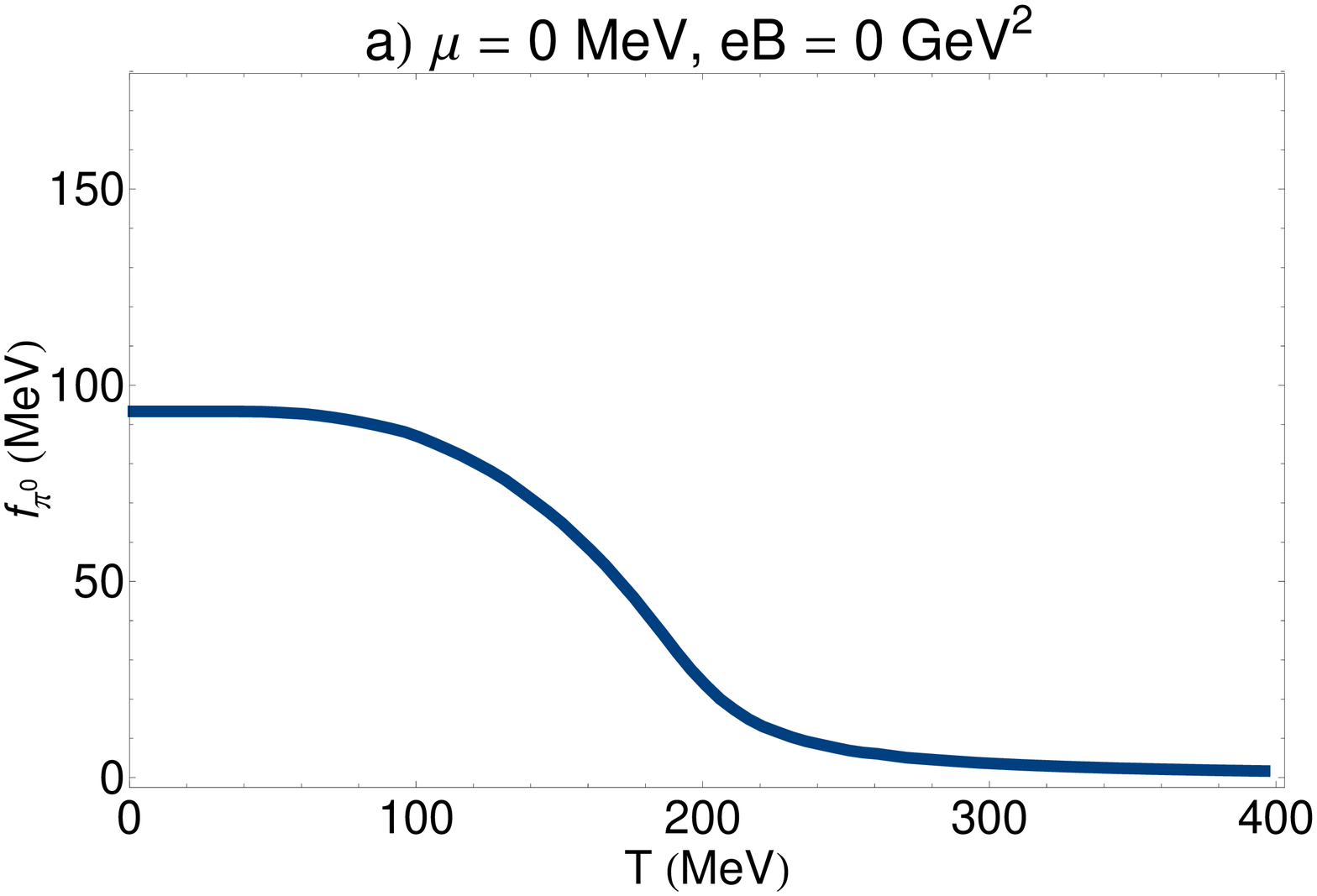} 
\includegraphics[width=5.5cm,height=4cm]{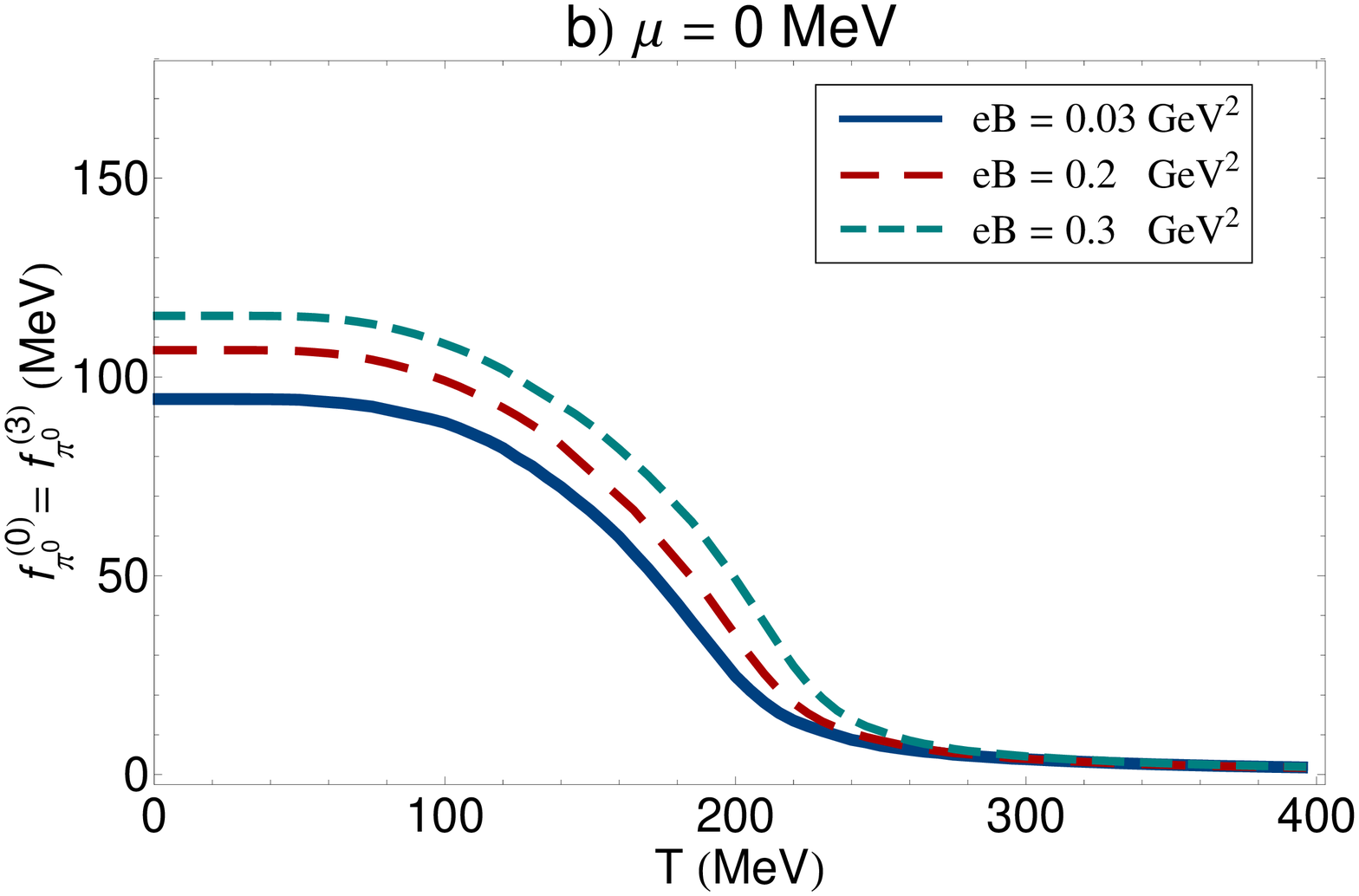}
\includegraphics[width=5.5cm,height=4cm]{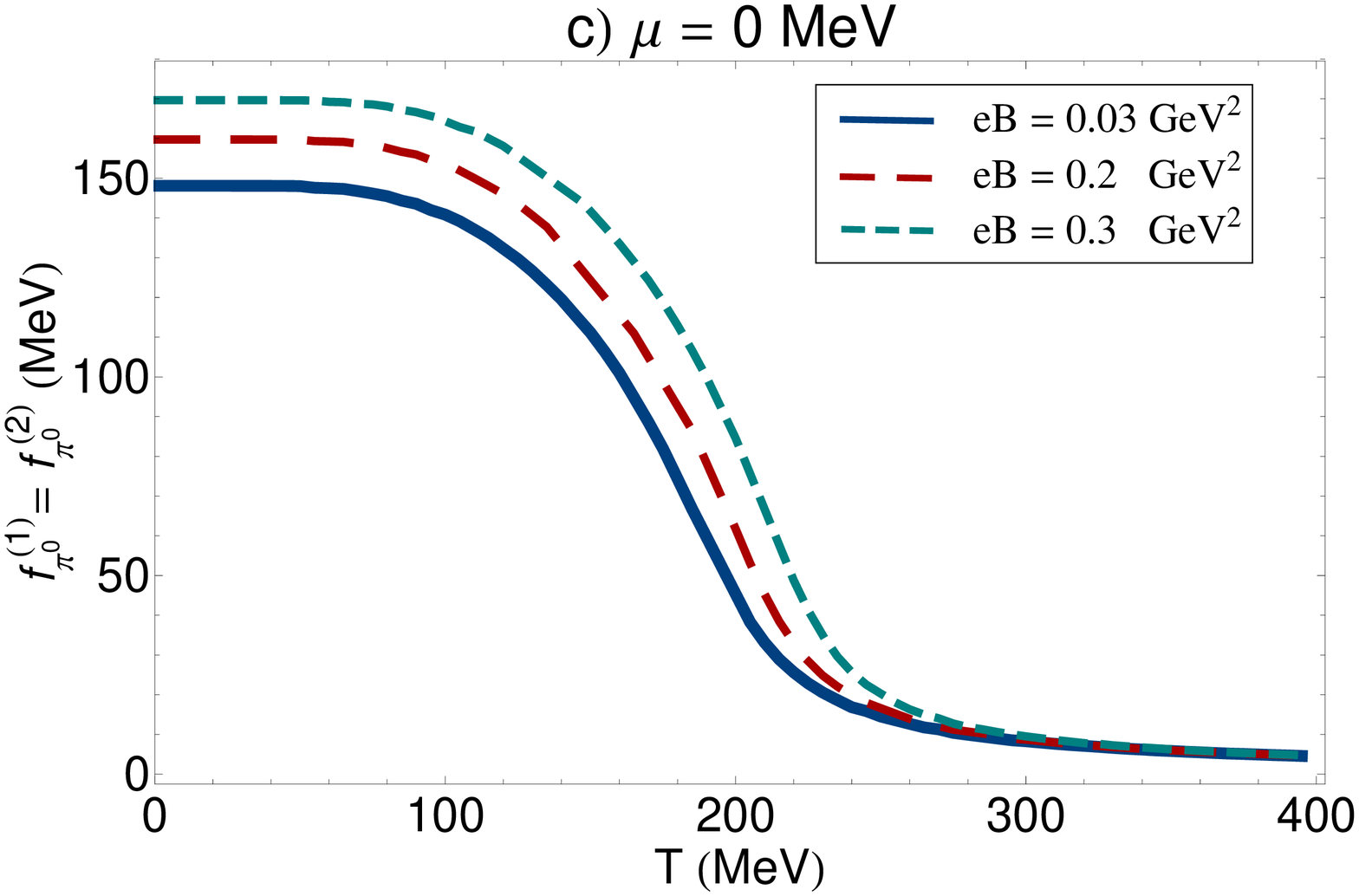}
\caption{[Panel (a)] The $T$ dependence of $f_{\pi^{0}}$ for
vanishing chemical potential and magnetic field. Panels (b) and (c):
The $T$ dependence of longitudinal [panel (b)] and transverse [panel
(c)] decay constants of neutral pions are plotted for zero chemical
potential and nonvanishing $eB=0.03, 0.2, 0.3$
GeV$^{2}$.}\label{fig11a}
\end{figure*}
\begin{figure*}[t]
\includegraphics[width=5.5cm,height=4cm]{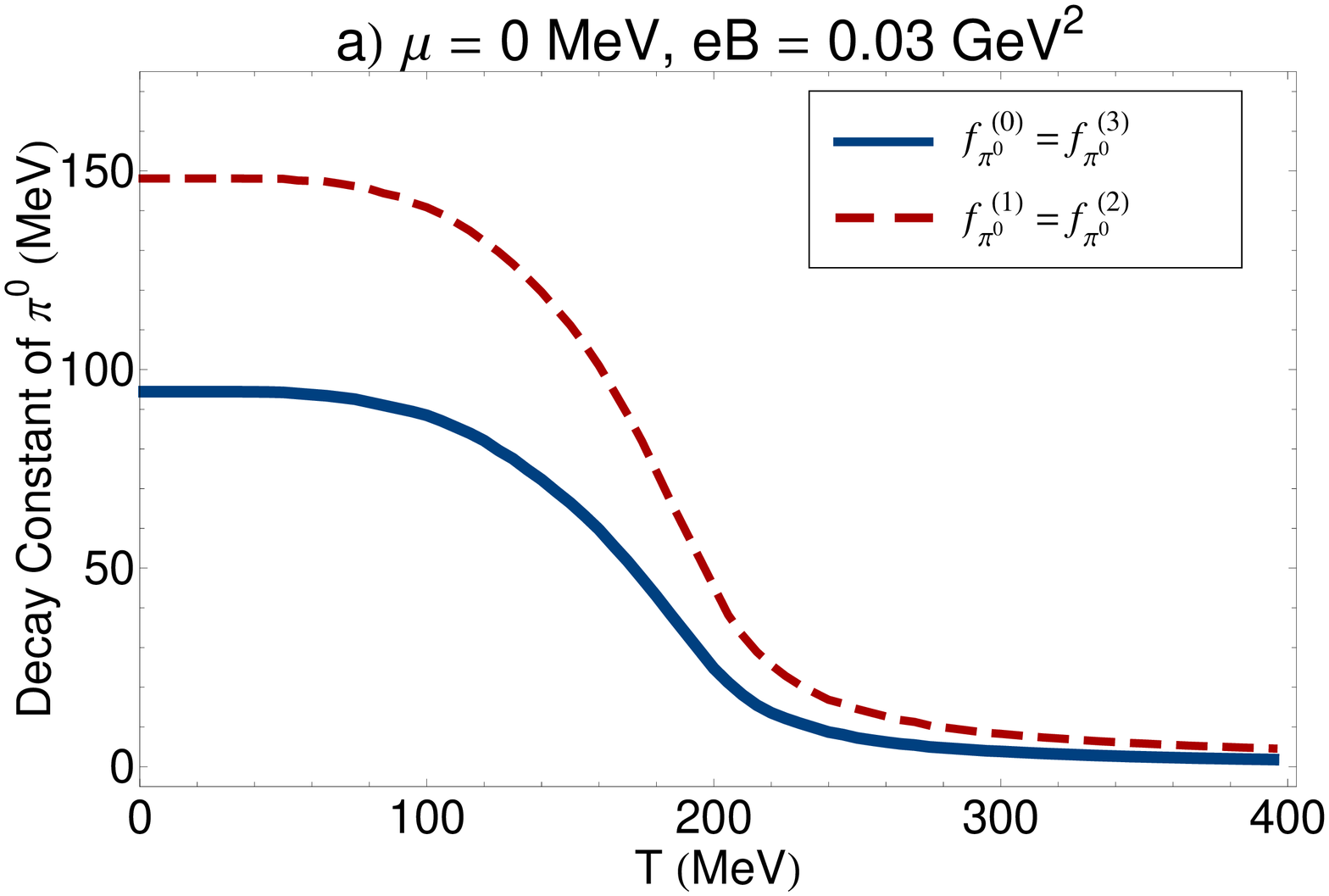}
\includegraphics[width=5.5cm,height=4cm]{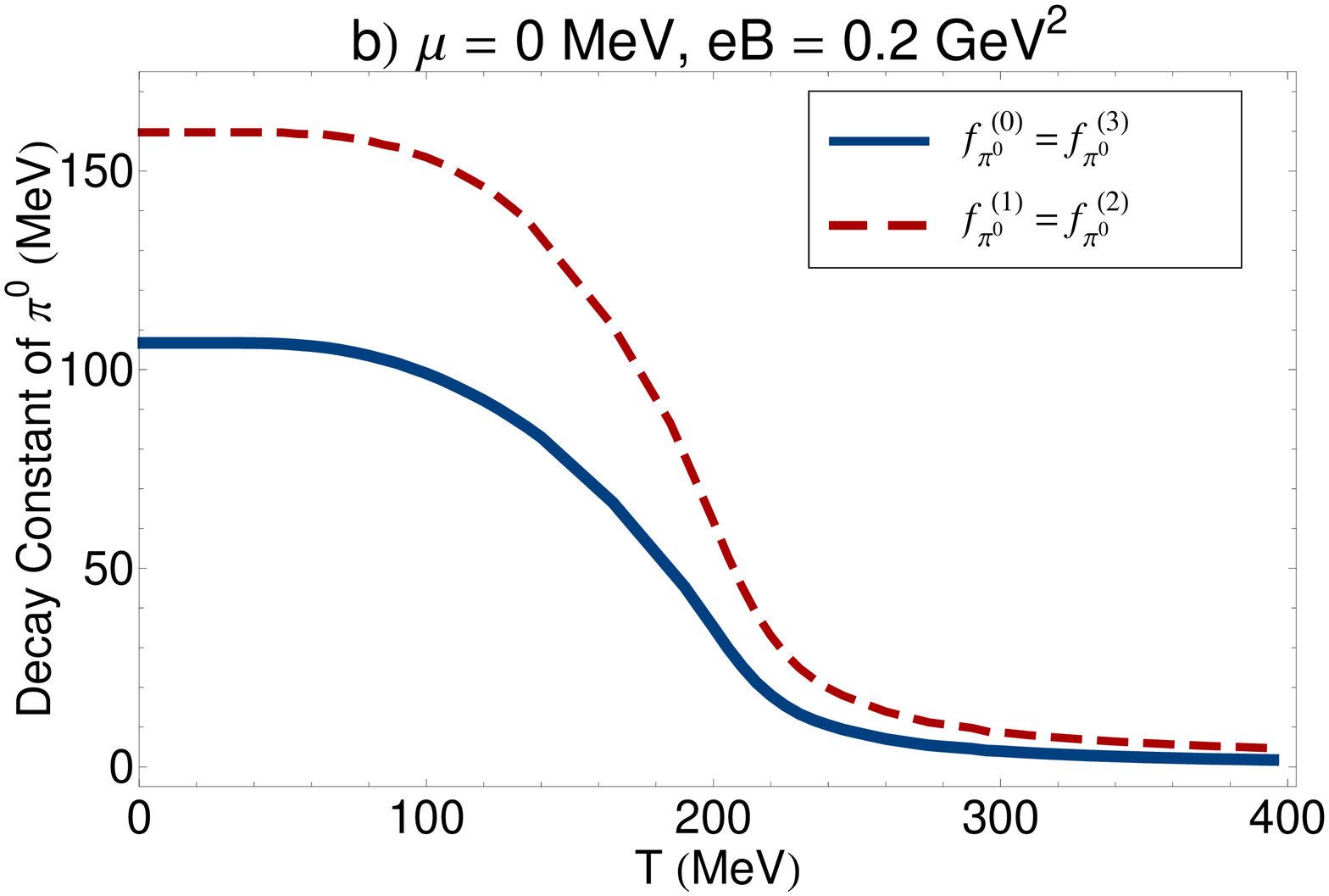}
\includegraphics[width=5.5cm,height=4cm]{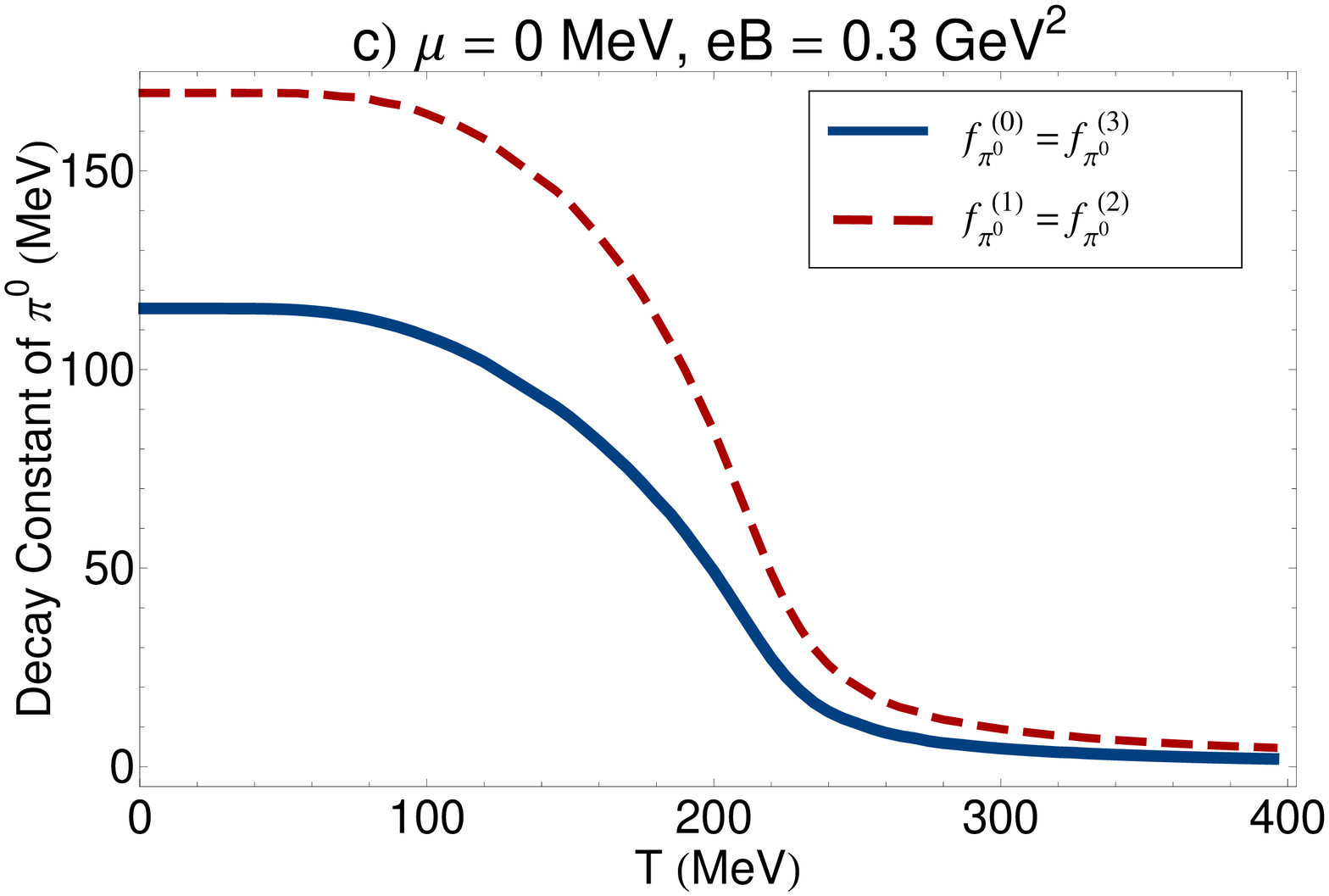}
\caption{The $T$ dependence of longitudinal (blue solid curves) and
transverse (red dashed curves) decay constants of neutral pions,
$f_{\pi^{0}}^{(0)}$ and $f_{\pi^{0}}^{(1)}$, are compared for
vanishing chemical potential and nonvanishing $eB=0.03,0.2,0.3$
GeV$^{2}$. As it turns out, $f_{\pi^{0}}^{(0)}<f_{\pi^{0}}^{(1)}$.
}\label{fig12a}
\end{figure*}
\begin{figure}[hbt]
\includegraphics[width=8cm,height=5.5cm]{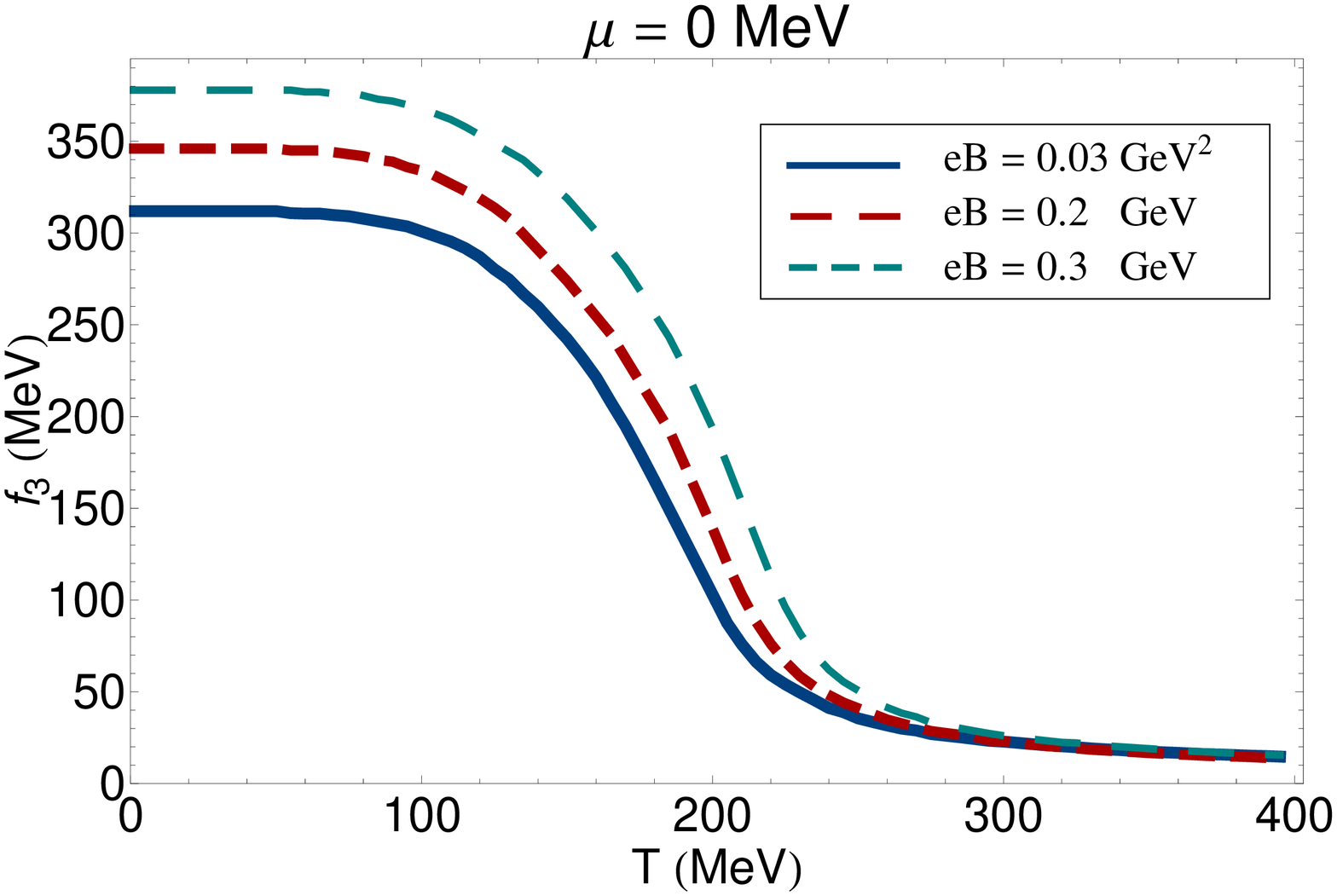}
\caption{The $T$ dependence of $f_{3}$ appearing in the PCAC
relation (\ref{NE27c}). Here, $f_{3}$ is determined using
$f_{3}=f_{\pi^{0}}^{(0)}|{\cal{F}}_{33}^{00}|^{-1/2}$ from
(\ref{NE31c}) with $\ell=3$. Comparing these results with the
corresponding data from the $T$ dependence of the constituent quark
mass $m$, it turns out that $f_{3}=m$ for an arbitrary value of
$m_{\pi^{0}}$.}\label{fig13a}
\end{figure}
\begin{figure*}[hbt]
\includegraphics[width=5.5cm,height=4cm]{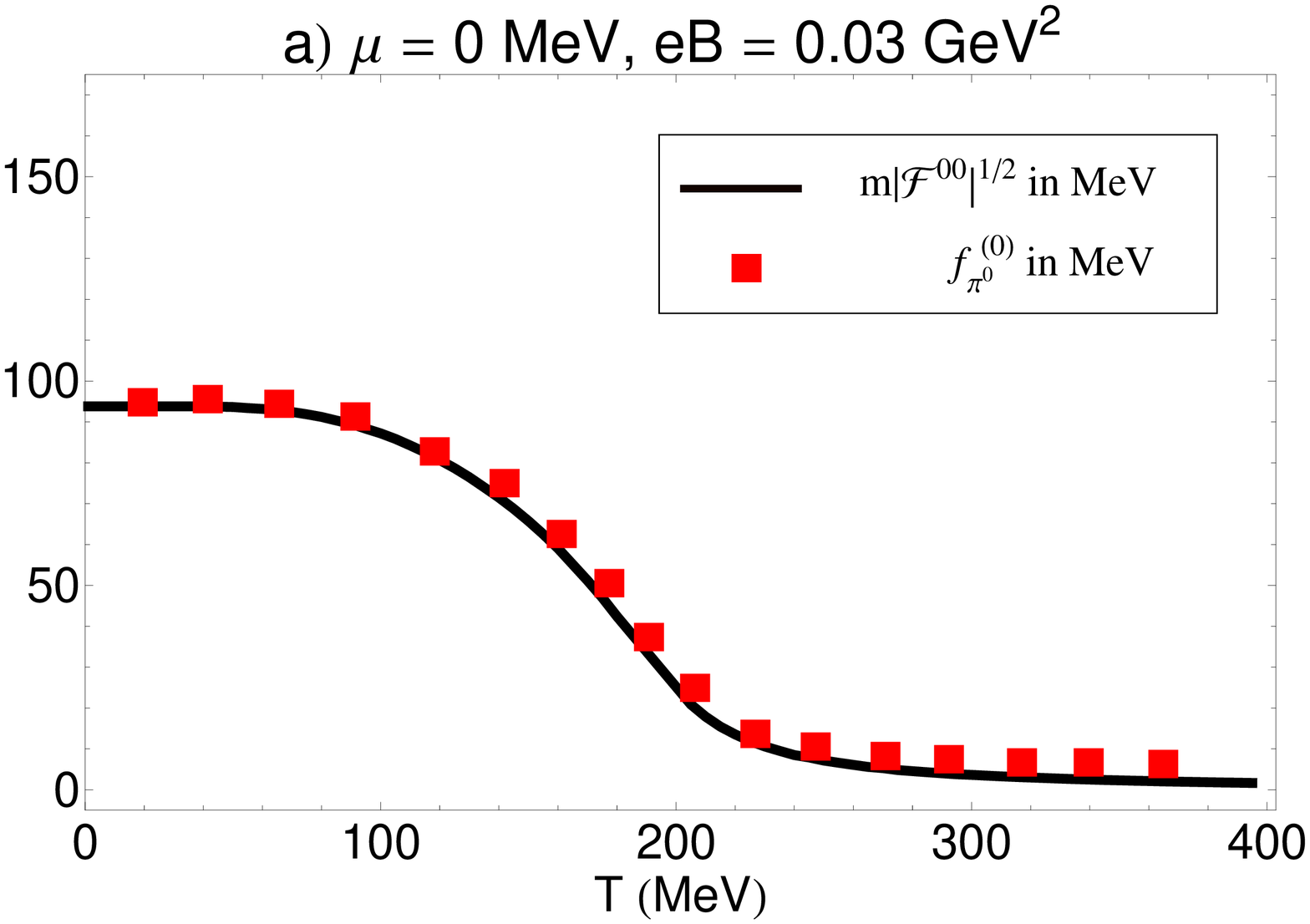}
\includegraphics[width=5.5cm,height=4cm]{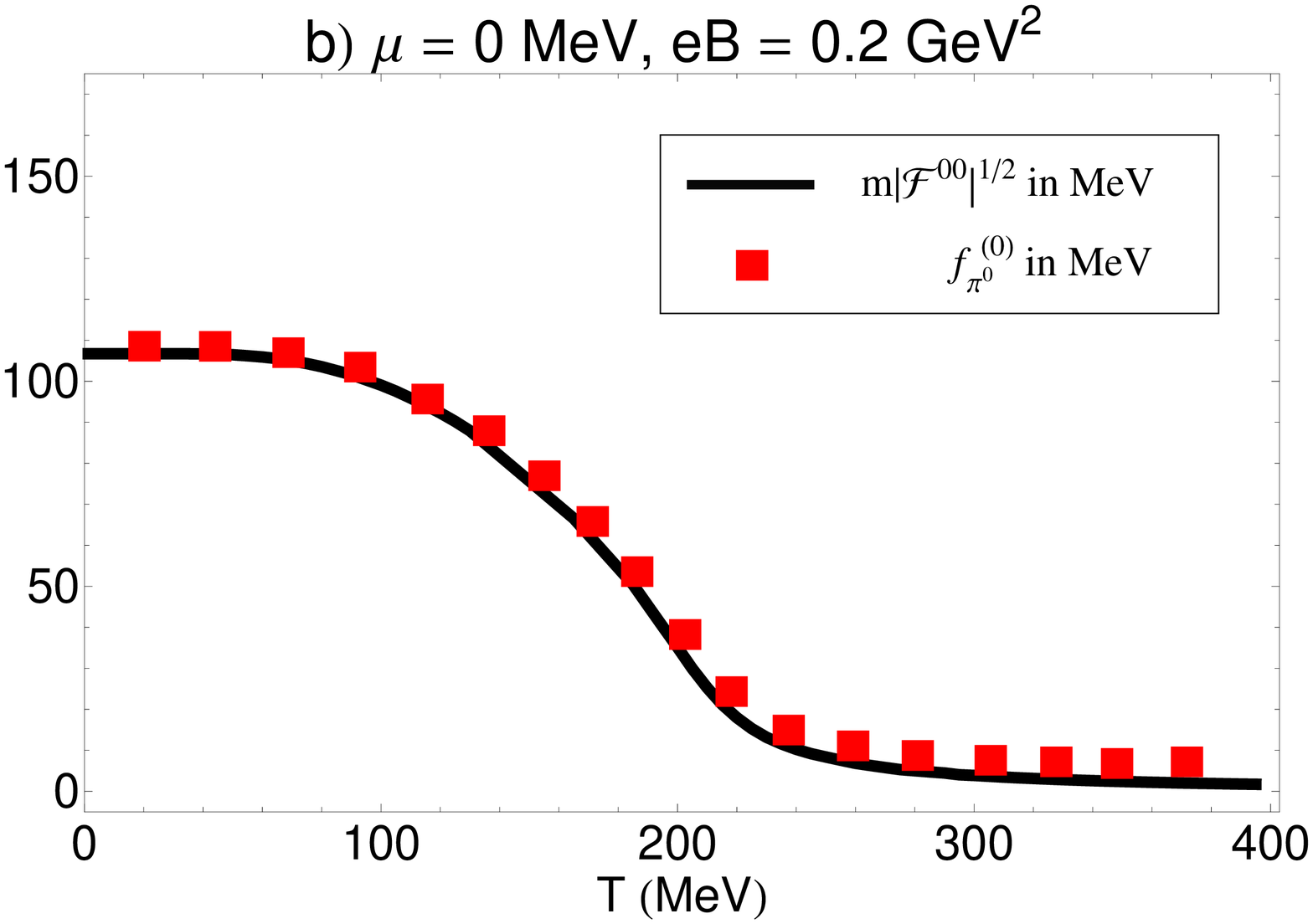}
\includegraphics[width=5.5cm,height=4cm]{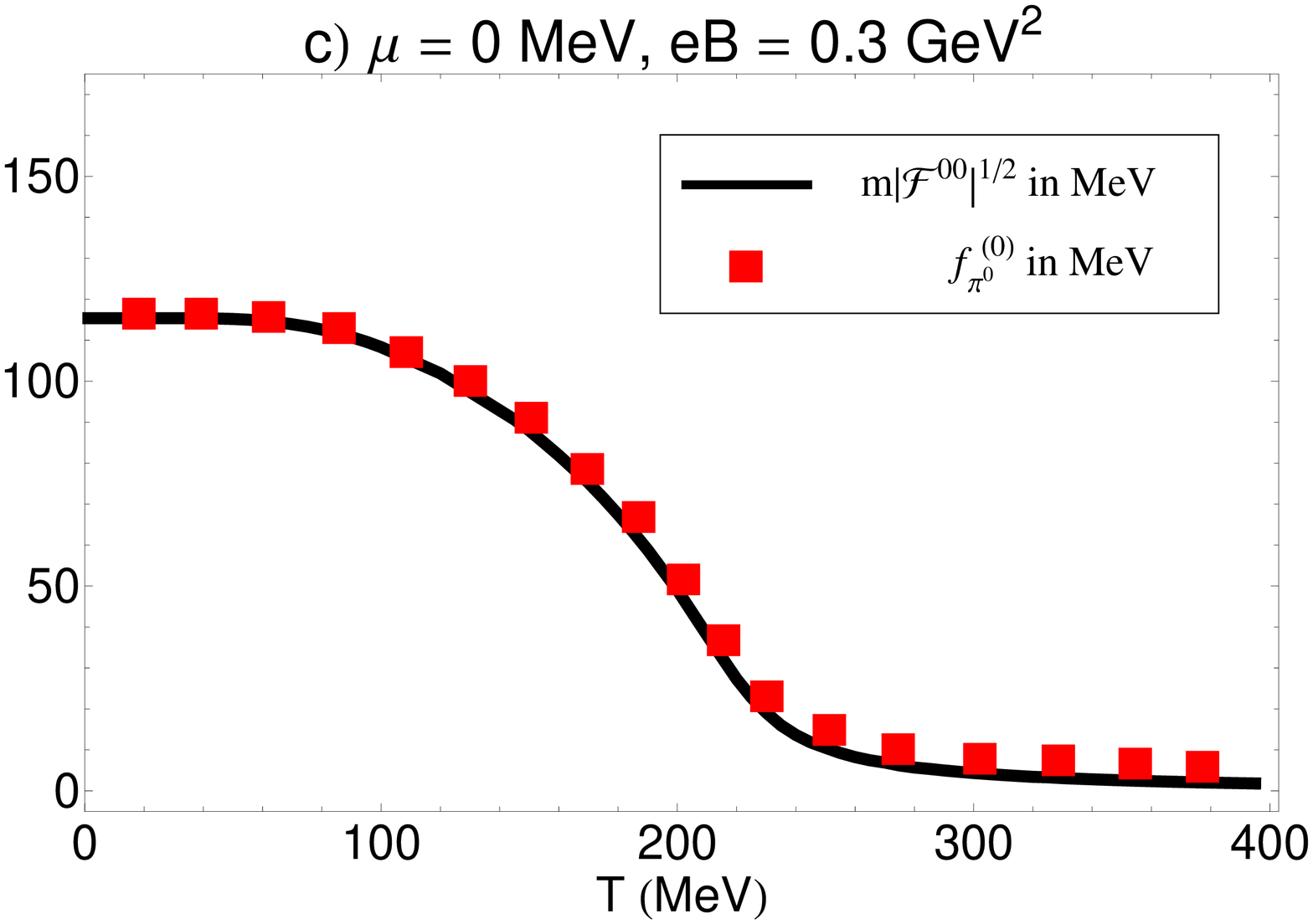}
\par
\includegraphics[width=5.5cm,height=4cm]{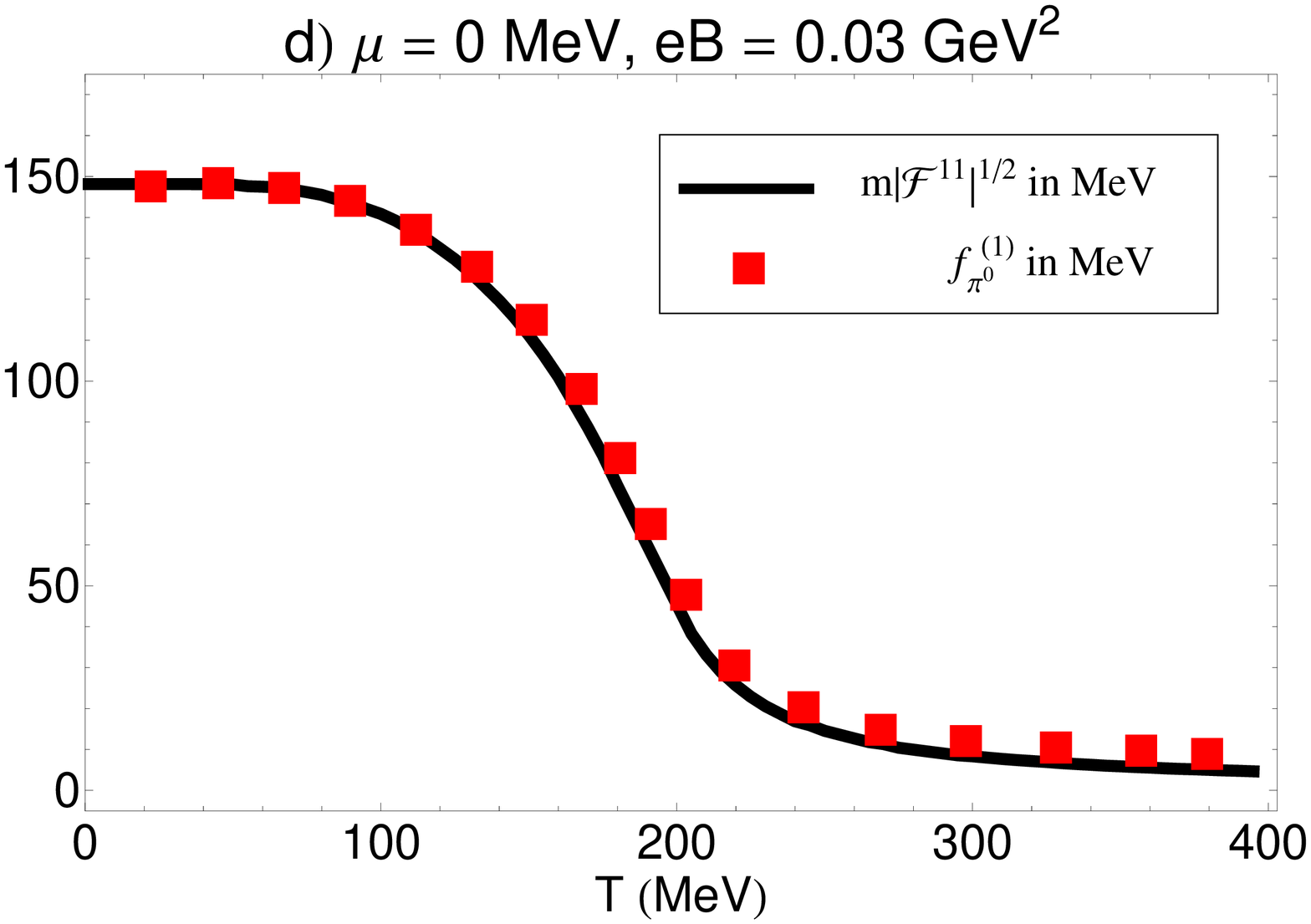}
\includegraphics[width=5.5cm,height=4cm]{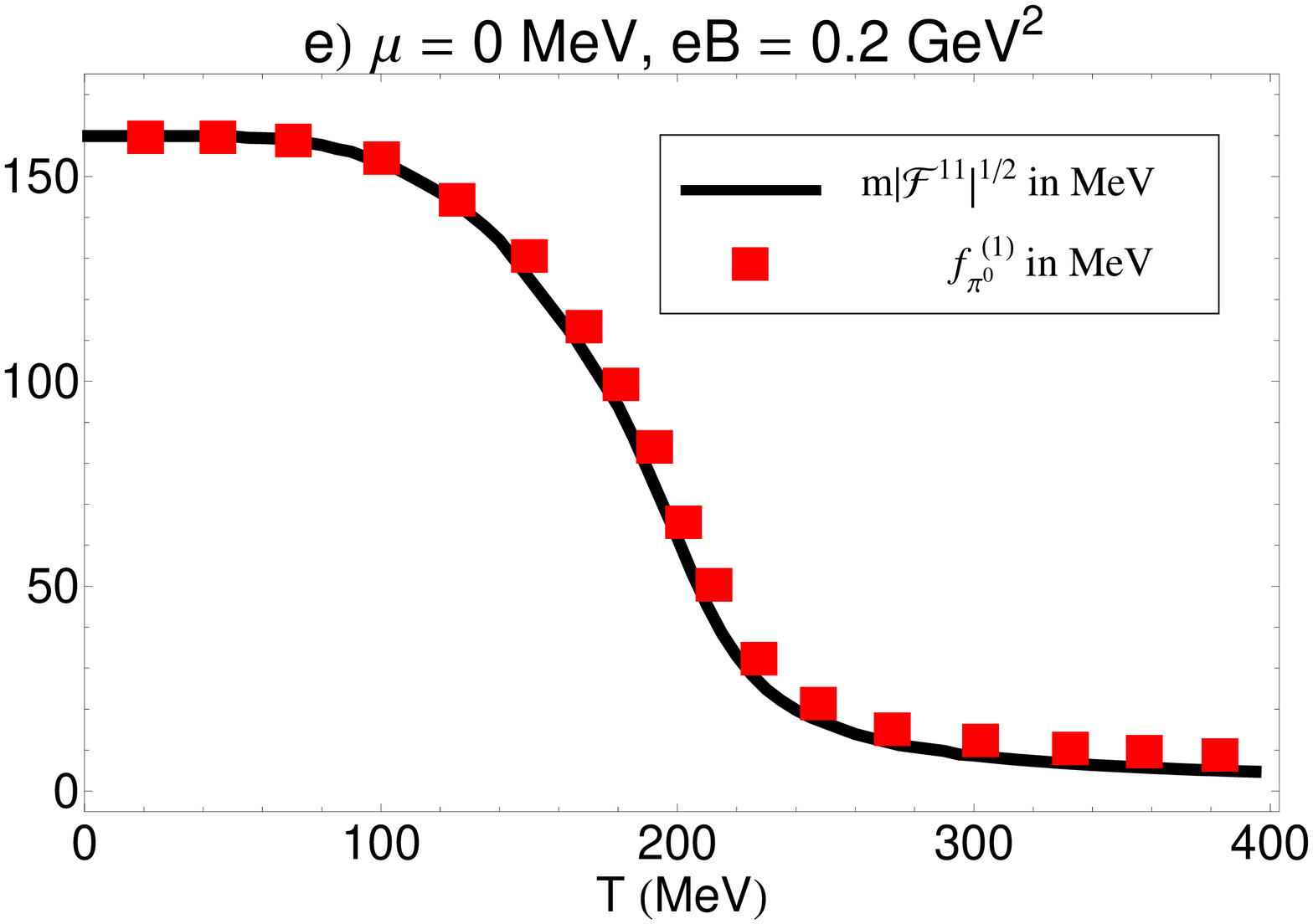}
\includegraphics[width=5.5cm,height=4cm]{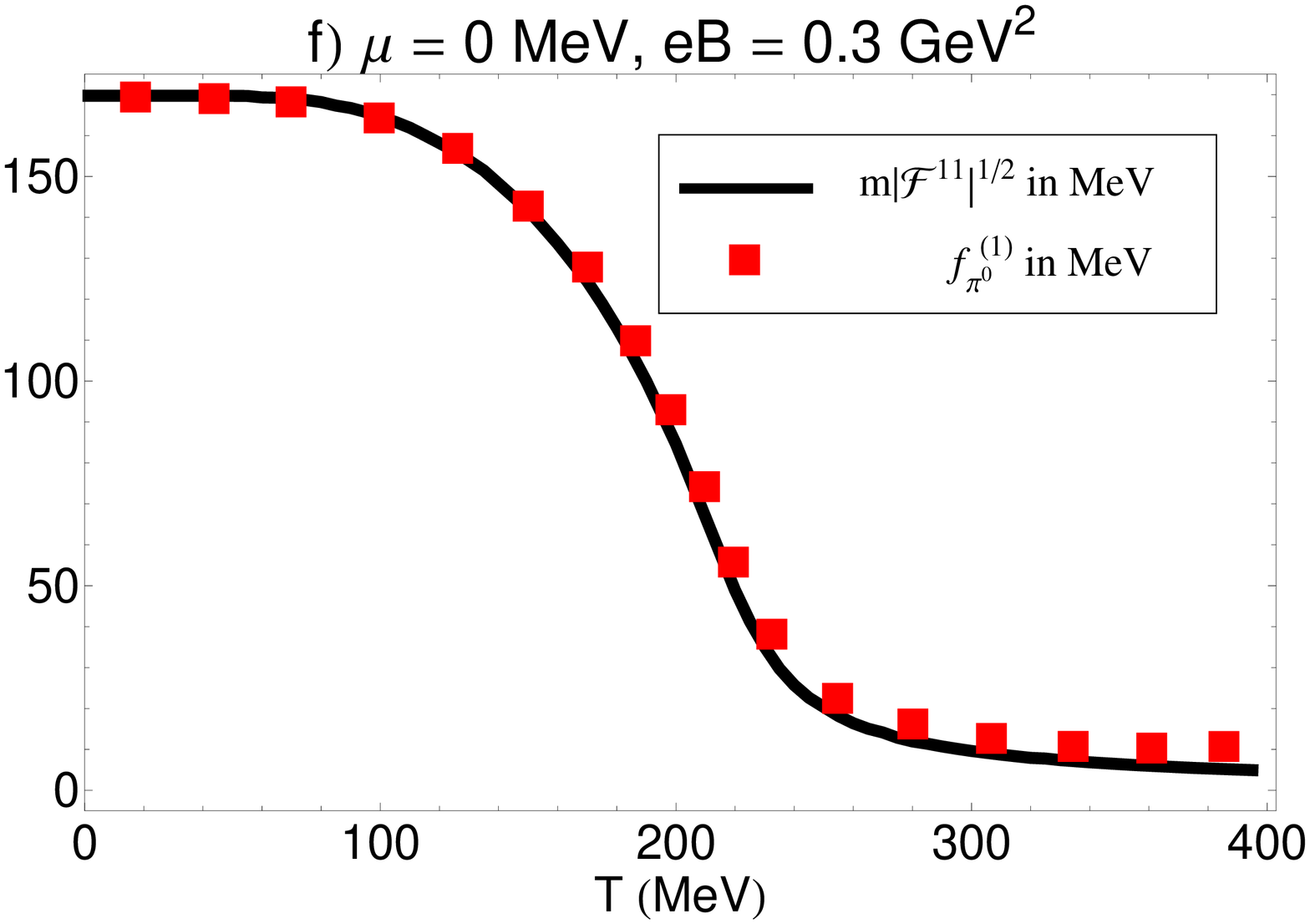}
\caption{The $T$ dependence of $f_{\pi^{0}}^{(\mu)}$ (red squares)
is compared with the $T$ dependence of
$m|{\cal{F}}_{33}^{\mu\mu}|^{1/2}$ (black solid lines) for
longitudinal $\mu=0,3$ [panels (a)-(c)] and transverse directions
$\mu=1,2$ [panels (d)-(e)]. Together with the definition
$f_{\pi^{0}}^{(\mu)}= f_{3}|{\cal{F}}_{33}^{\mu\mu}|^{1/2}$ from
(\ref{NE31c}) with $\ell=3$, we obtain $f_{3}=m$, as expected also
from the comparison of the data of $f_{3}$ in Fig. \ref{fig13a} and
the constituent quark mass $m=m_{0}+\sigma_{0}$ in Fig.
\ref{fig1}.}\label{fig14a}
\end{figure*}
\begin{figure*}[hbt]
\includegraphics[width=5.5cm,height=4cm]{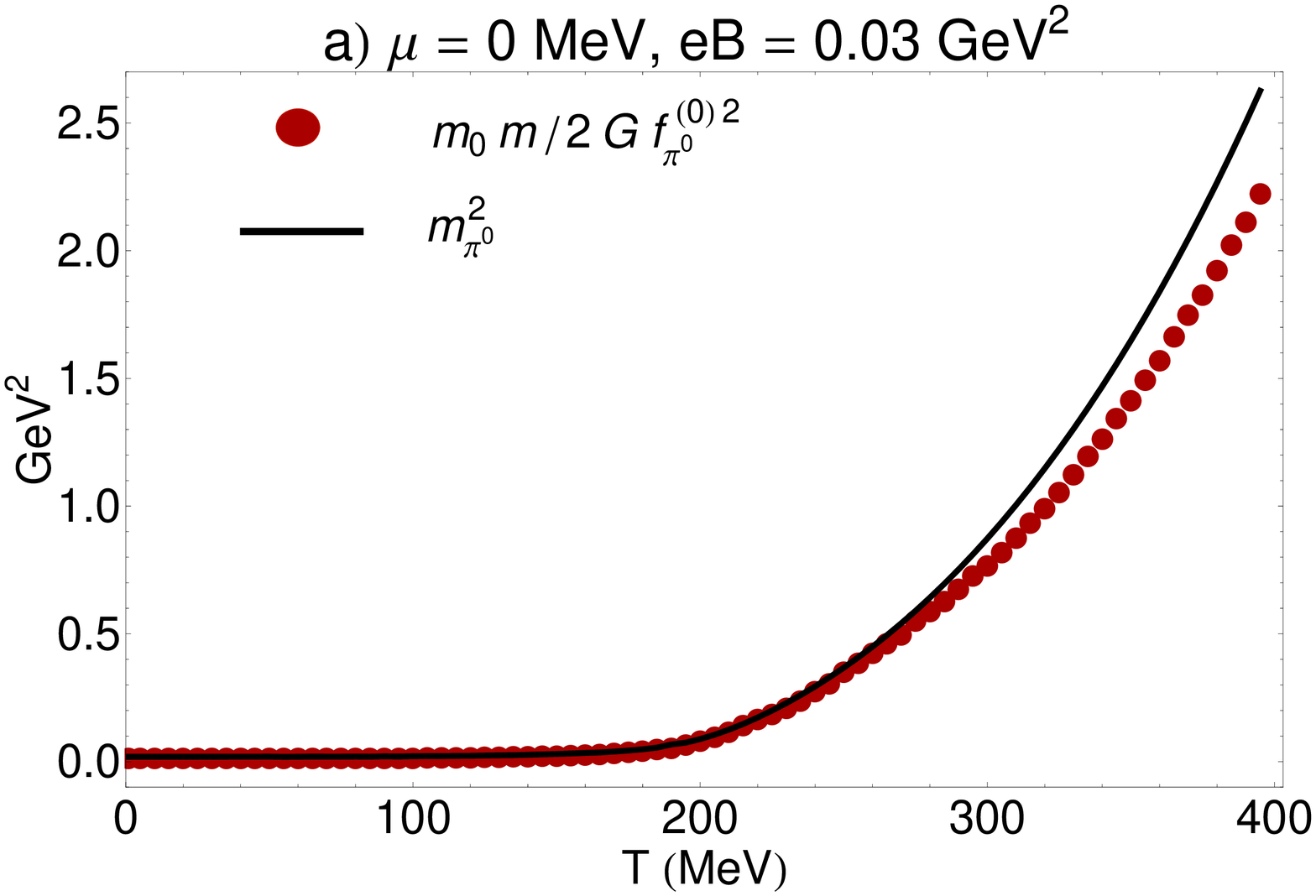}
\includegraphics[width=5.5cm,height=4cm]{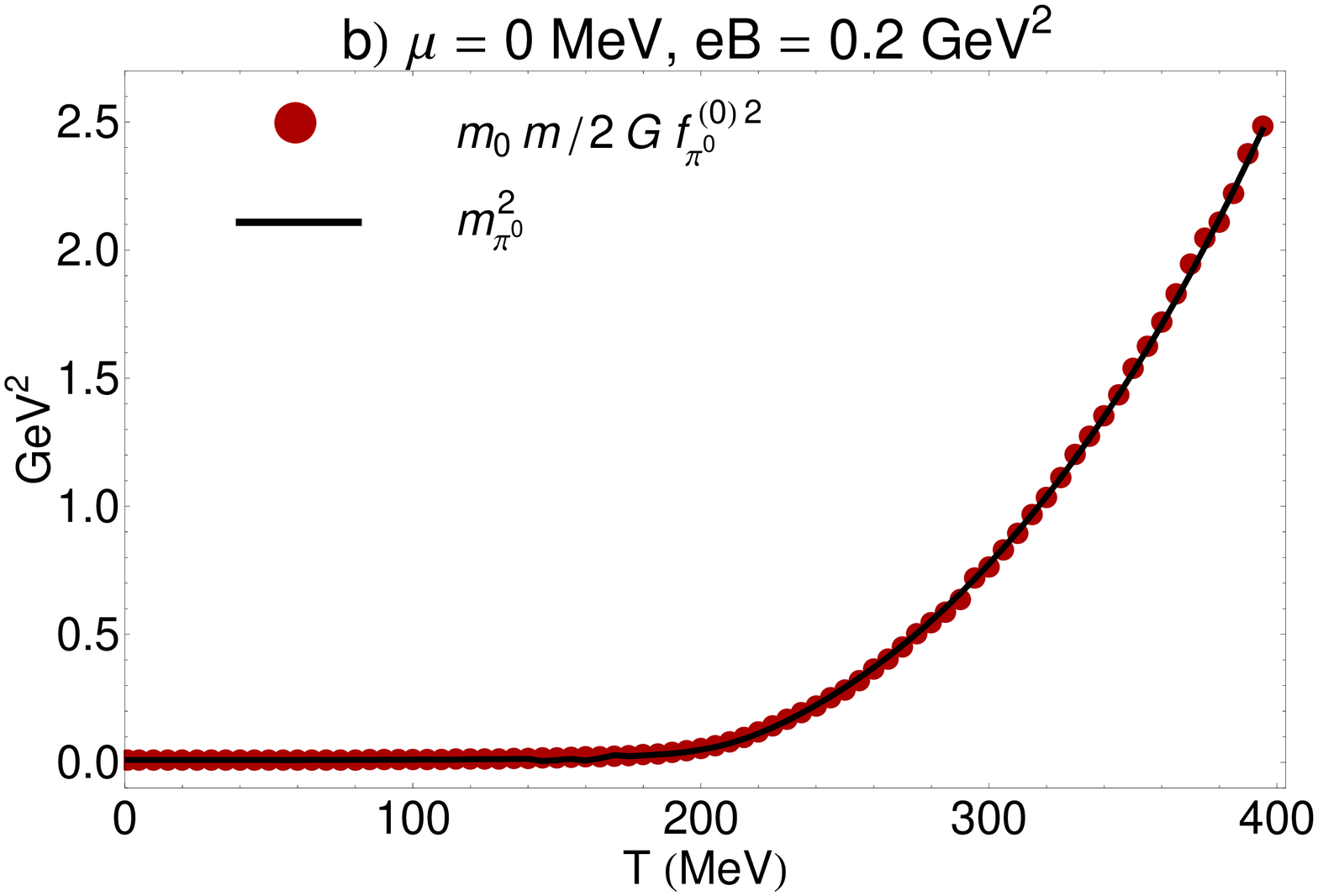}
\includegraphics[width=5.5cm,height=4cm]{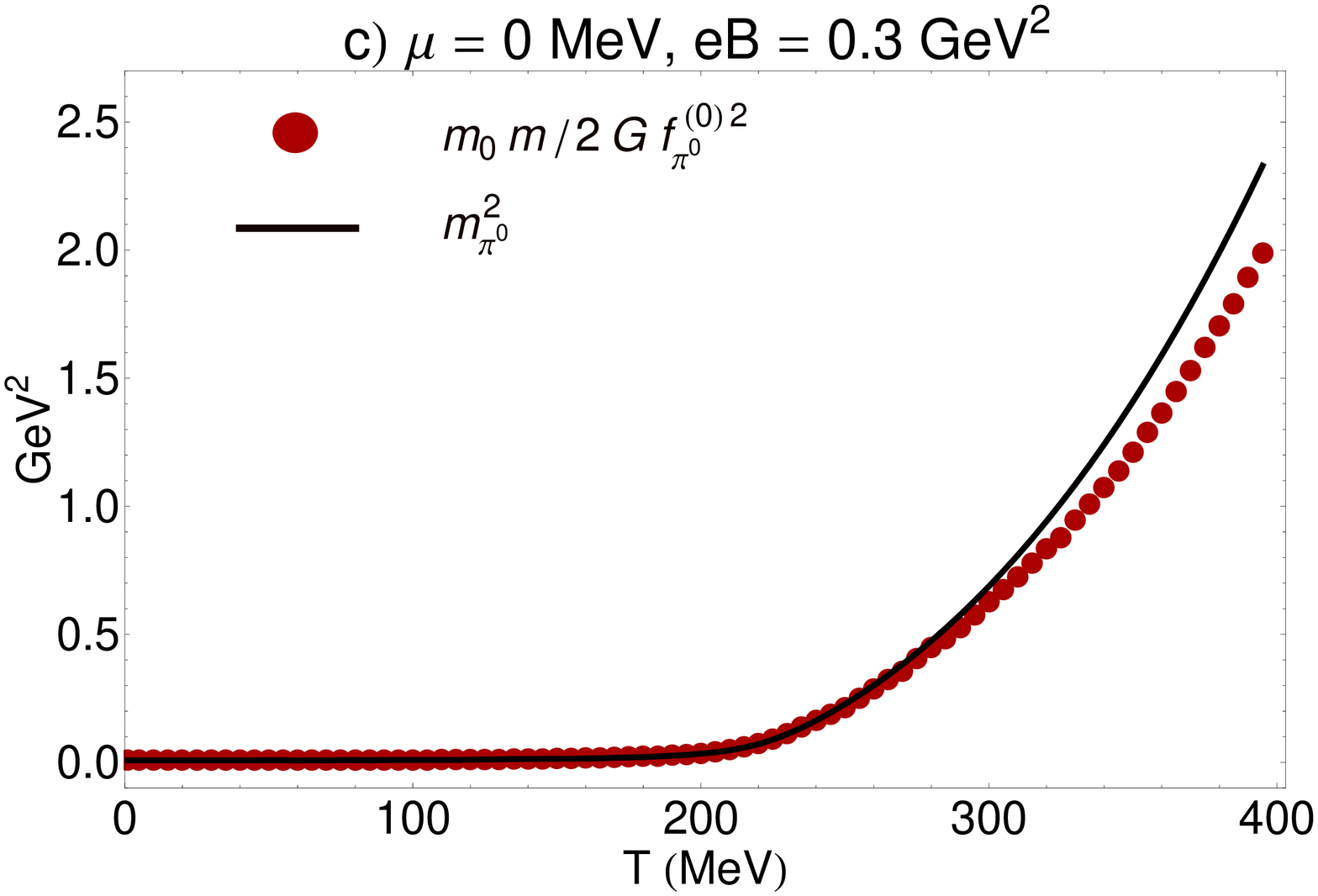}

\includegraphics[width=5.5cm,height=4cm]{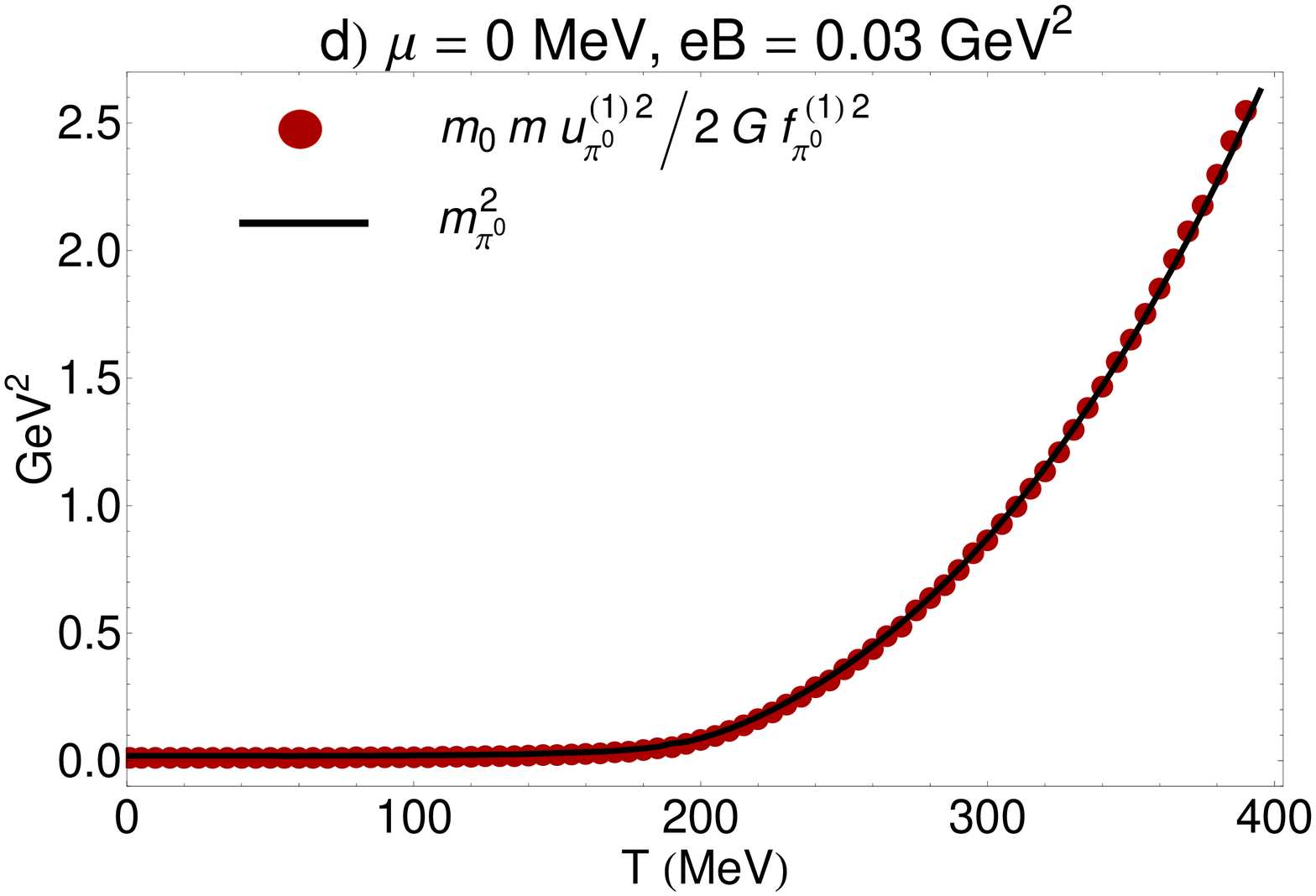}
\includegraphics[width=5.5cm,height=4cm]{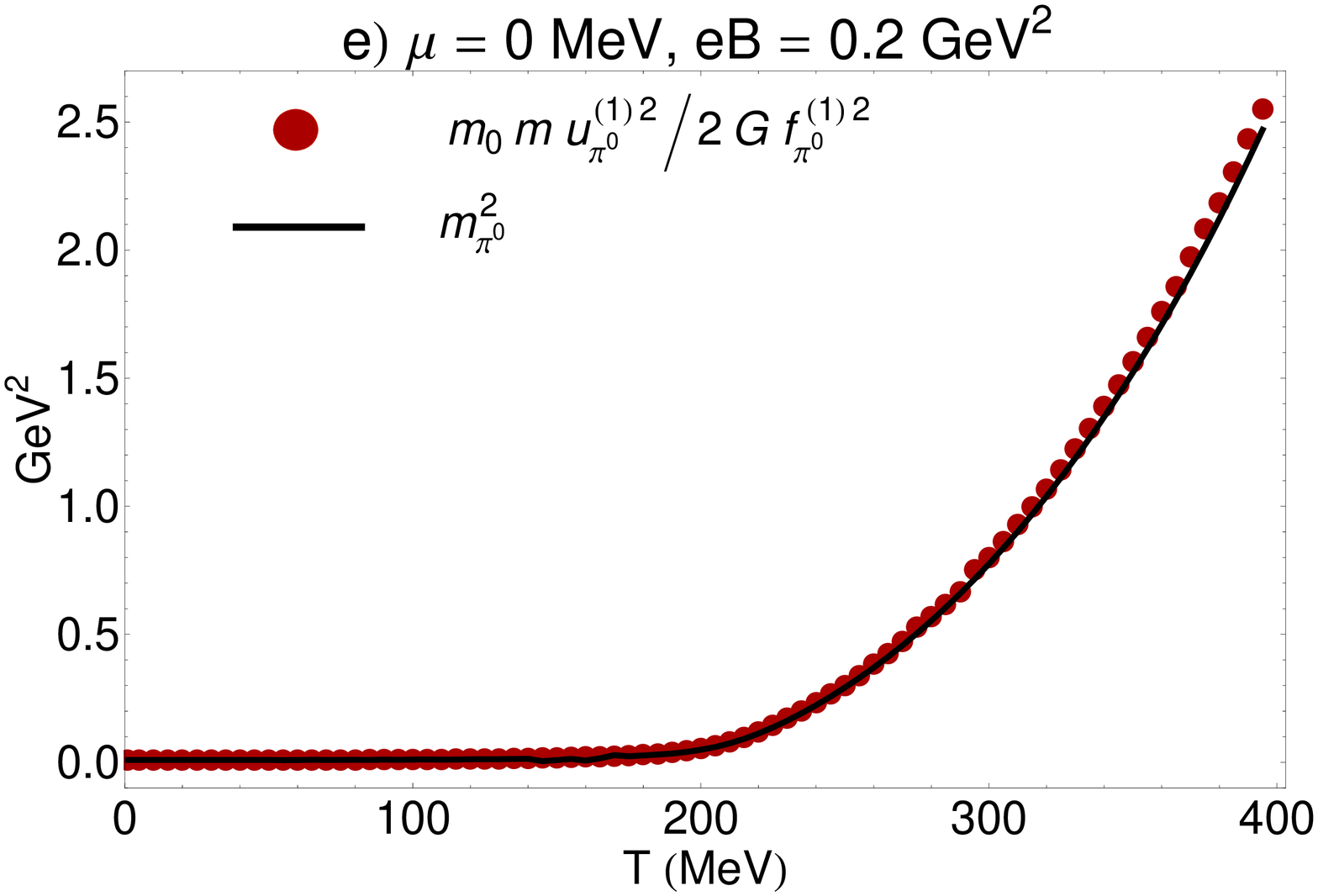}
\includegraphics[width=5.5cm,height=4cm]{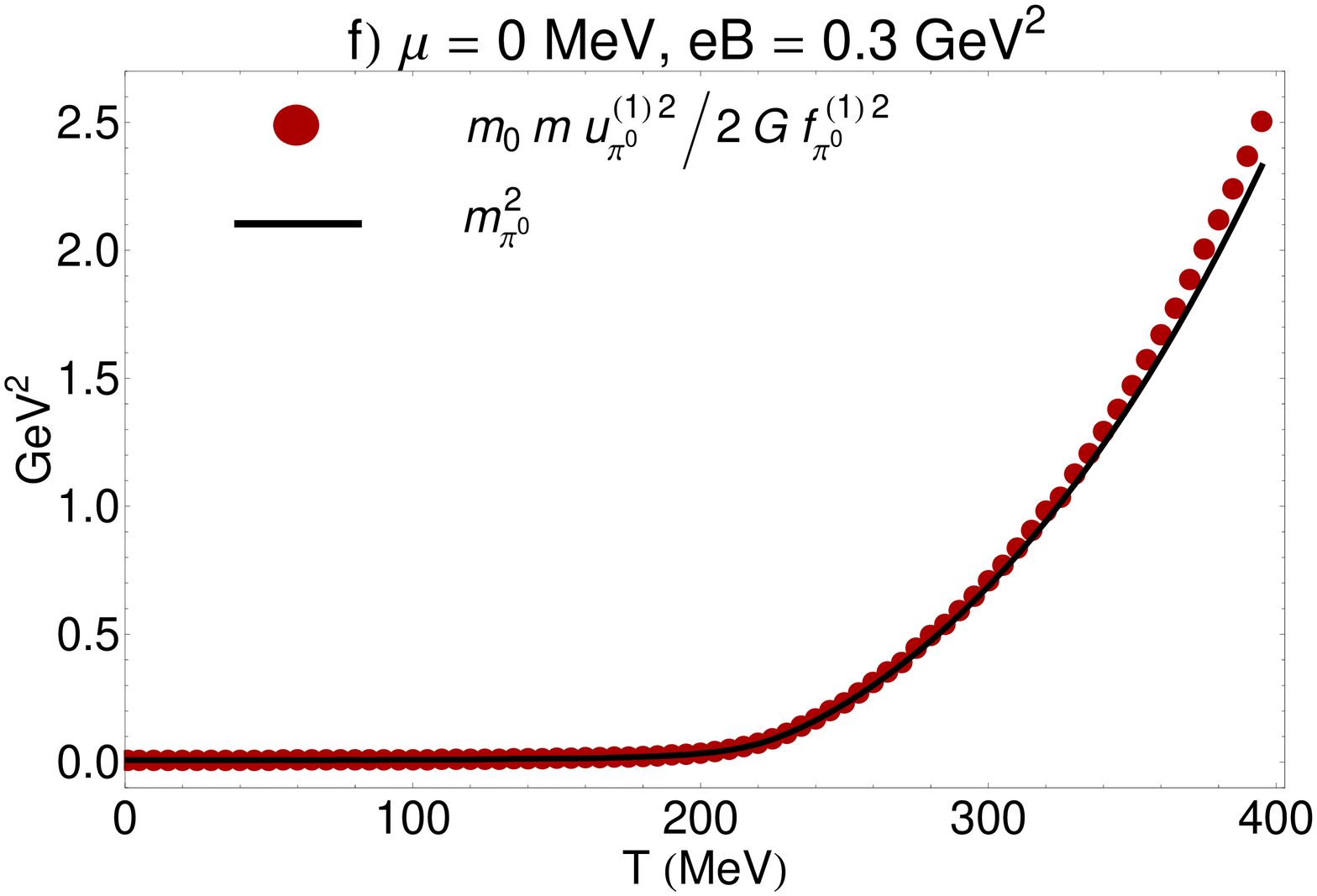}
\caption{To verify the relation (\ref{ND61}), suggested by the GOR
relation (\ref{ND55}), the $T$ dependence of $m_{\pi^{0}}^{2}$
(black solid lines) is compared with
$\frac{m_{0}m}{2Gf_{\pi^{0}}^{(0)2}}$ (red dots) [panels (a)-(c)]
and with $u_{\pi^{0}}^{(1)2}\frac{m_{0}m}{2Gf_{\pi^{0}}^{(1)2}}$
(red dots) [panels (d)-(f)]. Relation (\ref{ND61}) seems to be
exact, especially at temperatures below the critical temperature and
for small $eB$. }\label{fig15a}
\end{figure*}
\par\noindent
In the previous section, we determined analytically the directional
neutral mesons coupling and decay constants, $g_{qqM}^{(\mu)},
M\in\{\sigma,\pi^{0}\}$ and $f_{\pi^{0}}^{(\mu)}$, at finite
$(T,\mu,eB)$ up to an integration over $k_{3}$ momentum and a
summation over discrete Landau levels. Using these analytical
results, it is possible to prove the GT and GOR relations in the
limit of $m_{\pi^{0}}\to 0$. In the present section, we will
numerically determine the $T$ dependence of $g_{qqM}^{(\mu)},
M\in\{\sigma,\pi^{0}\}$ and $f_{\pi^{0}}^{(\mu)}$ for zero chemical
potential and finite $eB=0.03,0.2,0.3$ GeV$^{2}$. We will compare
$g_{qqM}^{(0)}$ with $g_{qqM}^{(1)}$, which corresponds to
quark-meson couplings in the longitudinal and transverse directions
with respect to the direction of the external magnetic field. We
then present the results for the $T$ dependence of
$f_{\pi^{0}}^{(\mu)}$ for vanishing chemical potential and fixed
$eB$, and will compare $f_{\pi^{0}}^{(0)}$ with $f_{\pi^{0}}^{(1)}$.
Using these numerical results, and the results from our previous
paper \cite{fayazbakhsh2012}, we will numerically determine the
dimensionful constant $f_{3}$, appearing in the definition
(\ref{NE31c}) with $\ell=3$. We will show that, for arbitrary
$m_{\pi^{0}}$, it is given by the constituent quark mass $m$.
Finally, a numerical verification of the GOR relation will be
presented.
\par
Let us start by fixing the parameters of our specific model,
$\Lambda,G$ and $m_{0}$. As in
\cite{fayazbakhsh2010,fayazbakhsh2012}, we have used
\begin{eqnarray}\label{ND59}
\Lambda=0.6643&\mbox{GeV},& \qquad G=4.668~\mbox{GeV}^{2},\nonumber\\
\qquad m_{0}=5&\mbox{MeV}.&
\end{eqnarray}
The UV momentum cutoff $\Lambda$ is necessary to perform the $k_{3}$
integrations in the results for $g_{qqM}^{(\mu)},
M\in\{\sigma,\pi^{0}\}$ and $f_{\pi^{0}}^{(\mu)}$ from previous
sections. It is also used to determine the upper limit $c$ for
the summation over Landau levels. We will fix $c$ by building the
ratio $\lfloor\frac{\Lambda^{2}}{|q_{f}eB|}\rfloor\equiv c$, where
$\lfloor z\rfloor$ is the greatest integer less than or equal to
$z$.  To perform the momentum integrations over ${\mathbf{k}}$ and
$k_{3}$, we use, as in \cite{fayazbakhsh2010, fayazbakhsh2012},
smooth cutoff functions
\begin{eqnarray}\label{ND60}
f_{\Lambda,0}&=&\frac{1}{1+\exp\left(\frac{|{\mathbf{k}}|-\Lambda}{A}\right)}, \nonumber\\
f_{\Lambda,B}^{k}&=&\frac{1}{1+\exp\left(\frac{\sqrt{k_{3}^{2}+2|q_{f}eB|k}-\Lambda}{A}\right)},
\end{eqnarray}
which correspond to integrals with vanishing and nonvanishing
magnetic fields. In $f_{\Lambda,B}^{k}$, the upper index $k$ labels
the Landau levels. In (\ref{ND60}), $A$ is a free parameter, that
determines the sharpness of the cutoff. It is fixed to be
$A=0.05\Lambda$, with $\Lambda$ from (\ref{ND59}).  As we have also
mentioned in Sec. \ref{sec2}, fixing the free parameters of the
model as in (\ref{ND59}), the constituent mass $m$ at zero
$(T,\mu,eB)$ turns out to be $m\sim 308$ MeV, as expected. Moreover,
the pion mass and decay constant for vanishing $(T,\mu,eB)$ are
given by $m_{\pi}\sim 139.6$ MeV and $f_{\pi}=93.35$ MeV,
respectively. These values are expected from the phenomenology of a
two-flavor NJL model \cite{buballa2004}. We will limit all our
numerical computations in this section in the interval $T\in
[0,400]$ MeV for vanishing $\mu$ and $eB=0,0.03, 0.2,0.3$ GeV$^{2}$.
As we have shown in \cite{fayazbakhsh2010, fayazbakhsh2012}, in this
interval the chiral symmetry of the theory is broken and the quark
condensate $\sigma_{0}\sim \langle\bar{\psi}\psi\rangle$, appearing
in $m=m_{0}+\sigma_{0}$, plays the role of the order parameter of
the transition from a chirally broken to a chirally symmetric phase.
\par
In Figs. \ref{fig8a}(a)-(c), we have plotted the $T$ dependence of
the directional quark-meson coupling $g_{qq\sigma}^{(\mu)}$ (solid
blue curves) and $g_{qq\pi^{0}}^{(\mu)}$ (red dashed curves) for
$eB=0$ [Fig. \ref{fig8a}(a)] and $eB=0.2$ GeV$^{2}$ [Figs.
\ref{fig8a}(b) and \ref{fig8a}(c)]. Whereas for $eB=0$, there is no
difference between $g_{qqM}^{(\mu)}, M\in\{\sigma,\pi^{0}\}$ for
different directions $\mu=0,\cdots,3$, for $eB\neq 0$, the
longitudinal and transverse quark-meson couplings for $\sigma$ and
$\pi^{0}$ are different [compare the solid blue and dashed red
curves in Figs. \ref{fig8a}(b) and \ref{fig8a}(c)]. At certain
temperature $T_{\mbox{\tiny{cross}}}$, the two curves cross, i.e.,
$g_{qq\sigma}^{(\mu)}$ becomes equal to $g_{qq\pi^{0}}^{(\mu)}$. For
$T<T_{\mbox{\tiny{cross}}}$,
$g_{qq\sigma}^{(\mu)}<g_{qq\pi^{0}}^{(\mu)}$ and for
$T>T_{\mbox{\tiny{cross}}}$,
$g_{qq\sigma}^{(\mu)}>g_{qq\pi^{0}}^{(\mu)}$. Moreover, as it turns
out, for $eB=0$,
$T_{\mbox{\tiny{diss.}}}<T_{\mbox{\tiny{cross}}}<T_{\mbox{\tiny{Mott}}}$.
For $eB\neq 0$, however, we have to separate the longitudinal and
transverse cases. Whereas for longitudinal couplings, we have
$T_{\mbox{\tiny{diss.}}}<T_{\mbox{\tiny{cross}}}<T_{\mbox{\tiny{Mott}}}$,
for transverse couplings, we get
$T_{\mbox{\tiny{cross}}}<T_{\mbox{\tiny{diss.}}}<T_{\mbox{\tiny{Mott}}}$.
Whether this temperature is of certain relevance, in particular, in
the scattering length and the total cross section of meson-meson
scattering, where the thermodynamic properties of quark-meson
couplings play an important role is under investigation.
\par
In Fig. \ref{fig9a}, the $T$ dependence of directional quark-pion
couplings, $g_{qq\pi^{0}}^{(\mu)}$, is plotted for vanishing
chemical potential and various $eB=0,0.03,0.2,0.3$ GeV$^{2}$. The
longitudinal quark-pion coupling
$g_{qq\pi^{0}}^{(0)}=g_{qq\pi^{0}}^{(3)}$ from (\ref{ND13}) with
$M=\pi^{0}$ is plotted in Fig. \ref{fig9a}(a), and the transverse
quark-pion coupling $g_{qq\pi^{0}}^{(1)}=g_{qq\pi^{0}}^{(2)}$ from
(\ref{ND25}) with $M=\pi^{0}$ is plotted in Fig. \ref{fig9a}(b).
Whereas the behavior of longitudinal and transverse couplings is
different by increasing the strength of the external magnetic field
for a fixed temperature, it turns out to be similar when we increase
the temperature from $T=0$ to $T=400$ MeV. An inflection point
arises in directional quark-pion coupling constants around the
critical temperature $T\sim 200$ MeV. Comparing the longitudinal and
transverse quark-pion couplings in Fig. \ref{fig10a}, it turns out
that $g_{qq\pi^{0}}^{(0)}>g_{qq\pi^{0}}^{(1)}$  in the interval
$T\in[0,400]$ MeV and for fixed chemical potential and magnetic
fields. This is a direct consequence of the fact that the transverse
refraction index of pions is larger than unity: As we have seen in
Fig. \ref{fig4}, $u_{\pi^{0}}^{(1)}=u_{\pi^{0}}^{(2)}>1$. Using
$g_{qq\pi^{0}}^{(\mu)}=g_{qq\pi^{0}}|{\cal{F}}_{33}^{\mu\mu}|^{-1/2}$
from (\ref{NE24c}) with $\ell=3$, and
$u_{\pi^{0}}^{(\mu)}=\big|{{\cal{F}}_{33}^{\mu\mu}/{\cal{F}}_{33}^{00}}\big|^{1/2}$
from (\ref{NE17c}), we will get
$g_{qq\pi^{0}}^{(0)}=g_{qq\pi^{0}}|{\cal{F}}_{33}^{00}|^{-1/2}$ as
well as $g_{qq\pi^{0}}^{(1)}=g_{qq\pi^{0}}^{(0)}/u_{\pi^{0}}^{(1)}$.
For $u_{\pi^{0}}^{(1)}>1$, it turns out that
$g_{qq\pi^{0}}^{(0)}>g_{qq\pi^{0}}^{(1)}$, as is observed in Fig.
\ref{fig10a}.
\par
Let us now consider the longitudinal and transverse pion decay
constants, $f_{\pi^{0}}^{(0)}=f_{\pi^{0}}^{(3)}$ and
$f_{\pi^{0}}^{(1)}=f_{\pi^{0}}^{(2)}$ from (\ref{ND42}) and
(\ref{ND46}), respectively. In Fig. \ref{fig11a}(a), the $T$
dependence of $f_{\pi^{0}}$ is plotted for vanishing chemical
potential and magnetic fields. In Figs. \ref{fig11a}(b) and
\ref{fig11a}(c), the $T$ dependence of longitudinal and transverse
decay constants of neutral pions, $f_{\pi^{0}}^{(0)}$ and
$f_{\pi^{0}}^{(1)}$, is plotted for zero chemical potential and
nonvanishing magnetic fields $eB=0.03,0.2,0.3$ GeV$^{2}$. In both
cases, $f_{\pi^{0}}^{(\mu)}$ is large below the critical temperature
and decreases with increasing temperature.  At a fixed temperature,
the directional pion decay constants increase as the strength of the
magnetic field increases. In Figs. \ref{fig12a}(a)-\ref{fig12a}(c),
we have compared the $T$ dependence of the longitudinal and
transverse decay constants for $\mu=0$ and $eB=0.03, 0.2,0.3$
GeV$^{2}$. As it turns out, $f_{\pi^{0}}^{(0)}<f_{\pi^{0}}^{(1)}$.
This behavior is again related to $u_{\pi^{0}}^{(1)}>1$ from Fig.
\ref{fig4}: Using the definition
$f_{\pi^{0}}^{(\mu)}=f_{3}|{\cal{F}}_{33}^{\mu\mu}|^{1/2}$ from
(\ref{NE31c}) with $\ell=3$, and
$u_{\pi^{0}}^{(\mu)}=\big|{{\cal{F}}_{33}^{\mu\mu}/{\cal{F}}_{33}^{00}}\big|^{1/2}$
from (\ref{NE17c}), it turns out that
$f_{\pi^{0}}^{(0)}=f_{3}|{\cal{F}}_{33}^{00}|^{1/2}$ and
$f_{\pi^{0}}^{(1)}=f_{\pi^{0}}^{(0)}u_{\pi^{0}}^{(1)}$. For
$u_{\pi^{0}}^{(1)}>1$, we have therefore
$f_{\pi^{0}}^{(0)}<f_{\pi^{0}}^{(1)}$, as observed in Fig.
\ref{fig12a}. Note that combining
$g_{qq\pi^{0}}^{(1)}=g_{qq\pi^{0}}^{(0)}/u_{\pi^{0}}^{(1)}$ and
$f_{\pi^{0}}^{(1)}=f_{\pi^{0}}^{(0)}u_{\pi^{0}}^{(1)}$ yields the GT
relation (\ref{NE32c}).
\par
Using the corresponding data of $f_{\pi^{0}}^{(\mu)}$ and the
definition of directional decay constant $f_{\pi^{0}}^{(\mu)}$ from
(\ref{NE31c}) with $\ell=3$, it is possible to determine the $T$
dependence of $f_{3}$ defined by
$f_{\pi^{0}}^{(\mu)}=f_{3}|{\cal{F}}_{33}^{\mu\mu}|^{1/2}$. To do
this, we have also used the data corresponding to the $T$ dependence
of ${\cal{F}}_{33}^{\mu\mu}$ from our previous paper
\cite{fayazbakhsh2012}. In Fig. \ref{fig13a}, we have presented the
$T$ dependence of $f_{3}$ for vanishing chemical potential and
various $eB=0.03,0.2,0.3$ GeV$^{2}$. Comparing these results with
the corresponding data from the $T$ dependence of the constituent
quark mass $m$, it turns out that $f_{3}=m$ for an arbitrary value
of $m_{\pi^{0}}$. According to our arguments from
\cite{fayazbakhsh2012}, for a fixed $T$ and $\mu$, the constituent
quark mass increases with increasing strength of the magnetic field
(see also Fig. \ref{fig1}). This is because of the phenomenon of
magnetic catalysis \cite{klimenko1992, miransky1995}. We conclude
therefore that the behavior of longitudinal and transverse pion
decay constants in external magnetic fields at fixed temperature and
chemical potential is mainly affected by this phenomenon.

Note that another way to verify $f_{3}=m$ is to compare the $T$
dependence of $f_{\pi^{0}}^{(\mu)}$ on the l.h.s. of the relation
$f_{\pi^{0}}^{(\mu)}=m|{\cal{F}}_{33}^{\mu\mu}|^{1/2}$ with the $T$
dependence of $m|{\cal{F}}_{33}^{\mu\mu}|^{1/2}$ on the r.h.s. of
this relation for longitudinal  directions $\mu=0,3$ [Figs.
\ref{fig14a}(a)-\ref{fig14a}(c)], and transverse directions
$\mu=1,2$ [Figs. \ref{fig14a}(d)-\ref{fig14a}(f)]. The red squares
denote the data for $f_{\pi^{0}}^{(\mu)}, \mu=0,1$ and the black
solid lines the combination $m|{\cal{F}}^{00}|^{1/2}$ for
longitudinal $\mu=0,3$ and $m|{\cal{F}}^{11}|^{1/2}$ for transverse
$\mu=1,2$ directions.
\par
Let us now consider the GOR relation (\ref{ND55}) for neutral pions.
In Sec. \ref{subsec4-A}, we have analytically proved the GOR
relation (\ref{ND55}) in the limit of vanishing $m_{\pi^{0}}$. In
Fig. \ref{fig15a}, however, we have compared the $T$ dependence of
the  expression $u_{\pi^{0}}^{(\mu)2}\frac{m_{0}m}{2G
f_{\pi^{0}}^{(\mu)}}$ (red dots) with the $T$ dependence of
$m_{\pi^{0}}^{2}$ (black solid lines) for any value of $m_{\pi^{0}}$
for various $eB=0.03,0.2,0.3$ GeV$^{2}$.  In Figs.
\ref{fig15a}(a)-\ref{fig15a}(c), we present the data for
longitudinal directions, i.e. for $\mu=0$ with
$u_{\pi^{0}}^{(0)}=u_{\pi^{0}}^{(3)}=1$ [see Fig. \ref{fig4}(b) for
the longitudinal refraction index of neutral pions], and in Figs
\ref{fig15a}(d)-\ref{fig15a}(f) the data for transverse directions
are presented. Note that, according to Fig. \ref{fig4}
\cite{fayazbakhsh2012}, the transverse refraction index of neutral
pions is larger than unity. As it is shown in Fig. \ref{fig15a}, the
relation
\begin{eqnarray}\label{ND61}
m_{\pi^{0}}^{2}=u_{\pi^{0}}^{(\mu)2}\frac{m_{0}m}{2G f_{\pi^{0}}^{(\mu)}},
\end{eqnarray}
seems to be exact in the interval below the critical temperature. As
we have explained in \cite{fayazbakhsh2012}, the critical
temperature is defined by the temperature from which the neutral
pion mass starts to increase. The deviations of the expression on
the r.h.s. of (\ref{ND61}) from $m_{\pi^{0}}^{2}$ on the l.h.s. of
this relation depend on $eB$ and $T$.
\section{Summary and Conclusions}\label{sec5}\par\noindent
Uniform magnetic fields affect the properties of mesons in a hot and
dense quark matter. In our previous paper \cite{fayazbakhsh2012}, we
systematically studied these effects on the pole and screening mass
as well as the refraction index of pions. In the present paper, we
continued with exploring the effect of constant magnetic fields on
the pion weak decay constant. One of our first observations was that
external magnetic fields break the isospin symmetry of pions,
insofar as charged and neutral pions behave differently in the
external magnetic fields. In contrast to the studies performed in
\cite{andersen2011-pions, anderson2012-2}, where chiral perturbation
theory is used to study the effect of constant magnetic fields on
the properties of charged pions, we focused on the properties of
neutral pions in external magnetic fields.
\par
In this paper, following the same method as in our previous paper
\cite{fayazbakhsh2012}, we used an appropriate derivative expansion
up to second order and derived the effective action of a two-flavor
bosonized NJL model at finite $(T,\mu,eB)$ in this approximation.
The resulting effective action is given as a functional of $\sigma$
and $\pi^{0}$ mesons and includes a kinetic and a potential part. To
derive the mass and refraction indices of neutral mesons, we
focused, in particular, on the nontrivial kinetic part of the
effective action, including nontrivial form factors. They were
evaluated in \cite{fayazbakhsh2012} using the method presented in
Sec. \ref{sec2}. As it turns out, they play an essential role in
defining the pole and screening mass as well as the directional
refraction indices of neutral pions. As concerns the pions'
refraction index, it turns out that in the presence of a constant
magnetic field, aligned in a fixed direction, the refraction indices
of neutral pions parallel to the direction of the external magnetic
field and in the plane perpendicular to this direction are
different. Within our approximation, where the contributions of
pion-pion interactions are neglected, and at $T\neq 0$ and $eB\neq
0$, the longitudinal refraction index is unity for all nonvanishing
$T$ and $eB$, while the transverse refraction index is larger than
unity. This is in contrast to the case $T=0$ and $eB=0$, where the
refraction index for neutral pions in all spatial directions is
equal to unity. This is also in contrast to the case $T\neq 0$ and
$eB=0$, where all spatial refraction indices are equal to and
smaller than unity \cite{pisarski1996, son2000,ayala2002}. One of
the possibilities to improve the results in this paper is to include
pion self-interaction terms in the original Lagrangian for pions and
to study the $T$ dependence of directional refraction indices in the
longitudinal and transverse directions in that setup.
\par
Our central analytical result in this paper is presented in Sec.
\ref{sec3}. In Sec. \ref{subsec3-A}, we briefly reviewed the method
introduced in \cite{buballa2004} and derived the weak decay constant
of pions at zero $T$ and $eB$. In this case, the pions satisfy an
ordinary isotropic energy dispersion relation
$E_{\pi}^{2}={\mathbf{q}}^{2}+m_{\pi}^{2}$, and the weak decay
constant $f_{\pi}$ is given by combining the ordinary PCAC relation
and the Feynman integral corresponding to a one-pion-to-vacuum
amplitude. As a by-product, the momentum-dependent quark-pion
coupling constant $g_{qq\pi}(q)$ is also defined. Our starting point
in Sec. \ref{subsec3-B}, however, was an anisotropic energy
dispersion relation,
$E_{\pi}^{2}=u_{\pi}^{(i)2}q_{i}^{2}+m_{\pi}^{2}$, including
directional refraction indices $u_{\pi}^{(i)}, i=1,2,3$. Note that,
according to the above arguments, the case of nonvanishing magnetic
field and/or finite temperature can be viewed only as special cases
of the general assumption of Sec. \ref{subsec3-B}.\footnote{Remember
that at finite temperature, the refraction indices satisfy
$u_{\pi}^{(1)}=u_{\pi}^{(2)}=u_{\pi}^{(3)}=u$ [see (\ref{int4})]. In
the nonvanishing magnetic field, however, they satisfy
$u_{\pi}^{(1)}=u_{\pi}^{(2)}\neq u_{\pi}^{(3)}$ [see (\ref{int5})].}
Introducing a modified PCAC relation with nontrivial four-vector
$u_{\pi_{a}}^{(\mu)}=(1,u^{(i)}_{\pi_{a}})$, as given in
(\ref{NE27c}), and following the same arguments as in Sec.
\ref{subsec3-A}, we derived the modified GT and GOR relations,
(\ref{NE32c}) and (\ref{NE33c}), including the directional pion weak
decay constant $f_{\pi^{0}}^{(\mu)}$ and the directional quark-pion
coupling constant $g_{qq\pi^{0}}^{(\mu)}$.
\par
In Sec. \ref{sec4}, we then used the results arising from the
general treatment presented in Sec. \ref{subsec3-B} and determined
$g_{qq\pi^{0}}^{(\mu)}$ as well as $f_{\pi^{0}}^{(\mu)}$ at finite
$T$ and $eB$ first analytically in Sec. \ref{subsec4-A}, and then
numerically in Sec. \ref{subsec4-B}. As it turns out, the
longitudinal and transverse quark-pion coupling and weak decay
constants satisfy $g_{qq\pi^{0}}^{(0)}=g_{qq\pi^{0}}^{(3)}\neq
g_{qq\pi^{0}}^{(1)}=g_{qq\pi^{0}}^{(2)}$ and
$f_{\pi^{0}}^{(0)}=f_{\pi^{0}}^{(3)}\neq
f_{\pi^{0}}^{(1)}=f_{\pi^{0}}^{(2)}$. We determined numerically
their $T$ dependence for various $eB$. We showed that, whereas
$g_{qq\pi^{0}}^{(0)}> g_{qq\pi^{0}}^{(1)}$, we have
$f_{\pi^{0}}^{(0)}< f_{\pi^{0}}^{(1)}$. This is a direct consequence
of the fact that $1=u_{\pi^{0}}^{(0)}<u_{\pi^{0}}^{(1)}$, as it is
shown in Fig. \ref{fig4}. Moreover, for a fixed $\mu$ and $eB$,
whereas the mass of neutral pions increases with increasing $T$ [see
Fig. \ref{fig3}(b)], the directional decay constants of pions
decrease with $T$. At fixed $\mu$ and $T$, they increase with
increasing strength of the external magnetic field (see Fig.
\ref{fig11a}). As concerns the behavior of pions, directional decay
constants $f_{\pi^{0}}^{(\mu)}, \mu=0,\cdots,3$ near the chiral
transition point $T_{c}$, it turns out, that, independent of the
direction, they are almost constant but large at $T\ll T_{c}$, start
to decrease at $T\simeq T_{c}$ and remain constant but very small at
$T\gg T_{c}$. Moreover, the slope of this decrease as a function of
$T$ depends on the strength of the external magnetic field. The
behavior of $g_{qq\pi^{0}}^{(\mu)}$ and $f_{\pi^{0}}^{(\mu)}$ in the
external magnetic field and at finite $T$ is strongly related to the
behavior of the constituent quark mass $m$ at finite $T$ and $eB$.
This relationship is reflected in the low energy relations of
neutral pions, the GT and GOR relations. We not only derived these
relations analytically in Sec. \ref{subsec4-A}, but also verified
them numerically in Sec. \ref{subsec4-B}. The behavior of the
constituent quark mass $m$ at finite $T$ and $eB$ was discussed
extensively in our previous paper \cite{fayazbakhsh2012}. Comparing
Figs. \ref{fig1} and \ref{fig11a}(a)-\ref{fig11a}(c), it turns out
that $m$ and $f_{\pi^{0}}^{(\mu)}$ have similar $T$ dependence for a
fixed $eB$. The phenomenon of magnetic catalysis is reflected in the
fact that at fixed $\mu$ and $T$, they both increase with increasing
$eB$. As we have shown in Sec. \ref{sec3}, $f_{\pi^{0}}^{(\mu)}$ and
$m$ are proportional, and the proportionality factor is, according
to the GT relation (\ref{NE32c}), $1/g_{qq\pi^{0}}^{(\mu)}$. Let us
finally note that our results can be improved by improving the
approximation we have used to determine the kinetic coefficients to
higher order derivative expansion, or by making use of the
functional renormalization group method, which was recently used in
\cite{skokov2012}.
\begin{appendix}
\section*{Appendix: Useful relations}\label{appA}
\setcounter{section}{1} \setcounter{equation}{0} \par\noindent
In this appendix, we present a number of useful relations leading to the analytical results presented in Sec. \ref{sec4}. Let us start with the orthonormality
 relations of $f_{p}^{\pm s}$ appearing in (\ref{NA26a}),
\begin{eqnarray}\label{A1}
\int dx_{1}\ f^{+s}_{p}(x)f^{+s}_{k}(x)\bigg|_{p_2=k_2}&=&\delta_{pk},\nonumber\\
\int dx_1\ x_1\ f^{+s}_{p}(x)f^{+s}_{k}(x)\bigg|_{p_2=k_2}&=&\ell_B\bigg\{\Pi_k C_{k}\delta_{p,k-1}\nonumber\\
&&\hspace{-4.7cm}+\Pi_{k+1}C_{k+1}\delta_{p,k+1}+s\ell_{B}k_{2}\delta_{p,k}\bigg\},\nonumber\\
\int dx_1\ x_1^2\ f^{+s}_{p}(x)f^{+s}_{k}(x)\bigg|_{p_2=k_2}&&\nonumber\\
&&\hspace{-5cm}=\ell_B^2\bigg\{(2C_{2k+1}^{2}+\ell_B^2k_2^2)\delta_{pk}
+\Pi_{k}\Pi_{k-1}C_{k}C_{k-1} \delta_{p,k-2}\nonumber\\
&&\hspace{-4.7cm}+C_{k+1}C_{k+2}\delta_{p,k+2}+2\Pi_{k}C_{k}s\ell_{B}k_{2}\delta_{p,k-1}\nonumber\\
&&\hspace{-4.7cm}+2\Pi_{k+1}C_{k+1}s\ell_{B}k_{2}\delta_{p,k+1}\bigg\},\hspace{-0.3cm}
\end{eqnarray}
and
\begin{eqnarray}\label{A2}
\int dk_2\ f^{+s}_{p}(0)f^{+s}_{k}(0)\bigg|_{p_2=k_2}&=&
\frac{\delta_{pk}}{\ell_B^2},\nonumber\\
\int dk_2\ k_2\ f^{+s}_{p}(0)f^{+s}_{k}(0)\bigg|_{p_2=k_2}&=&-\frac{s}{\ell_B^3}\bigg\{C_{k+1}\delta_{p,k+1}\nonumber\\
&&\hspace{-4.7cm}+\Pi_{k}C_{k}\delta_{p,k-1}\bigg\},\nonumber\\
\int dk_2\ k_2^2\
f^{+s}_{p}(0)f^{+s}_{k}(0)\bigg|_{p_2=k_2}&=&\frac{1}{\ell_B^4}
\bigg\{2C_{2k+1}^{2}\delta_{pk}\nonumber\\
&&\hspace{-4.7cm}+\Pi_k\Pi_{k-1}
C_{k}C_{k-1}\delta_{p,k-2}+
C_{k+1}C_{k+2}\delta_{p,k+2}\bigg\}.\nonumber\\
\end{eqnarray}
Here, $C_{k}\equiv \sqrt{\frac{k}{2}}$ and $\ell_{B}\equiv
\frac{1}{\sqrt{|q_{f}eB|}}$. To derive these relations, we have used
\begin{eqnarray}\label{A3}
x_{1}f_{p}^{+s}(x)&=&\ell_{B}\bigg\{\Pi_{p+1}C_{p+1}f_{p+1}^{+s}(x)+\Pi_{p}C_{p}f_{p-1}^{+s}(x)\nonumber\\
&&+s\ell_{B}p_{2}f_{p}^{+s}(x)\bigg\},\nonumber\\
p_{2}f_{p}^{+s}(x)
&=&-\frac{s}{\ell_{B}}\bigg\{\Pi_{p+1}C_{p+1}f_{p+1}^{+s}(x)\nonumber\\
&&+\Pi_{p}C_{p}f_{p-1}^{+s}(0)\bigg\}.
\end{eqnarray}
Apart from (\ref{ND44}), we have
\begin{eqnarray}\label{A4}
\lefteqn{\hspace{-0.5cm}
\int dx_{1}dk_{2}x_{1}A_{pk}^{+(1)}\alpha_{pk}^{+}=\frac{1}{4\ell_{B}}\bigg\{2\Pi_{k}C_{k}\alpha_{k}
}\nonumber\\
&&-\Pi_{k}\delta_{p,k-1}\left(C_{k}+C_{k-1}\right)\nonumber\\
&&-\Pi_{k+1}\delta_{p,k+1}\left(C_{k+1}-C_{k}\right)\bigg\},
\end{eqnarray}
as well as
\begin{eqnarray}\label{A5}
\lefteqn{\hspace{-0.5cm}
\int dx_{1}dk_{2}x_{1}A_{pk}^{-(1)}\alpha_{pk}^{-}
}\nonumber\\
&&=-\frac{1}{4\ell_{B}}\bigg\{\Pi_{k}\delta_{p,k-1}\left(C_{k}-C_{k-1}\right)\nonumber\\
&&-\Pi_{k+1}\delta_{p,k+1}\left(C_{k+1}-C_{k}\right)\bigg\},
\end{eqnarray}
which can be used to derive (\ref{ND48}) as well as
\begin{eqnarray}\label{A6}
\lefteqn{\hspace{-0.5cm}
\int dx_{1}dk_{2}\frac{d}{dk_{2}}\bigg[\alpha_{pk}^{+}A_{pk}^{+(1)}+\alpha_{pk}^{-}A_{pk}^{-(1)}\bigg]\bigg|_{p_{2}=k_{2}}
}\nonumber\\
&=&-\frac{s}{2\ell_{B}^{2}}\big[\Pi_{k}\delta_{p,k-1}+\delta_{p,k-1}\big].
\end{eqnarray}
Similarly, we have
\begin{eqnarray}\label{A7}
\lefteqn{\hspace{-0.5cm}
\int dx_{1}dk_{2}\frac{d}{dk_{2}}\bigg[\alpha_{pk}^{+}A_{pk}^{+(0)}+\alpha_{pk}^{-}A_{pk}^{-(0)}\bigg]\bigg|_{p_{2}=k_{2}}
}\nonumber\\
&=&-\frac{s}{2\ell_{B}^{2}}\big[\Pi_{k}\delta_{p,k-1}+\delta_{p,k-1}\big].
\end{eqnarray}
Finally, we present the method leading to $g_{qq\pi^{0}}^{(1)}$ from
(\ref{ND25}) and to $f_{\pi^{0}}^{(1)}$ from (\ref{ND45}). The
relations leading to (\ref{ND25}) and (\ref{ND49}) have the
following general structure:
\begin{eqnarray}\label{A8}
I&\equiv& \sum\limits_{p,k=0}^{\infty}\int_{0}^{1}dx f(p,k,x)\nonumber\\
&&\times \big[\Pi_{k}\delta_{p,k-1}F(p,k)+\delta_{p,k+1}G(p,k)\big].
\end{eqnarray}
The aim is to show
\begin{eqnarray}\label{A9}
I=\sum\limits_{k=0}\int_{0}^{1}dx f(k+1,k,x)\big[F(k,k)+G(k+1,k)\big].\nonumber\\
\end{eqnarray}
Starting from (\ref{A8}) and summing over $p$, we arrive first at
\begin{eqnarray}\label{A10}
I&=&\sum\limits_{k=0}^{\infty}\int_{0}^{1}dx\big[\Pi_{k}f(k-1,k,x)F(k-1,k)\nonumber\\
&&+f(k+1,k,x)G(k+1,k)\big].
\end{eqnarray}
Changing the variables in the first term and using the property of $f(p,k,x)$
\begin{eqnarray*}
f(k,k+1,x)=f(k,k+1,1-x),
\end{eqnarray*}
we arrive at
\begin{eqnarray*}
I&=&\sum\limits_{k=0}^{\infty}\int_{0}^{1}dx\big[f(k,k+1,1-x)F(k)\nonumber\\
&&+f(k+1,k,x)G(k+1)\big].
\end{eqnarray*}
Then using the symmetry property
\begin{eqnarray*}
\hspace{-0.5cm}\int_{0}^{1}dx f(k+1,k,1-x)=\int_{0}^{1}dx
f(k+1,k,x),
\end{eqnarray*}
we arrive at (\ref{A9}). In all our examples in Sec. \ref{sec4},
$f(p,k,x)$ has the general form
\begin{eqnarray*}
f(p,k,x)=\frac{A+B\sqrt{pk}}{[C+x(1-x)m^{2}+D(px+k(1-x))]^{2}}.
\end{eqnarray*}
\end{appendix}

\end{document}